\documentclass[a4paper,11pt]{article}
\pdfoutput=1
\usepackage{jheppub}
\usepackage{graphicx}
\usepackage{subcaption}
\usepackage{dsfont}
\usepackage[dvipsnames]{xcolor}
\usepackage{etoolbox}
\usepackage[T1]{fontenc}
\usepackage[utf8]{inputenc}
\usepackage{amsmath}
\usepackage{amssymb}
\usepackage{bbm}
\usepackage{mathtools}
\usepackage{hyperref}
\usepackage{amsthm}
\usepackage{physics}
\usepackage{tensor}
\usepackage{marginnote}
\usepackage{comment}
\usepackage{appendix}

\usepackage{float}

\newcommand{\pd}{\partial}
\newcommand{\nn}{\nonumber\\}
\usepackage[english]{babel}
\usepackage{enumitem}  

\usepackage{color}
\definecolor{darkgreen}{rgb}{0,0.5,0}
\definecolor{darkblue}{rgb}{0,0,0.6}
\definecolor{purple}{rgb}{0.4,.2,0.7}

\usepackage{slashed}
\usepackage{leftidx}
\usepackage{esint}
\usepackage{amsfonts,bm}

\usepackage{tikz}
\usetikzlibrary{arrows.meta,decorations.markings,patterns}

\newcommand{\anyonstrip}{%
\begin{scope}[line width=0.9pt]
  \draw (0,-1.4) rectangle (1,1.4);
  \draw[-{Latex[length=2.2mm,width=2mm]}] (0.3,1.4) -- (0.62,1.4);
  \draw[-{Latex[length=2.2mm,width=2mm]}] (0.3,-1.4) -- (0.62,-1.4);
\end{scope}}

\newcommand{\anyonpair}[1]{%
\begin{scope}[line width=0.9pt]
  \draw (0,1.4) -- (0,-1.4);
  \draw[-{Latex[length=2.2mm,width=2mm]}] (0,0.6) -- (0,0.25);
  \node[above] at (0,1.4) {$#1$};
  \draw (1,1.4) -- (1,-1.4);
  \draw[-{Latex[length=2.2mm,width=2mm]}] (1,0.0) -- (1,0.35);
  \node[above] at (1,1.4) {$#1^{*}$};
  \draw[dashed] (0,-0.05) -- (1,-0.05);
\end{scope}}

\newcommand{\qMPStorusoutline}{%
  \draw (0,0) ellipse (3 and 1.7);
  \draw (-1.3,0.15) .. controls (-0.6,0.75) and (0.6,0.75) .. (1.3,0.15);
  \draw (-0.75,0.0) .. controls (-0.35,-0.4) and (0.35,-0.4) .. (0.75,0.0);
}

\newcommand{\qMPStorus}{%
\begin{scope}[scale=0.6]
  \qMPStorusoutline
  \draw[red, thick] (0,0.05) ellipse (2.35 and 1.15);
  \node[red, above] at (0,1.15) {\small $R$};
\end{scope}}

\newcommand{\qMPSnet}{%
\begin{scope}[scale=0.6]
  \draw (-2.5,-0.3) rectangle (2.5,0.3);
  \node at (0,0) {\small $R$};
  \draw (-2.5,0) -- (-2.9,0) to[out=180,in=180] (-2.9,1.5) -- (2.9,1.5)
        to[out=0,in=0] (2.9,0) -- (2.5,0);
  \draw (0,0.3) -- (0,0.9);
  \draw (0,-0.3) -- (0,-0.8);
  \draw (-2.5,-0.8) rectangle (2.5,-1.5);
  \node at (0,-1.15) {\small $\ket{0}_{T^{2}}$};
\end{scope}}

\newcommand{\qMPStoruscut}{%
\begin{scope}[scale=0.6]
  \qMPStorusoutline
  \begin{scope}[red, thick, shift={(0,0.05)}]
    \foreach \a/\b in {100/168, 192/250, 268/348, 12/80}
      \draw[domain=\a:\b, smooth, variable=\t, samples=30]
        plot ({2.35*cos(\t)}, {1.15*sin(\t)});
  \end{scope}
\end{scope}}

\newcommand{\qMPSfactnet}{%
\begin{scope}[scale=0.6]
  \foreach \x in {-2.1,-0.7,0.7,2.1} {
    \draw (\x-0.45,-0.3) rectangle (\x+0.45,0.3);
    \node at (\x,0) {\small $R$};
    \draw (\x,0.3) -- (\x,0.9);
    \draw (\x,-0.3) -- (\x,-0.8);
  }
  \foreach \x in {-1.65,-0.25,1.15}
    \draw (\x,0) -- (\x+0.5,0);
  \draw (-2.55,0) -- (-2.95,0) to[out=180,in=180] (-2.95,1.5) -- (2.95,1.5)
        to[out=0,in=90] (3.5,0.95) -- (3.5,0.8);
  \draw (3.05,0.2) rectangle (3.95,0.8);
  \node at (3.5,0.5) {\scriptsize $D_{R}$};
  \draw (3.5,0.2) to[out=-90,in=0] (2.95,0) -- (2.55,0);
  \draw (-2.55,-0.8) rectangle (2.55,-1.5);
\end{scope}}

\newcommand \mathtikz[1] {\quad \vcenter{\hbox{\tikz{#1}}} \quad}

\usetikzlibrary{shapes.geometric}
\tikzset{
  ribbon/.style={red!80!black, line width=0.6pt, double=white, double distance=2.2pt},
  marked/.style={circle, fill=black, inner sep=1.1pt},
  sphere/.pic={
    \draw[thick] (0,0) circle (2);
    \draw[thick] (-2,0) arc[start angle=180,end angle=360,x radius=2,y radius=0.45];
    \draw[thick,dashed] (2,0) arc[start angle=0,end angle=180,x radius=2,y radius=0.45];
  },
  upper cap/.pic={
    \draw[thick] (-2,0.3) arc[start angle=180,end angle=0,radius=2];
    \draw[thick] (-2,0.3) arc[start angle=180,end angle=360,x radius=2,y radius=0.45];
    \draw[thick,dashed] (2,0.3) arc[start angle=0,end angle=180,x radius=2,y radius=0.45];
  },
  lower cap/.pic={
    \draw[thick] (-2,-0.3) arc[start angle=180,end angle=360,radius=2];
    \draw[thick] (0,-0.3) ellipse (2 and 0.45);
  },
  hookarrow/.style={thick, {Hooks[right]}-{Stealth}}
}

\tikzset{
  marked/.style={circle, fill=black, inner sep=1.2pt},
  cilium/.style={thick, black},
  wilsonline/.style={red!85!black, thick},
  tbox/.style={draw, thick, minimum size=0.5cm, inner sep=1pt},
  tleg/.style={thick}
}

\newcommand{\tensorU}[3]{%
\vcenter{\hbox{\begin{tikzpicture}
  \node[tbox] (T) at (0,0) {$#1$};
  \draw[tleg] (T.north) -- ++(0,0.35);
  \draw[tleg] (T.south) -- ++(0,-0.35);
  \draw[tleg] (T.west) -- ++(-0.4,0) node[left] {$#2$};
  \draw[tleg] (T.east) -- ++(0.4,0) node[right] {$#3$};
\end{tikzpicture}}}}
\newcommand{\tensorUU}[5]{%
\vcenter{\hbox{\begin{tikzpicture}
  \node[tbox] (T1) at (0,0.4) {$#1$};
  \node[tbox] (T2) at (0,-0.4) {$#1$};
  \draw[tleg] (T1.north) -- ++(0,0.35);
  \draw[tleg] (T1.south) -- (T2.north);
  \draw[tleg] (T2.south) -- ++(0,-0.35);
  \draw[tleg] (T1.west) -- ++(-0.4,0) node[left] {$#2$};
  \draw[tleg] (T1.east) -- ++(0.4,0) node[right] {$#3$};
  \draw[tleg] (T2.west) -- ++(-0.4,0) node[left] {$#4$};
  \draw[tleg] (T2.east) -- ++(0.4,0) node[right] {$#5$};
\end{tikzpicture}}}}
\newcommand{\fatannulus}[2]{
\begin{scope}[shift={(#1,#2)}]
  \draw (0,0) circle (1.5);
  \draw (0,0) circle (0.5);
  \draw[red!75!black] (-0.07,-0.5) -- (-0.07,-1.5);
  \draw[red!75!black] (0.07,-0.5) -- (0.07,-1.5);
  \node[red!75!black,star,star points=5,star point ratio=2.25,fill,inner sep=1.3pt] at (0,-1.5) {};
  \node[red!75!black,star,star points=5,star point ratio=2.25,fill,inner sep=1.3pt] at (0,-0.5) {};
\end{scope}}

\renewcommand{\tensorU}[3]{%
\begin{scope}
  \node[tbox] (T) at (0,0) {$#1$};
  \draw[tleg] (T.north) -- ++(0,0.35);
  \draw[tleg] (T.south) -- ++(0,-0.35);
  \draw[tleg] (T.west) -- ++(-0.4,0) node[left] {$#2$};
  \draw[tleg] (T.east) -- ++(0.4,0) node[right] {$#3$};
\end{scope}}

\renewcommand{\tensorUU}[5]{%
\begin{scope}
  \node[tbox] (T1) at (0,0.4) {$#1$};
  \node[tbox] (T2) at (0,-0.4) {$#1$};
  \draw[tleg] (T1.north) -- ++(0,0.35);
  \draw[tleg] (T1.south) -- (T2.north);
  \draw[tleg] (T2.south) -- ++(0,-0.35);
  \draw[tleg] (T1.west) -- ++(-0.4,0) node[left] {$#2$};
  \draw[tleg] (T1.east) -- ++(0.4,0) node[right] {$#3$};
  \draw[tleg] (T2.west) -- ++(-0.4,0) node[left] {$#4$};
  \draw[tleg] (T2.east) -- ++(0.4,0) node[right] {$#5$};
\end{scope}}

\newcommand{\tensorUV}[4]{%
\begin{scope}
  \node[tbox] (T1) at (0,0) {$#1$};
  \node[tbox] (T2) at (1.3,0) {$#2$};
  \foreach \t in {T1,T2} {
    \draw[tleg] (\t.north) -- ++(0,0.35);
    \draw[tleg] (\t.south) -- ++(0,-0.35);
  }
  \draw[tleg] (T1.east) -- (T2.west);
  \draw[tleg] (T1.west) -- ++(-0.4,0) node[left] {$#3$};
  \draw[tleg] (T2.east) -- ++(0.4,0) node[right] {$#4$};
\end{scope}}

\renewcommand{\qMPSnet}{%
\begin{scope}[scale=0.6]
  \draw (-2.5,-0.3) rectangle (2.5,0.3);
  \node at (0,0) {\small $R$};
    \draw (-2.5,0) -- (-2.9,0) to[out=180,in=180] (-2.9,1.5) -- (2.9,1.5)
        to[out=0,in=90] (3.5,1.1) -- (3.5,0.95);
  \draw (2.95,0.1) rectangle (4.05,0.95);
  \node at (3.5,0.52) {\scriptsize $D_{R}^{-1}$};
  \draw (3.5,0.1) to[out=-90,in=0] (2.9,0) -- (2.5,0);
  \draw (0,0.3) -- (0,0.9);
  \draw (0,-0.3) -- (0,-0.9);
\end{scope}}

\renewcommand{\qMPSfactnet}{%
\begin{scope}[scale=0.6]
  \foreach \x in {-2.1,-0.7,0.7,2.1} {
    \draw (\x-0.45,-0.3) rectangle (\x+0.45,0.3);
    \node at (\x,0) {\small $R$};
        \draw (\x,0.3) -- (\x,0.9);
    \draw (\x,-0.3) -- (\x,-0.9);
  }
  \foreach \x in {-1.65,-0.25,1.15}
    \draw (\x,0) -- (\x+0.5,0);
  \draw (-2.55,0) -- (-2.95,0) to[out=180,in=180] (-2.95,1.5) -- (2.95,1.5)
        to[out=0,in=90] (3.5,1.1) -- (3.5,0.95);
    \draw (2.95,0.1) rectangle (4.05,0.95);
  \node at (3.5,0.52) {\scriptsize $D_{R}^{-1}$};
    \draw (3.5,0.1) to[out=-90,in=0] (2.95,0) -- (2.55,0);
\end{scope}}

\newcommand{\fatdisk}[3]{
\begin{scope}[shift={(#1,#2)}]
  \draw (0,0) circle (1.5);
  \fill (0,0.15) circle (1.6pt);
  \node[above] at (0,0.15) {\footnotesize $#3$};
  \draw[red!75!black] (-0.12,-1.5) .. controls (-1.6,-0.6) and (-1.3,1.25) .. (0,1.25)
                       .. controls (1.3,1.25) and (1.6,-0.6) .. (0.12,-1.5);
  \draw[red!75!black] (-0.05,-1.5) .. controls (-1.15,-0.6) and (-0.95,0.9) .. (0,0.9)
                       .. controls (0.95,0.9) and (1.15,-0.6) .. (0.05,-1.5);
  \node[red!75!black,star,star points=5,star point ratio=2.25,fill,inner sep=1.3pt] at (0,-1.5) {};
\end{scope}}

\newcommand\nablaAflip[2]{ 
\begin{scope}[xshift=#1,yshift=#2,yscale=-1]
\draw (-0.25,-0.5) -- (0.25,0) -- (0.75,0) -- (0.25,-0.5) --(0.25,-1) -- (-0.25,-1) -- (-0.25,-0.5) -- (-0.75,0) -- (-0.25,0) -- (0,-0.25);
\draw[dashed] (-0.25,-0.5) -- (0.25,-0.5);
\end{scope}
}

\newcommand\idA[2] { 
\begin{scope}[xshift=#1,yshift=#2]
\filldraw[fill=white,draw=black] (-0.25,0) rectangle (0.25,-1);
\end{scope}
}

\newcommand\muA[2]{ 
\begin{scope}[xshift=#1,yshift=#2]
\draw (-0.75,0) -- (-0.25,0) to [out=-90,in=180] (0,-0.33) to [in=-90,out=0] (0.25,0) -- (0.75,0) to [in=90,out=-90] (0.25,-1);
\draw (-0.25,-1) -- (0.25,-1);
\draw (-0.75,0) to [in=90,out=-90] (-0.25,-1);
\end{scope}
}

\newcommand\pairA[2]{ 
\begin{scope}[xshift=#1,yshift=#2]
\draw (-0.75,0) -- (-0.25,0) to [out=-90,in=180] (0,-0.33) to [in=-90,out=0] (0.25,0) -- (0.75,0) to [out=-90,in=0] (0,-0.83) to [out=180,in=-90] (-0.75,0);
\end{scope}
}

\newcommand\pairSAnyons[2]{ 
\begin{scope}[xshift=#1,yshift=#2]
  \draw 
      (-0.75,0)
      to[out=-90,in=180] (0,-0.83)
      to[out=0,in=-90] (0.75,0);
\end{scope}
}

\newcommand\copairSAnyons[2]{ 
  \begin{scope}[xshift=#1,yshift=#2]
\draw (-0.75,0)  to [out=90,in=180]  (0,0.83) to [in=90,out=0] (0.75,0)  ;
\end{scope}
}

\newcommand\copairA[2]{ 
\begin{scope}[xshift=#1,yshift=#2]
\draw (-0.75,0) -- (-0.25,0) to [out=90,in=180] (0,0.33) to [in=90,out=0] (0.25,0) -- (0.75,0) to [out=90,in=0] (0,0.83) to [out=180,in=90] (-0.75,0);
\end{scope}
}

\newcommand\deltaA[2]{ 
\begin{scope}[xshift=#1,yshift=#2]
\draw (-0.75,-1) -- (-0.25,-1) to [out=90,in=180] (0,-0.66) to [in=90,out=0] (0.25,-1) -- (0.75,-1) to [in=-90,out=90] (0.25,0) -- (-0.25,0) to [in=90,out=-90] (-0.75,-1);
\end{scope}
}

\newcommand\zipper[2]{ 
\begin{scope}[xshift=#1,yshift=#2]
\draw (-0.25,-1) -- (0.25,-1);
\filldraw[right color=white,left color=lightgray] (-0.25,0) to (-0.25,-1) to [out=90,in=225] (0,-0.5) to [out=-45,in=90] (0.25,-1) to (0.25,0);
\filldraw[left color=white,right color=lightgray] (0,0) ellipse (0.25 and 0.1);
\end{scope}
}

\newcommand\cozipper[2]{ 
\begin{scope}[xshift=#1,yshift=#2]
\draw (-0.25,0) -- (0.25,-0);
\filldraw[right color=white,left color=lightgray] (-0.25,-1) to (-0.25,0) to [out=-90,in=135] (0,-0.5) to [out=45,in=-90] (0.25,0) to (0.25,-1) to [in=-90,out=-90] (-0.25,-1);
\draw[dotted] (0.25,-1) arc (0:180:0.25 and 0.1);
\end{scope}
}

\newcommand\epsilonC[2]{ 
\begin{scope}[xshift=#1,yshift=#2]
\filldraw[right color=white,left color=lightgray] (-0.25,0) to [out=-90,in=180] (0,-0.33) to [in=-90,out=0] (0.25,0);
\filldraw[left color=white,right color=lightgray] (0,0) ellipse (0.25 and 0.1);
\end{scope}
}

\newcommand\etaC[2] { 
\begin{scope}[xshift=#1,yshift=#2]
\filldraw[right color=white,left color=lightgray] (-0.25,0) to [out=90,in=180] (0,0.33) to [in=90,out=0] (0.25,0) to [in=-90,out=-90] (-0.25,0);
\draw[dotted] (0.25,0) arc (0:180:0.25 and 0.1);
\end{scope}
}

\newcommand\epsilonA[2] {
\begin{scope}[xshift=#1,yshift=#2]
\draw (-0.25,0) -- (0.25,0);
\draw (-0.25,0) to [out=-90,in=180] (0,-0.33) to [in=-90,out=0] (0.25,0);
\end{scope}
}

\newcommand\etaAX[2] {
\begin{scope}[xshift=#1,yshift=#2]
\draw (-0.25,0) -- (0.25,0);
\draw (-0.25,0) to [out=90,in=180] (0,0.33) to [in=90,out=0] (0.25,0);
\node at (0,0.16) {\footnotesize $\times  $};
\end{scope}
}

\newcommand\etaA[2] {
\begin{scope}[xshift=#1,yshift=#2]
\draw (-0.25,0) -- (0.25,0);
\draw (-0.25,0) to [out=90,in=180] (0,0.33) to [in=90,out=0] (0.25,0);
\end{scope}
}

\newcommand\rightbraidA[2]{ 
  \begin{scope}[xshift=#1,yshift=#2]
    \draw (-0.75,0) -- (-0.25,0) 
      to [out=-90,in=90,looseness=0.5] (0.75,-1) -- (0.25,-1) 
      to [out=90,in=-90,looseness=0.5] (-0.75,0);

    \filldraw[fill=white,draw=black] (0.75,0) -- (0.25,0) 
      to [out=-90,in=90,looseness=0.5] (-0.75,-1) -- (-0.25,-1) 
      to [out=90,in=-90,looseness=0.5] (0.75,0);
  \end{scope}
}

\newcommand\leftbraidA[2]{ 
  \begin{scope}[xshift=#1,yshift=#2]
    \draw (0.75,0) -- (0.25,0) 
      to [out=-90,in=90,looseness=0.5] (-0.75,-1) -- (-0.25,-1) 
      to [out=90,in=-90,looseness=0.5] (0.75,0);

    \filldraw[fill=white,draw=black] (-0.75,0) -- (-0.25,0) 
      to [out=-90,in=90,looseness=0.5] (0.75,-1) -- (0.25,-1) 
      to [out=90,in=-90,looseness=0.5] (-0.75,0);
  \end{scope}
}

\newcommand\twistA[2]{ 
  \begin{scope}[xshift=#1,yshift=#2]
    \draw (-0.25,0) to [out=-90,in=90] (0.25,-1);
    
    \draw[line width=2mm, white] (0.25,0) to [out=-90,in=90] (-0.25,-1);
    
    \draw (-0.25,0) -- (0.25,0) to [out=-90,in=90] (-0.25,-1) -- (0.25,-1);
  \end{scope}
}

\usepackage{feynmf}
\usepackage{braket}

\usepackage{prettyref}

\newcommand{\amat} {\begin{pmatrix} 1 & 0 \\ 0 & 0 \end{pmatrix}}
\newcommand{\bmat} {\begin{pmatrix} 0 & 1 \\ 0 & 0 \end{pmatrix}}
\newcommand{\cmat} {\begin{pmatrix} 0 & 0 \\ 1 & 0 \end{pmatrix}}
\newcommand{\emat} {\begin{pmatrix} 0 & 0 \\ 0 & 1 \end{pmatrix}}

\numberwithin{equation}{section}
\numberwithin{figure}{section}
\numberwithin{table}{section}

\title{\Large \bf  Topological subregions in Chern Simons theory and topological string theory  \\[.7cm] }
\author{Gabriel Wong}
\affiliation{Mathematical Institute, University of Oxford, Andrew Wiles Building, Radcliffe Observatory Quarter, Woodstock Road, Oxford, OX2 6GG, U.K.}
\emailAdd{gabrielwon@gmail.com}

\date{Created: May 30 2025 (Last updated: \today)}

\abstract{ The standard, gapless entanglement boundary condition in Chern--Simons theory breaks the topological invariance of the theory by introducing a complex structure on the entangling surface.  This produces an infinite-dimensional subregion Hilbert space, a non-trivial modular Hamiltonian, and a UV-divergent entanglement entropy that is a universal feature of local quantum field theories.  In this work, we appeal to the combinatorial quantization of Chern--Simons theory to define a purely topological notion of a subregion.   The subregion operator algebras are spaces of functions on a quantum group.  We develop a diagrammatic calculus for the associated $q$-deformed entanglement entropy, which arises from the entanglement of anyonic edge modes.   The $q$-deformation regulates the divergences of the QFT, producing a finite entanglement entropy associated to a $q$-tracial state.   We explain how these ideas provide an operator algebraic framework for the entanglement entropy computations in topological string theory \cite{Donnelly:2020teo,Jiang:2020cqo,Wong:2026ftn} where a large-$N$ limit of the $q$-deformed subregion algebra plays a key role in the stringy description of spacetime.

 }

\begin{document}

\maketitle


\tableofcontents


\section{Introduction}

Defining subregions is subtle in theories with diffeomorphism symmetry as a gauge symmetry. 
This is particularly acute in gravity, where the location of the entangling surface itself fails to be diff-invariant. 
Topological field theories provide a simplified arena to explore this issue.   Indeed, pure gravity in 3D can be formulated as a Chern--Simons-like topological field theory \cite{Witten:1988hc,Achucarro:1986vz,Verlinde:1989ua,Collier:2023fwi}, 
while topological string theory arises from the large-$N$ limit of $U(N)$ Chern--Simons theory \cite{GopakumarVafa1998}.  

In this work, we revisit the question of subsystems in $U(N)$ Chern--Simons theory as a springboard for understanding subregions in low-dimensional quantum gravity and string theory. The standard Chern Simons entanglement boundary condition breaks diffeomorphism invariance along the entangling surface, which leads to an infinite entanglement entropy.   
In contrast, we will define a \emph{purely topological subregion} using the combinatorial quantization of Chern--Simons theory.  The associated subregion operator algebra produces a finite entropy
\footnote{Similar subregions were considered in \cite{Mertens:2025ydx}, for the case of annuli with one topological and one gapless boundary.}. 
We will focus on elements that are relevant to the computations of entanglement entropy in topological string theory, which was initiated in \cite{Donnelly:2020teo,Jiang:2020cqo}, and will be expanded upon in \cite{Wong:2026ftn}.  Our definition of subregion algebras also applies to q-deformed 2DYM, which provides a non-perturbative definition of topological strings \cite{Aganagic:2004js}.

\subsection*{The factorization problem }
Our work is best understood in the broader context of Hilbert space factorization in QFT.   In a local QFT, the Hilbert space
fails to factorize into subregions because the vacuum is infinitely entangled, due to the presence of entanglement at all length scales.  To disentangle the vacuum into a product state  would cost infinite energy, and such states are excluded from the physical Hilbert space.

In a theory with diffeomorphisms as a gauge symmetry, there are no local operators because the insertion point is not diff invariant.  This means there cannot be entanglement up to arbitrary small scales.  Thus, we should be able to avoid the UV divergent entanglement in local QFTs, allowing the Hilbert space to factorize.  Indeed, for non-chiral Chern Simons theory one could impose topological, gapped boundary conditions that avoid UV infinities.  But then the entanglement entropy is zero, because there are simply no subregion degrees of freedom to entangle!  This misses the long range entanglement present in topological phases described by Chern Simons theory.

The problem can be understood in terms of the ``shrinkability'' of the entanglement boundary condition.  In any QFT, given a UV regulator that separates two subregions $V,\bar{V} \subset\Sigma $ by a thin corridor of size $\epsilon$, one can define a factorization map: 
\begin{align}
i_{\epsilon}: \mathcal{H}_{\Sigma}   \to \mathcal{H}_{V}^{\epsilon} \otimes \mathcal{H}_{\bar{V} }^{\epsilon} 
\end{align}
from which we can compute a UV regulated EE. However, 
this must satisfy the isometry condition
\begin{align}
\lim_{\epsilon \to 0} i_{\epsilon}^{\dagger}i_{\epsilon}  = 1, 
\end{align}
in order for the correlation functions to be preserved in the factorized state.   

In a generic QFT,  we can only achieve isometry up to an infinite subtraction that is independent of subsystem size  \cite{Ball:2024hqe}.  Nevertheless, this still imposes a strong constraint: it can be interpreted in the path integral language, where  $i_{\epsilon}$ is
implemented by a Euclidean evolution that cuts open the spatial slice: 
\begin{align} \label{factor}
\includegraphics[scale=.2]{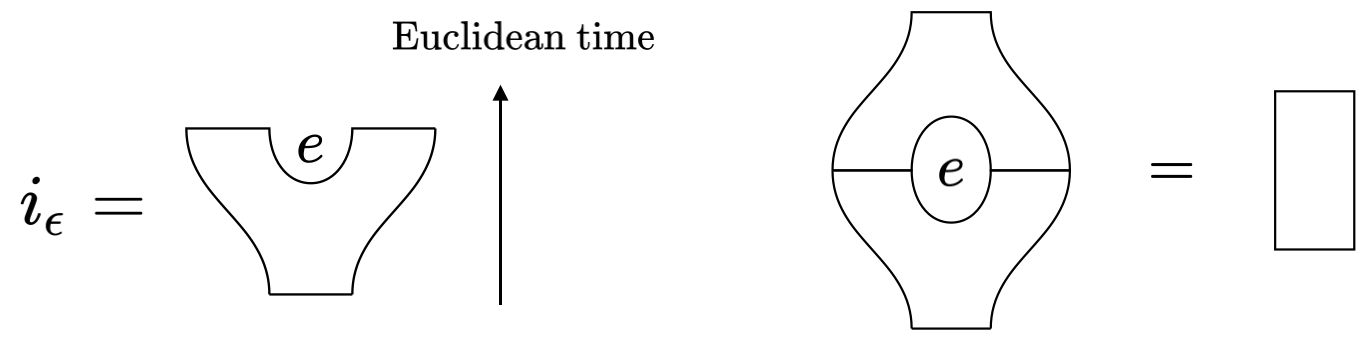}
\end{align}
The isometry condition implies that the hole created by the splitting and fusion of the spatial slice can be removed in the limit $\epsilon \to 0$.  This defines the shrinkable boundary condition\footnote{ This boundary condition has been determined in various settings \cite{Ball:2024hqe,Donnelly:2016jet,Donnelly:2018ppr},
and given a rigorous mathematical formulation as a ``cloaking'' boundary condition in conformal field theories \cite{Brehm:2024zun,Chen:2024unp,Hung:2024gma, Cheng:2023kxh,Brehm:2021wev,Hung:2019bnq} }  $e$  on the  stretched entangling surface.  
\paragraph{Local edge modes in Chern Simons theory }
The gapped boundary condition in Chern Simons theory gave a trivial EE because it fails to be even approximately shrinkable in the sense described above.  On the other hand,  Chern Simons does have a gapless boundary condition that is shrinkable up to an infinite subtraction  \cite{Wong:2017pdm}. The infinity arises because diffeomorphisms along the entangling surface are broken, leading to local degrees of freedom on the surface.  These edge modes are a crucial holographic feature of the shrinkable boundary condition, and for Chern Simons theory their entanglement produces all of the EE.
 For a single connected entangling surface of length $l$, this can be expressed in terms of the modular S matrix element $S_{00}$ and the quantum dimension $\dim_{q}R$ of a Wilson line crossing the entangling surface \cite{KitaevPreskill2006,LevinWen2006}: 
\begin{align} \label{topS}
    S = \frac{l}{\epsilon} + \log S_{00} + \log \dim_{q} R 
\end{align}
The leading area term proportional to $l$,  which appears due to the breaking of topological invariance, is an important feature of gapped topological phases. Thus, the diffeomorphism breaking boundary condition is appropriate when Chern Simons theory is viewed as the low energy limit of such systems.


\subsection*{Finite entropy, quantum group edge modes, and tensor networks}
 Our work aims to provide an intrinsic notion of a subregion in Chern Simons theory.   We propose an \emph{exactly shrinkable} factorization map that preserves topological invariance 
and yields a finite entropy capturing the long range entanglement of anyonic excitations:
\begin{align} \label{logdimR}
S = \log \dim_q R \, .
\end{align}
Finiteness is achieved without introducing a rigid UV cutoff: the regularization is intrinsic, arising from replacing classical holonomies by 
\emph{quantum group holonomies}—the fundamental observables in the combinatorial quantization of Chern--Simons theory 
\cite{Alekseev:1994pa,Alekseev:1994au}.  
The resulting subregion Hilbert space is defined via a quantum-group analogue of the GNS construction, producing a q deformed version of  $L^2(G)$.
Its edge modes transform under a quantum group symmetry, and their entanglement yields the $\log \dim_q R$ contribution.
This $q$-deformation acts as a \emph{topological} UV regulator that closely parallels the role of gravitational fluctuations that render entanglement entropy finite \cite{wonglocal} in 3d gravity.
Indeed, the deformation parameter $q$ in 3d gravity is determined by the gravitational coupling.  Moreover we will see that the $q$ deformation in Chern Simons theory can be motivated by a nonlocal condition at the entangling surface, reminiscent of those encountered in quantum gravity 
\cite{Jafferis:2019wkd}.  

Our factorization map provides a tensor network simulation of quantum states in Chern Simons theory.  As we explain in \ref{sec:tensor}, iterating the factorization map on an element of the Chern Simons Hilbert space  produces a matrix product state (MPS) whose virtual edges support the Hilbert space of quantum group edge modes.   These are finite dimensional  in contrast to the Kac--Moody edge modes associated to a gapless boundary condition.    We emphasize that the MPS is composed of ordinary, finite dimensional tensors:  the generically non-integer quantum dimension in \eqref{logdimR} arises from modifying the trace on the edge mode Hilbert space.
\subsection*{Quantum trace and the q-tracial state}

Our construction incorporates an algebraic fusion rule that glues together subregions, providing a toy model for emergent spacetime built from the entanglement of anyons 
\cite{Mertens:2022ujr,Wong:2022eiu}.   The entangling of subregions can be described in terms of spacetime ribbon diagrams, which describe the representation category of the quantum group $U(N)_q$.  In this context, 
 entanglement entropy is defined with respect to the categorical trace, also known as the \emph{quantum} trace \cite{Bonderson:2007ci,Bonderson:2017osr, Couvreur_2017}.   We will identify exact shrinkability with the $q$-tracial property of a Hartle Hawking state,  produced by the hemisphere in the diagrammatic relation below:
\begin{align}\label{eshrink}
 \mathtikz{ \epsilonC{0cm}{0cm} \etaC{0cm}{0cm} }=
    \mathtikz{ \pairA{0cm}{0cm} \copairA{0cm}{0cm} } 
\end{align}
The LHS describes the norm of a $q$ tracial state, which is exactly equal to a quantum trace on an interval Hilbert space given by the RHS.  While this is reminiscent of the path integral shrinkability described earlier,  it is now defined purely in terms of category theory.   The large N limit of this quantum trace  plays a crucial role in \cite{Wong:2026ftn,Donnelly:2020teo, Jiang:2020cqo} where it captures the backreaction of stringy edge modes, and implements the geometric transition of entanglement branes \cite{wonglocal}.  In this context, \eqref{eshrink} becomes an open-closed string duality.  Our work provides an operator-algebraic perspective on this duality, which forms the basis for the EE calculations in \cite{Donnelly:2020teo,Jiang:2020cqo,Wong:2026ftn}.
To provide a physical interpretation to these computations, we explain the operational meaning of the entanglement entropy defined with respect to the categorical trace, closely following \cite{Kato:2013ava}.

\section{Shrinkable boundary condition for Chern Simons theory} \label{sec:shrinkable}
Consider level $k$ Chern--Simons theory with gauge group $U(N)$, with bulk Lagrangian
\begin{equation}\label{lag}
    L = \frac{k}{4 \pi} \tr \Big(A\, dA + \frac{2}{3} A^3\Big).
\end{equation}
The associated quantum group parameter is
\begin{align}\label{qdef}
    q=\exp\Big(\frac{2 \pi i}{k+N}\Big).
\end{align}
In some arguments, in particular those concerning positivity of the Haar state on a subregion \ref{sec:factorization}, we analytically continue to real $q$, where the quantum group is a compact matrix quantum group in the sense of \cite{woronowicz1987compact}. We will indicate explicitly when this continuation is used.

We begin by deriving the exact, shrinkable boundary state from the path integral quantization of Chern Simons theory.  We will see that the resulting boundary condition is non-local, and produces a density matrix that violates the $U(N)$ gauge symmetry. 
 This problem will motivate us to appeal to the combinatorial quantization \cite{Alekseev:1994pa,Alekseev:1994au} of Chern Simons theory, which resolves this problem by q-deforming the gauge symmetry and introduces quantum group algebras into a subregion.  We explain how shrinkability is formally restored by the quantum trace on the algebra of q-deformed holonomies. 

\subsection{A holonomy constraint on the shrinkable boundary } \label{shrink}

In 2+1 D, each connected component of the entangling surface has the topology of a circle. The shrinkable boundary condition imposes a nontrivial holonomy around the cycle that links with this circle. 
This can be illustrated in the simplest case of a state $\ket{B^3,R}$ on $S^2$ prepared by the path integral on a 3-ball $B^3$, with a Wilson line in the representation $R$. 
\begin{align}\label{B3state}
\ket{B^{3},R} \;=\;
\vcenter{\hbox{%
\begin{tikzpicture}[scale=1]
  \draw[thick] (0,0) circle (1);
  \draw[dashed] (1,0) arc (0:180:1 and 0.3);
  \draw (-1,0) arc (180:360:1 and 0.3);
  \draw[wilsonline] (0,-0.75) -- (0,0.75);
  \fill[red!85!black] (0,0.75) circle (1.4pt);
  \fill[red!85!black] (0,-0.75) circle (1.4pt);
  \node[red!85!black] at (0.28,0.78) {$R$};
  \node[red!85!black] at (0.28,-0.78) {$\bar R$};
\end{tikzpicture}}}
\;\in\; \mathcal{H}_{S^{2}}(R,\bar R).
\end{align}
The entangling surface is the equator.  As in \eqref{factor}, we define a factorization map $i_{\epsilon}$ by path integral evolution that splits $S^2$ into two hemispheres.  This produces a  state obtained from the path integral on a 3-ball with a toroidal neighborhood of size $\epsilon$ removed:
\begin{align}\label{B3factor}
i_{\epsilon}\ket{B^{3},R} \;=\;
\vcenter{\hbox{%
\begin{tikzpicture}[scale=1.3]
  \def\r{0.14}
  \draw[thick] (8.03:1) arc (8.03:171.97:1);
  \draw[thick] (188.03:1) arc (188.03:351.97:1);
  \fill[pattern=north east lines,pattern color=black!45]
       (8.03:1) arc[start angle=94.01,end angle=265.99,radius=\r] arc[start angle=-8.03,end angle=8.03,radius=1] -- cycle;
  \fill[pattern=north east lines,pattern color=black!45]
       (171.97:1) arc[start angle=85.99,end angle=-85.99,radius=\r] arc[start angle=188.03,end angle=171.97,radius=1] -- cycle;
  \draw[gray,densely dotted] (-8.03:1) arc (-8.03:8.03:1) (171.97:1) arc (171.97:188.03:1);
  \draw[thick] (8.03:1) arc[start angle=94.01,end angle=265.99,radius=\r];
  \draw[thick] (171.97:1) arc[start angle=85.99,end angle=-85.99,radius=\r];
  \draw[dashed] (0.990,0.1397) arc (0:180:0.990 and 0.28);
  \draw (-0.990,0.1397) arc (180:360:0.990 and 0.28);
  \draw[dashed] (0.990,-0.1397) arc (0:180:0.990 and 0.28);
  \draw (-0.990,-0.1397) arc (180:360:0.990 and 0.28);
  \draw[{Stealth[length=1.2mm]}-{Stealth[length=1.2mm]}] (1.3,0.1397) -- (1.3,-0.1397);
  \node[right] at (1.32,0) {$\epsilon$};
  \draw[wilsonline] (0,-0.75) -- (0,0.75);
  \fill[red!85!black] (0,0.75) circle (1.4pt);
  \fill[red!85!black] (0,-0.75) circle (1.4pt);
  \node[red!85!black] at (0.25,0.78) {$R$};
  \node[red!85!black] at (0.25,-0.78) {$\bar R$};
  \node at (-0.95,0.85) {$V$};
  \node at (-0.95,-0.85) {$\bar V$};
\end{tikzpicture}}}
\;\in\; \mathcal{H}_{V}\otimes\mathcal{H}_{\bar V}.
\end{align}
where $\mathcal{H}_{V}$ is the Hilbert space supported on the punctured disk. 

Since gluing two $B^3$ makes $S^3$, the norm  of the factorized state is a path integral on $S^3$  with a solid torus $\mathcal{T}_{\epsilon}$ removed:
\begin{align}\label{normB3}
\braket{B^{3},R|i^{\dagger}_{\epsilon}i_{\epsilon}|B^{3},R} \;
&=\vcenter{\hbox{\includegraphics[scale=.6]{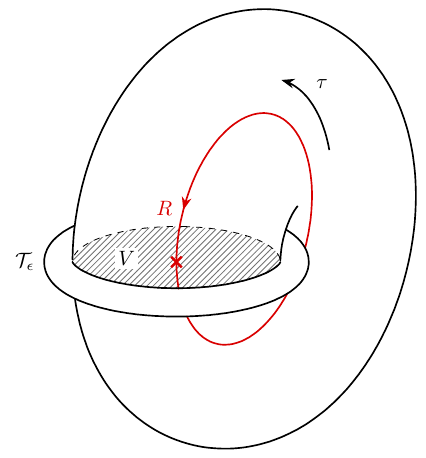}}}=Z(S^3,R)
\end{align}
The two tori drawn are the ones appearing in the Heegaard splitting of $S^3$.
The last equality follows from imposing an exactly shrinkable boundary condition,  so that the path integral on the $V \times S^1$ reproduces the norm of the original unfactorized state.

We can probe this boundary condition by performing the Chern Simons path integral on $\mathcal{T}$, which produces a state $\ket{\mathcal{T}}$ on its boundary.   The wavefunction for this state provides gauge invariant information about the holonomy $U=P \exp \oint A$ around  $\tau$.  Specifically, the conjugacy class of $U$ makes up a configuration space variable on the torus $\pd \mathcal{T}$.  Denoting a conjugacy class basis by $\ket{U}$ we can learn about $U$ by studying the overlap $\braket{U|\mathcal{T}}$.
This wavefunction satisfies two important properties:
\begin{enumerate}
    \item It is independent of $\epsilon$, since the boundary condition obtained from an exact path integral over $\mathcal{T}_{\epsilon}$ is topological.  Thus, the factorization map and resulting EE are also $\epsilon$ independent, in contrast to the gapless QFT boundary condition. We omit the $\epsilon$ label from now on.
    \item It is supported on a single conjugacy class of a non-trivial holonomy around the contractible cycle, i.e.  along the  modular time $\tau$.
\end{enumerate}
Let us elaborate on the second statement.  
Classically, the $\tau$ circle bounds a disk in $\mathcal{T}$ with no Wilson line punctures, so the Gauss law $F=0$ sets the holonomy to the identity. Quantum mechanically, we would expect a general gauge theory path integral over a solid torus to describe a superposition of states with different holonomies.  
Remarkably, in the case of Chern Simons theory, the torus wavefunction is localized on the conjugacy class of a single holonomy, but it is  not equal to the identity.  
Indeed, \cite{Elitzur:1989nr} showed that one-loop effects shift the holonomy around a Wilson line in  the $\lambda$ representation to
\begin{align}\label{Ulambda}
    U_{\lambda}=\exp\big(\tfrac{2\pi i(\lambda+\rho) }{k+N}\big)
\end{align} where $\lambda$ in the exponent refers to the highest weight vector of the representation, and $\rho$ is the Weyl vector\footnote{$\lambda $ in this formula is a linear combination  $\lambda= \sum_{i} \lambda_{i} h_{i} $ of Cartan generators $h_{i}$}.  Due to the shift by $\rho$, the holonomy in the $\tau$ circle is nontrivial even though it links with the trivial Wilson loop.  Up to a conjugation, we must have 
\begin{align}\label{Dcons}
    P \exp \oint A_{\tau} d\tau &= D
\end{align}
where $D$ is the so-called balancing element of $U_{q}(\mathfrak{u}(N))$:
\begin{align} 
    D&= q^{\rho},\qquad  D_{ij}=\delta_{ij}\,q^{-i+\frac{N+1}{2}}, \qquad i=1,\dots,N .
\end{align}
When evaluated on the representation $\lambda$ of $\mathfrak{u}(N)$, this matrix produces the quantum dimensions for $\lambda$:
\begin{align}
\tr_{\lambda} (D) = \dim_{q} \lambda 
\end{align}
The distinguished holonomy $D$ carries a deeper significance beyond this formula.  Mathematically, it is defined in terms of the braiding data of a quantum group, thus hinting at the need for quantum group algebras  in defining a topological subregion.    In the context of topological string theory, the large N limit of this holonomy represents moduli data that determines the configuration of D branes at a shrinkable boundary \cite{Donnelly:2020teo}.  In this work,  we will focus on its role in defining a quantum trace on the operator algebra of a topological subregion.
In the most pedestrian terms, this trace can be defined on a finite dimensional matrix $O$ by absorbing $D$ into the ordinary trace. 
\begin{align}\label{qmatrixtr}
\widetilde{\tr}_{\lambda} (O)= \tr_{\lambda} D O
\end{align} 
With this definition, the quantum dimension is just the quantum trace of the identity matrix in the representation $\lambda$.   For matrix representations of quantum groups, \eqref{qmatrixtr} provides an operational definition of the quantum trace.

\paragraph{The exact entanglement boundary state}
 The results of \cite{Elitzur:1989nr} imply that the wavefunction for the boundary state $\ket{\mathcal{T}}$ is proportional to a delta function 
\begin{align}
\braket{U|\mathcal{T}} =  \delta (U,D),
\end{align}
which is defined on a  \emph{finite} set of conjugacy classes that form a basis for the Hilbert space on  $\pd \mathcal{T}$.   The exact expression will be of interest. 
The easiest way to obtain this is to observe that, in terms of the representation basis $\ket{\lambda_{b}}$ labelled by Wilson  in the non contractible $b$ cycle of   $\mathcal{T}$, $\ket{\mathcal{T}}= \ket{0_{b}}$ is the state with the trivial Wilson loop.  We can then expand  this in the representation basis in the dual a-cycle  using the modular S-matrix elements $S_{0\lambda}$. 
This gives the wavefunction 
\begin{align}\label{cons}
\braket{U_{a}|\mathcal{T}} &= \sum_{\lambda} S_{0\lambda}   \braket{U_{a}|\lambda_{a}} 
\nn  
&= S_{00}\sum_{\lambda } \dim_{q}(\lambda)\tr_{\lambda}(U) ,
\end{align}
where we used the relation $S_{0\lambda}= S_{00} \dim_{q}\lambda$, $\braket{U_{a}|\lambda_{a}}= \tr_{\lambda}(U)$, and the sum over $\lambda$ is over a finite number of integrable representations. Since $\tr_{\lambda}(D) = \dim_{q}\lambda$ the last expression can be interpreted as a delta function expressing the completeness of characters: 
\begin{align}
    \delta(U',U)=S_{00} \sum_{ \lambda} \tr_{\lambda} (U')  \tr_{\lambda}(U)
\end{align}
If the sum over $\lambda$ ranged over all representations of $U(N)$, this is a delta function on an infinite dimensional space of conjugacy classes.   But due to the truncation to integrable representations, the characters span a finite dimensional space, and $S_{00}= \frac{1}{\sum_{\lambda} ( \dim_{q}\lambda )^2}$ plays the role of an inverse  volume factor normalizing the delta function. 

\subsection{Naive quantization of a subregion }
We now attempt a standard canonical quantization of Chern Simons theory on $V$ that incorporates the constraint \eqref{Dcons}.  We will find that this approach is inadequate, but it is instructive to understand why it fails. 

From the point of view of the subregion theory, the $\tau$-circle  defines a Euclidean thermal cycle.  Let us fix this length to be $\beta$ and quantize the 
Chern Simons theory action  on $M(\beta)=V \times S^{1}(\beta)$.  The Euclidean bulk action and its variation are given by 
\begin{align}
  S_{\text{bulk}}&=  \frac{k}{4 \pi} \int_{M(\beta)}\tr \Big(A\wedge  dA + \frac{2}{3} A\wedge A \wedge A\Big) , \nn
    \delta S_{\text{bulk}} &= \frac{k}{2\pi}\int_{M(\beta) } \tr  (\delta A\wedge  F) + \frac{k}{4\pi}\int_{\partial V \times S^{1}(\beta) } d\tau\, d\phi\;\tr (A_{\tau}\, \delta A_{\phi } -  A_{\phi}\, \delta A_{\tau}) .
\end{align}
The variational principle for $S_{bulk}$ admits the gapless boundary condition $A_{\tau}=0$ \cite{Elitzur:1989nr}, or more commonly, $A_{\tau}=vA_{\phi}$. Both cases lead to a boundary Hilbert space on $\pd V$  given by the infinite dimensional Kac--Moody module specified by the puncture in $\pd V$ , with the boundary current $J=A_{\phi}|_{\pd V}$ capturing the edge mode degrees of freedom.  However, neither is consistent with constraint \eqref{Dcons}.   Instead, a boundary condition that satisfies the constraint is:
\begin{align}\label{local}
    A_{\tau}\big|_{\partial V} = \frac{1}{\beta}  \log D \equiv \frac{1}{\beta}\,\mu, \qquad \mu =\mathrm{Diag}\{x_1,\dots,x_N\}, 
\end{align}
where 
\begin{align}
    x_{i}   = \frac{2\pi i}{k+N} \left( -i + \frac{N +1}{2}\right) 
\end{align}

This boundary condition is compatible with the variational principle only if we add the boundary term
\begin{align}
S_{\partial} =-\frac{k}{4\pi}\int_{\partial V \times S^{1}(\beta) }d\tau\,d\phi\;\tr (A_{\tau} A_{\phi } ),
\end{align}
after which the boundary variation reduces to $-\frac{k}{2\pi}\int\tr(A_\phi\,\delta A_\tau)$, which vanishes for fixed $A_\tau$. The subregion action is then
\begin{align}
    S_{V} &= S_{\text{bulk}} - \frac{k}{4\pi\beta} \int_{\partial V \times S^{1}(\beta) }d\tau\,d\phi\;\tr (\mu A_{\phi } ) \nn
    &= S_{\text{bulk}} -  \frac{1}{2} Q(\mu),
\end{align}
where we identified the boundary term  with the edge mode charge
\begin{align}
Q(\mu) =  \frac{k}{2\pi}\oint_{\partial V  }d\phi\;\tr (\mu A_{\phi } ) .
\end{align}
On the physical phase space where the constraints are satisfied, the subregion Hamiltonian derived from $S_{V}$ is given by\footnote{ \cite{Banados:1994tn} showed that \begin{align}
H=H_{bulk}+ \frac{1}{\beta}\,Q(\mu) 
\end{align}
where the bulk term generates bulk gauge transformations and therefore vanishes on the constraint surface.   This is the same phenomenon as in gravity \cite{Regge:1974zd}, where global charges live on the boundary.  The charge is twice the boundary term in the action because the bulk term also contains half of it.  } \cite{Banados:1994tn}:
\begin{align}
    H =\frac{1}{ \beta}\, Q(\mu),
\end{align}
so the subregion  density matrix is
\begin{align} \label{rhoV}
    \rho_{V} =S_{00} e^{-\beta H} &= S_{00}e^{- Q(\mu)},
\end{align}
where we have introduced an overall constant $S_{00}$ arising from  normalization of the boundary state $\ket{\mathcal{T}}$ in \eqref{cons}.
As in the case of the gapless boundary conditions described above, quantizing  $A_{\phi}$ on $\pd_{V}$ gives an infinite dimensional Kac--Moody module $\mathcal{H}_{R}$ labelled by the bulk puncture $R$.     On this Hilbert space,  the density matrix is equal to  the balancing element in the representation $R$: it has an infinite dimensional degeneracy corresponding to the non-zero modes of the current $A_{\phi}$. This leads to an un-normalizable density matrix
\begin{align}
\tr_{V} \rho_{V} = S_{00} \tr_{\mathcal{H}_{R}} (D) =\infty.
\end{align} 

This divergence arises from the local degrees of freedom on the entangling surface which should be absent in a topological theory, as was recently emphasized \cite{Mertens:2025ydx}.   To remove them,  we set
\begin{align}
    \partial_{\phi} A_{\phi}= 0 
\end{align}
in addition to \eqref{local}. This reduces the Kac--Moody representation to the zero mode subspace $V_{R}$, which describes a representation of the universal enveloping algebra $U(\mathfrak{u}(N))$.   With these boundary conditions, we find
\begin{align}
\tr_{V} \rho_{V} = S_{00} \tr_{V_{R}} (D) = S_{00} \dim_{q}R
\end{align} 
This agrees with the partition function on $S^3$:
\begin{align}
Z(S^3 ,R) = S_{0R}
\end{align}
once we account for the fact that $\dim_{q}R=\frac{S_{0R}}{S_{00}}$.   This confirms the shrinkability of our boundary condition.

\subsection{The need for quantum groups}
Our pedestrian approach for incorporating the holonomy constraint in the previous subsection is useful for providing intuition about topological subregions: it shows that with a shrinkable boundary, the edge mode Hilbert space for a punctured disk effectively reduces to the Hilbert space of a particle in the representation $R$, with a modular Hamiltonian given by the edge mode charge Q in that representation.  

However, this analysis is problematic for two related reasons.  First, the density matrix $\rho_{V}$ is not invariant under the gauge transformation $\rho_{V} \to R(g) \rho_{V} R(g^{-1}) $, since $D$ is not central.  Given that the original state \eqref{B3state} is gauge invariant, this would imply that gauge invariance was broken by the factorization map defined by the path integral \eqref{B3factor}, or by the subsequent gluing which defines the partial trace over $\bar{V}$.  

The second, related issue is that the shrinkable boundary condition is non-local in $\tau$.  
This is because the boundary condition \eqref{local} on the gauge field  depends on the length $\beta$ of the modular time circle, which ultimately led to a $\beta$ dependent Hamiltonian.    
This non-locality affects  the entanglement entropy obtained from the replica trick.  Indeed, the path integral for the replicated theory on $V$ is obtained by extending the Euclidean time circle from  $\beta \to n \beta$, which does not change the path  integral of a topological theory.  For the same reason, the entanglement boundary state remains the same for all $n$.  We thus find that the replicated path integral is \begin{align}
    Z(n) &\equiv \tr_{V} \rho_{V}'^n\nn
    &= \tr_{V} \rho_{V}',
\end{align} which implies that $\rho'_{V}$ is the identity, corresponding to a maximal state.  The path integral replica trick would then give a different entropy than the one obtained from the density matrix \eqref{rhoV}. 

One way to hide this problem is to absorb $D$  into the trace.   Equivalently, we can view $D$ as a ``defect operator" that is inserted only once irrespective of the replica number $n$ \footnote{ A completely analogous defect operator appears in JT gravity \cite{Jafferis:2019wkd}, and in topological string theory \cite{Donnelly:2020teo}}.
However, these ad hoc procedures do not fix the problem of gauge invariance: the quantum trace is not  conjugation invariant under ordinary gauge transformations: indeed it does not satisfy the usual cyclicity condition.  However, it is invariant under q-deformed gauge transformations, which we define in  \eqref{qgauge}.
Thus, gauge invariance can be restored by treating the quantum particle describing the Hilbert space of the punctured disk as an anyon, which transforms under a quantum group symmetry.
\paragraph{Quantum groups and Combinatorial quantization}
So far, our introduction of quantum groups into the path integral quantization of a subregion is ad hoc. However, the introduction of quantum group algebras can be given a
 rigorous formulation in   the combinatorial quantization of Chern Simons theory
 \cite{Alekseev:1994pa,Alekseev:1994au}, where quantum groups play a fundamental role.     We will explain how this framework provides an operator algebra for subregions in Chern Simons theory.  In this language, the shrinkable boundary condition is replaced by the statement 
 \begin{align}
     \braket{B^3,R|O|B^3,R} = \widetilde{tr}_{R}(O)
 \end{align}
 Once we have formulated the subregion algebra for a punctured disk and the associated Hilbert space $V_{R}$ via combinatorial quantization,   we can construct more general subregions with nontrivial topologies by entangling $V_{R}$. 

\paragraph{Relation to previous work}
To give a broader perspective on our work, let us first recall that quantum groups were introduced in Chern--Simons theory 
early on by Reshetikhin and Turaev \cite{ReshetikhinTuraev1990},
who replaced Witten's path integral \cite{Witten:1988hf} with a purely
algebraic construction of link and three-manifold invariants.  This formalism  is based on the
representation theory of $U_q(\mathfrak{g})$, organized into a ribbon
(modular tensor) category. It produces anyon diagrammatics that computes 
expectation values as topological invariants, but does not explain what a spatial subregion is and how entanglement arises.  On the other hand, the 
combinatorial quantization of Chern--Simons theory
\cite{Alekseev:1994pa,Alekseev:1994au,Buffenoir:1994fh}, quantizes the
phase space of flat connections on a discretized surface, leading to
operator algebras built from the quantum group, with quantum group
gauge transformations acting locally at vertices.

Our work draws heavily on these two frameworks to define subregion algebras and compute their $q$-deformed entanglement entropies. To explain the mathematical basis of our computations, we review the relevant aspects of combinatorial quantization in Section~\ref{QGholonomies} and of ribbon categories in Section~\ref{sec:categorical}. In these two sections, our main contribution is the application of these frameworks to factorization and q deformed entanglement entropy for the twice-punctured sphere, which resolves the puzzle posed by the path integral quantization discussed in this section. Section~\ref{sec:gluing} applies these ideas to new computations of $q$-deformed entanglement entropies and tensor network simulations in Chern Simons theory. The ribbon diagrammatics used there appeared in \cite{Donnelly:2020teo,Jiang:2020cqo} in the context of topological string theory, and one of the main purposes of our work is to put those computations on a firm mathematical footing.

\section{Topological subregions and combinatorial quantization} \label{QGholonomies}
Combinatorial quantization
\cite{Fock:1998nu,Alekseev:1994au,Alekseev:1994pa} is essentially a q-deformed lattice gauge theory description of Chern Simons theory, closely related to the one developed in \cite{buffenoir1995chernsimonstheorylatticenew}.  We will use this perspective to define subregions in Chern Simons theory, by q-deforming the extended Hilbert space framework in lattice gauge theory \cite{Donnelly2011}. 
\subsection{The Fock-Rosly phase space}
We begin with a description of the Chern Simons phase space underlying combinatorial quantization, which was formulated by Fock and Rosly \cite{Fock:1998nu}.
The usual phase space derived from the Lagrangian \eqref{lag} \cite{Elitzur:1989nr}
is a symplectic quotient of the infinite-dimensional space of connections on a surface
$\Sigma$. The symplectic form and associated Poisson brackets are
\begin{align}\label{compact_bracket}
  \omega &= \frac{k}{4\pi}\int_{\Sigma} \tr ( \delta A \wedge \delta A), \nn
  \{ A_{i}^{a} (x), A_{j}^{b}(y)\} &= \frac{2\pi}{k}\, \epsilon_{ij}\, \delta^{ab}\, \delta^{2}(x-y),
\end{align}
where $i,j$ are spatial indices and $a,b$ are Lie algebra indices.

In contrast, Fock and Rosly \cite{Fock:1998nu} constructed a finite-dimensional
phase space for $\Sigma$ before quotienting by gauge transformations.  It is defined on a fat graph (ribbon graph) onto which $\Sigma$
retracts.  Each puncture and boundary of $\Sigma$ belongs to a face of the fatgraph. For
example, the fat graphs for the 
torus and the three-holed sphere are
\begin{align}
\vcenter{\hbox{\includegraphics[scale=.2]{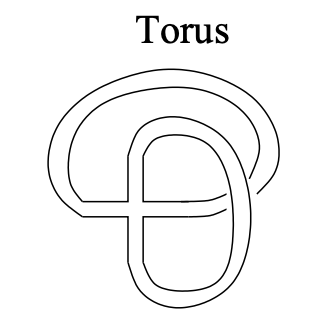} }} ,\qquad
\vcenter{\hbox{\includegraphics[scale=.07]{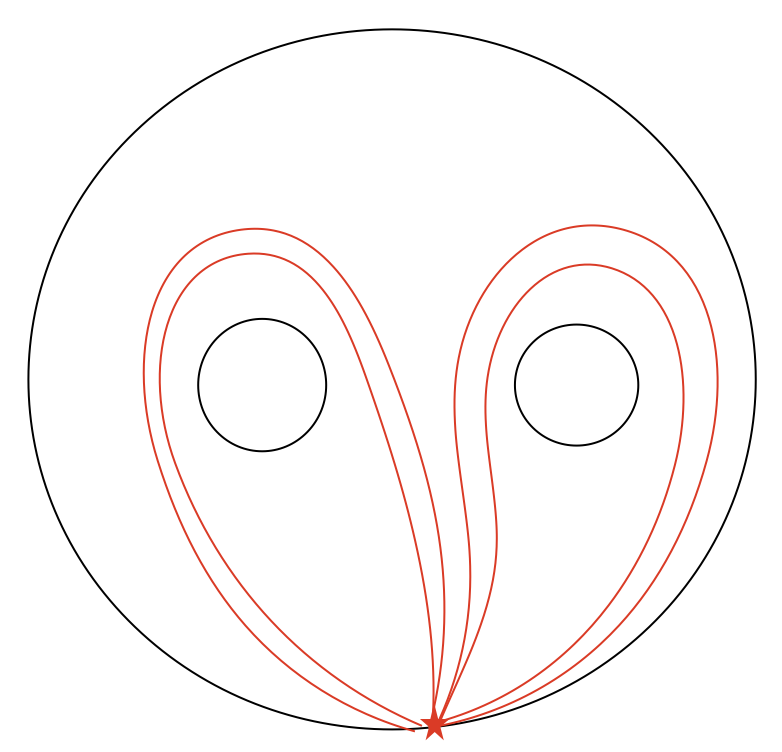} }}
\end{align}
As in lattice gauge theory, the degrees of freedom are  holonomies given by group elements on oriented
links, which are collectively described as a graph connection $U(i)$. This gives a finite-dimensional space $G^{E}$, where
$E$ is the number of links, which is then quotiented by gauge transformations
acting at the vertices:
\begin{align}\label{gauge}
  U(i) \to h_{s}\, U (i)\, h_{t}^{-1},
\end{align}
where $h_s$ and $h_t$ act at source and target endpoints of the oriented link $i$.  
\paragraph{Classical origin of the q-deformation}
The key difference between Chern Simons and Yang-Mills lattice gauge theory lies in the variables
conjugate to the holonomies. In Yang--Mills these are 
non-abelian electric fields valued in the Lie algebra\footnote{The phase space on each link is then $T^*G$}, which implement the infinitesimal form of the gauge transformation \eqref{gauge}. These are canonical transformations that preserve the Poisson brackets on the phase space.

In contrast, Chern--Simons theory does not contain an independent set of conjugate electric field degrees of freedom: instead 
the brackets \eqref{compact_bracket} pair different spatial components of the gauge
field, so the holonomies $U=P\exp \int A$ must be conjugate to other holonomies.  Thus, their matrix elements have nonzero
Poisson brackets among themselves, and become non-commuting operators
upon quantization. This differs from Yang--Mills, where
the only nontrivial brackets are between the holonomies  and the
electric field.

The non-commuting holonomies of the graph connection $U(i)$ imply that the gauge group should be endowed with a compatible Poisson structure, so that the gauge transformations do not change the brackets of the holonomies. 
The standard structure on such a Poisson-Lie group is 
given by a classical $r$ matrix $r\in\mathfrak g\otimes\mathfrak g$, which defines the Sklyanin bracket:
\begin{equation}\label{sklyanin}
  \{h^{1},h^{2}\} = \,\big[\,r,\; h^{1}h^{2}\,\big],
  \qquad h^{1}=h\otimes 1,\; h^{2}=1\otimes h.
\end{equation}
Here we used a condensed notation common in the quantum group literature, where the holonomies are embedded into tensor product of matrices ( See the appendix for an explicit example).  On the right hand side the commutator bracket is the one defined on the tensor product.
The Jacobi-identity requires that the classical $r$-matrix 
satisfies the classical Yang--Baxter equation
\begin{align}\label{rconditions}
  [r_{12},r_{13}]+[r_{12},r_{23}]+[r_{13},r_{23}] = 0,
\end{align}

Fock and Rosly (FR) showed the Poisson brackets of the graph connection are fixed by requiring the gauge group to act via Poisson maps that preserve the brackets on the phase space: this is necessary in order to perform the gauge  quotient.  Furthermore, the FR brackets are compatible with \eqref{compact_bracket} and lead to the same gauge invariant  phase space.
Upon quantization, \eqref{sklyanin} becomes the defining relation of the
quantum group, and the gauge transformations become $U_q(\mathfrak g)$
transformations. This is the classical origin of the $q$-deformed gauge
symmetry of combinatorial quantization.

\subsection{Subregion phase spaces}\label{FRsubregions}

We now present the Fock-Rosly brackets on the graph connection, which we will use to define subregion phase spaces.
Let $\xi^{(v,i)}_a$ be the vector field that performs an infinitesimal gauge
transformation on $U(i)$ by the generator $T_a$ at vertex $v$. This acts on functions of holonomies by sending
$$U(i)\to T^{a}U(i) \qquad \text{or}\qquad   U(i) \to -U(i)T^{a},$$ depending on whether the link is ingoing or outgoing. 
Given the $r$-matrix
\begin{align}
    r= r^{ab}T_{a}\otimes T_{b},
\end{align}
 the FR brackets are defined by the bivector 
\begin{equation}\label{FRbivector}
  \pi_{FR}=\sum_{v}\Big(\sum_{i<j\ \text{at}\ v} r_v^{ab}\,\xi^{(v,i)}_a\wedge\xi^{(v,j)}_b
  \;+\;\tfrac12\sum_{i\ \text{at}\ v} r_v^{ab}\,\xi^{(v,i)}_a\wedge\xi^{(v,i)}_b\Big)
\end{equation}
The first term, acting between different link holonomies, requires a choice of linear order for links that end at each vertex, specified by a ``cilium" marking the first link. It only depends on the symmetric part of r,
which is constrained to be the quadratic Casimir \cite{Fock:1998nu}:
\begin{align}
\frac{1}{2}(r^{ab} +r^{ba}) = \tr(T_{a}T_{b})
\end{align} 
This is the same for all vertices and is what reproduces the gauge invariant quantities\footnote{When acting on different links, it reproduces the Goldman bracket describing intersecting Wilson lines}.  

The second  term acts on holonomies on the same link, and depends only on the anti-symmetric part of r,  which can be different for different vertices, and affects gauge variant quantities. 

\paragraph{Subregions from fat graphs} 

Applying \eqref{FRbivector} to simple graphs gives the elementary phase
spaces from which more general phase spaces can be assembled  by gluing operations on the fatgraphs \cite{Fock:1998nu}.  These gluing operations are Poisson maps, and result in phase spaces that are products of $G$ and its dual $G^*$, possibly quotiented by gauge transformations.

We will focus on three fat graphs discussed in \cite{Fock:1998nu}, and identify the corresponding  topological subregions whose boundary supports the vertices of the graph \footnote{ Similar subregion algebras were derived in \cite{Mertens:2025ydx}}.  We present the action of the FR brackets on the link holonomies as derived in \cite{Fock:1998nu}, which determine the algebra of observables.

\begin{enumerate}
\item \emph{A single link}
        \begin{equation}\vcenter{\hbox{
\begin{tikzpicture}[scale=0.8]
  \fill[gray!15, even odd rule] (0,0) circle (2) (0,0) circle (0.8);
  \draw[thick] (0,0) circle (2);
  \draw[thick] (0,0) circle (0.8);
  \draw[ribbon] (0,2) -- (0,0.8);
  \node[marked] at (0,2) {};    \draw[cilium] (0,2) -- (0,2.3);
  \node[marked] at (0,0.8) {};  \draw[cilium] (0,0.8) -- (0,0.5);
  \node[font=\small] at (-0.25,1.45) {$U$};
\end{tikzpicture}}}\simeq G
  \end{equation}
 For a link $U$ with endpoints carrying $r$ and $r'$, defined by $r'_{ab}=r_{ba}$
\begin{equation}\label{FRlink}
  \{U^1,U^2\}=r\,U^1U^2-U^1U^2\,r'
\end{equation}
This endows the group $G$ with a Poisson structure, but it is not the same as the Sklyanin bracket due to $r'$.  It is designed to be preserved by conjugation of $U$ via a Poisson--Lie group 
$G$, carrying the Sklyanin  bracket.

We will identify this as the configuration space of the annulus subregion, which lives on a radial link as indicated above.

\item \emph{Two links between the same vertices.}
\begin{equation}\vcenter{\hbox{\begin{tikzpicture}[scale=0.8]
  \fill[gray!15, even odd rule] (0,0) circle (2) (0,0) circle (0.8);
  \draw[thick] (0,0) circle (2);
  \draw[thick] (0,0) circle (0.8);
  \draw[ribbon] (0,2) -- (0,0.8);
  \draw[ribbon] plot[smooth] coordinates
        {(0,2) (-0.9,1.35) (-1.4,0) (0,-1.4) (1.4,0) (1.0,1.0) (0.35,1.05) (0,0.8)};
  \node[marked] at (0,2) {};    \draw[cilium] (0,2) -- (0,2.3);
  \node[marked] at (0,0.8) {};  \draw[cilium] (0,0.8) -- (0,0.5);
  \node[font=\small] at (-0.25,1.45) {$U$};
  \node[font=\small] at (-1.75,-0.55) {$V$};
\end{tikzpicture}}} \simeq D_{+} G^*\end{equation}
 For links $U,V$ joining the same
two vertices with the same r-matrix, with a linear order specifying $U$ before $V$ at both ends, we have the algebra
\begin{equation}\label{FRbigon}
  \{U^1,V^2\}=r\,U^1V^2+U^1V^2\,r,
\end{equation}
together with \eqref{FRlink} for each link. This gives the
Heisenberg double $D_{+}G^*$, a symplectic space which is the
$q$-deformation of the lattice Yang--Mills link phase space $T^*G$.  In this context, the
holonomy $UV^{-1}$ around the face between the two links plays the role of
a group-valued electric field. 
\item \emph{A single loop.}
\begin{equation}\vcenter{\hbox{
\begin{tikzpicture}[scale=0.8]
  \draw[thick] (0,0) circle (2);
  \draw[ribbon] (0,-2) .. controls (-1.7,-0.6) and (-1.1,1.5) .. (0,1.5)
                      .. controls (1.1,1.5) and (1.7,-0.6) .. (0,-2);
  \node at (0,0.35) {$\times$};
  \node[font=\small] at (0.35,0.1) {$R$};
  \node[marked] at (0,-2) {};
  \draw[cilium] (0,-2) -- (0,-2.3);
  \node[font=\small] at (1.25,0.9) {$M$};
\end{tikzpicture}}}\simeq  G^*
\end{equation}
When both ends of a link sit at the same vertex, they
necessarily carry the same $r$-matrix, and the cross term between the two
ends adds to \eqref{FRlink}:
\begin{equation}\label{FRloop}
  \{U^1,U^2\}=r\,U^1U^2-U^1U^2\,r'+U^2\,r'\,U^1-U^1\,r\,U^2 .
\end{equation}
This is the Semenov-Tian-Shansky bracket on the group $G$. It generates the action of conjugation on  $U$, which leaves its conjugacy class invariant. The specific conjugacy class is determined by the puncture $R$, and forces $U$ to belong to the conjugacy class of $U_{R}$ \eqref{U_R}.   
\end{enumerate}

\paragraph{Gluing subregions}
More general graphs can be obtained by gluing along vertices.   Two vertices with the same number of incident edges can be glued if their $r$ matrices are related by a swap.  The basic rule is to remove the vertices and then connect the edges as specified by their linear order\footnote{ For $N$ incident  edges, the $k$ th edge glues to the $N-k$ th edge }. 
These gluings are Poisson maps between copies of $G$ \cite{Fock:1998nu} that describe the fusion of subregions. 

For example, gluing two single links
end to end produces the group multiplication $G\times G\to G$, which describes the fusion of two annuli into one annulus. Meanwhile, gluing two loops each encircling a puncture on a disk  creates a single loop on a sphere.  This is only allowed if the punctures have dual representations and  creates a trivial phase space because one side of the glued loop encloses no punctures.

\paragraph{Relation to the Lagrangian formalism} We close our discussion of the FR phase space with some comments on the relation to the Lagrangian formalism. As noted in \cite{Alekseev:1994au},  direct exponentiation of \eqref{compact_bracket} leads to singular brackets for the holonomies: while they give well defined brackets of crossing Wilson lines, they do not define the bracket of over-lapping Wilson lines corresponding to the links of the graph.   The FR brackets provide a regularization of this singularity via the $r$ matrix.  This is not part of the data defining the standard Lagrangian formulation of Chern Simons theory, but it is  data we need to define the Sklyanin bracket that leads to quantum group algebra upon quantization.  This is why the conventional path integral quantization fails to produce  the quantum group subregion algebras that we are seeking.    Interestingly, \cite{Dupuis:2020ndx} has succeeded in incorporating the $r$ matrix into a total derivative term in the Lagrangian of 3d gravity (with a cosmological constant) in the 1st order formalism.  This suggests one could do the same for compact Chern Simons theory.  Finally,  while there are in principle different solutions to the classical Yang Baxter equations, they cannot be distinguished by gauge invariant observables, which only depend on the symmetric part of the r matrix given by the quadratic Casimir \cite{Fock:1998nu}.

\subsection{Combinatorial quantization} \label{sec:observables}We now apply combinatorial quantization to the Fock--Rosly phase space defined on a fat graph, following
\cite{Alekseev:1994pa,Alekseev:1994au,Buffenoir:1994fh}.  The quantum commutators are obtained from the FR brackets $\{\,,\}_{\rm FR}$ by replacing the classical $r$-matrix with a quantum $\mathcal{R}$-matrix satisfying the quantum Yang Baxter equation.  The $\mathcal{R}$ matrix is part of the data defining the q deformation of the gauge group $G$.  For $q\equiv e^{\hbar}$, it is related to the  classical $r$ matrix by 
\begin{align}
    \mathcal{R} = 1+\hbar\, r + O(\hbar^2).
\end{align} The quantum algebra is then a q-deformation of the classical one in the sense that  
\begin{align}
U^1U^2 - U^2U^1 = \hbar\,\{U^1,U^2\}_{\rm FR} + O(\hbar^2),
\end{align}
The resulting quantum theory is defined by four ingredients.

\begin{enumerate}
    \item A lattice algebra $\mathcal{B}$, generated by link holonomies satisfying relations defined by the $\mathcal{R}$ matrix.  On each open link, it reduces to the non-commutative algebra Fun$_{q}(G)$, the q-deformation of the algebra Fun$(G)$ of functions on $G$ that makes up the link algebra of ordinary Yang--Mills.
    \item A q-deformed gauge algebra $U_{q}(\mathfrak{g})$ acting on Fun$_{q}(G)$ at the vertices.   It is the Hopf dual to $\mathrm{Fun}_q(G)$,  arising  from the quantization of $G^*$.   This is the q-deformation of the electric field algebra $U(\mathfrak{g})$ in Yang--Mills. 
    \item The Haar state $\omega_{q}$ on $\mathcal{B}$ which defines the GNS Hilbert space. This is an extended, gauge-variant state space, that reduces to a $q$-deformation of the Yang Mills Hilbert space  $L^2(G)$ on each link.  
    \item A gauge-invariant algebra $\mathcal{A}\subset\mathcal{B}$ of observables built from quantum traces and projectors onto the flatness constraints.  On this algebra, $\omega$ restricts to a positive trace.
\end{enumerate}
These ingredients allow us to define subregion algebras as the lattice algebras on their fatgraphs, which are quantizations of the classical subregion algebras \eqref{FRlink}, \eqref{FRloop} and \eqref{FRbigon}.  We review aspects of these ingredients relevant to the  entanglement entropy computations in the rest of the paper. 

\subsubsection{The lattice algebra $\mathcal{B}$.}
Consider the algebra $\mathcal{B}(I)$ on a single link. There are two cases, leading to two different algebras. 
\begin{itemize}
    \item  \textbf{Link algebra} The quantum holonomies on a single open link with different endpoints satisfy the RTT relations
    \begin{align}\label{RTT}
    \mathcal{R}\,U^1U^2 = U^2U^1\,\mathcal{R},
    \end{align}
which defines Fun$_{q}(G)$, and is fundamental to combinatorial Chern Simons theory.
This is the algebra on a single radial link joining the inner and outer boundaries; it quantizes the classical algebra \eqref{FRlink}.
Concretely, a quantum holonomy satisfying \eqref{RTT} is a matrix of algebra elements, called a quantum matrix: For $G=SL(2)$, this takes the form 
\begin{align}\label{qmatrix}
U= \begin{pmatrix} a & b\\ c & d\end{pmatrix},
\end{align} 
with the RTT commutation relations
\begin{align} \label{sl2commutators}
ab&= q ba, \quad  ac= q ca,\quad  bd= q db,\quad cd= q dc\quad  \nn
bc&=cb.
\end{align}
Fun$_q(SL(2))$ is then generated by the $a,b,c,d$ with these relations.   
In the quantum group setting,  the matrix inverse is replaced by the antipode $S$, an anti-automorphism with $S(U)U=US(U)=1$.  A deformed  $*$-structure can be defined, which determines the notion of unitarity. Finally, it will be useful to consider co-representation matrix elements $R_{ab}(U) \in \text{Fun}_{q}(G)$.  These can be constructed explicitly in terms of $U_{ij}$ by q-symmetrizing and anti q- symmetrizing the indices, which now produces an algebra element rather than a number.  Note that $R$ labels representations of  $U_q(\mathfrak{g})$, which act on $\text{Fun}_{q}(G)$ on the left and right by ordinary matrix multiplication.  The dual representation is  defined by $R^*=R\circ S$. 
    \item \textbf{The loop algebra}  Holonomies $M$ on a link that begins and ends at the same vertex obey the reflection equation
    \begin{align}\label{reflection}
    \mathcal{R}'M_1\mathcal{R}M_2 = M_2\mathcal{R}'M_1\mathcal{R} .
    \end{align}
    \cite{Alekseev:1995rn} showed that this loop algebra is equivalent to $U_{q}(\mathfrak{g})$.
To understand this, recall that any classical group element close to the identity can be factored
uniquely into a ``lower'' and an ``upper'' triangular piece,
$M = (L^-)^{-1} L^+$. In the quantum theory the entries of $M$ do not commute, but we can still make such a decomposition, with the matrix entries of $L^\pm$ parametrized by
 algebra elements.  It turns out the reflection equation implies  these matrix elements satisfy the algebra for   $U_{q}(\mathfrak{g})$.  For example, if $E,F,K$ are the generators of $U_{q}(sl(2))$, we can define the matrices
\begin{equation}
  L^+ = \begin{pmatrix} K^{1/2} & 0 \\ (q-q^{-1})\,K^{-1/2}E & K^{-1/2} \end{pmatrix},
  \qquad
  L^- = \begin{pmatrix} K^{-1/2} & -(q-q^{-1})\,F K^{1/2} \\ 0 & K^{1/2} \end{pmatrix},
\end{equation}
so that
\begin{equation}\label{gauss}
  M = (L^-)^{-1} L^+ =
  \begin{pmatrix}
    K + q^{-1}(q-q^{-1})^2\, FE & q^{-1}(q-q^{-1})\, F \\
    (q-q^{-1})\, K^{-1}E & K^{-1} 
  \end{pmatrix}
\end{equation}
The commutation relations of $E$, $F$, $K$ are exactly what makes $M$ obey the
reflection equation \cite{Alekseev:1995rn}. Any representation of $U_q(\mathfrak{g})$ therefore gives a
representation of the loop algebra.

\end{itemize}

There are also commutation relations between two different links that touch, which are again deformation quantizations of the FR bracket.  These brackets altogether define the lattice algebra $\mathcal{B}$  

\subsubsection{Quantum gauge transformations.}
The quantum gauge algebra  is $\bigotimes_v U_q(\mathcal{G})$. A  gauge transformation assigns an element of $U_q(\mathcal{G})$ to each vertex. Schematically, it acts on a link $I$ from $a$ to $b$ as
\begin{align}\label{qgauge}
U(I) \to h_a\,U(I)\, S(h_b),
\end{align}
where $h_{a}$ and $S(h_{b})$ are matrices with numerical matrix elements.
When $a,b$ are the same point, so that the link forms a closed loop, this is a q-deformed adjoint action, which is defined with the co-product\footnote{
The co-product must also be used to define the action on general products $U_{ab}U_{cd}\cdots$ of holonomies. 
} on 
 $U_q(\mathcal{G})$:
\begin{align} \label{Uqcoproduct}
\Delta : U_q(\mathcal{G}) \to U_q(\mathcal{G})\otimes U_q(\mathcal{G}),\qquad
\Delta(\xi) = \sum \xi_{(1)}\otimes\xi_{(2)} ,
\end{align}

The  $q$-deformed adjoint action is then
\begin{align}\label{qadjoint}
\xi\triangleright U(I) = \sum \xi_{(1)}\,U(I)\,S(\xi_{(2)}).
\end{align}

Since $\Delta$ is not cocommutative, the ordinary trace of a holonomy is no longer invariant under such a gauge transformation.  Instead, gauge invariant elements of the algebra are produced by the quantum trace \eqref{qmatrixtr}, in an arbitrary representation $R$
\begin{align}
    \widetilde{\tr}_{R}(U) \equiv \sum_{i}  (D)^{R}_{ij} R(U)_{ji},
 \end{align}
where $D \in U_{q}(\mathfrak{g})$ is the balancing element, and  $(D)^{R}_{ij}$ is its matrix elements in the representation $R$.

\subsubsection{The Haar state and the GNS Hilbert space}
To discuss entanglement and factorization it will be useful to have a description of the quantum Hilbert space.  In lattice Yang-Mills theory, the holonomies $U$ play the role of a configuration space variable, so a Hilbert space can be constructed in terms of the space of functions  on the group.  In Chern Simons theory, the holonomies do not commute, so such a function space does not exist.  Instead, combinatorial quantization produces a Hilbert space 
through the  GNS construction applied to Fun$_{q}(G)$. This takes a  $*$-algebra $\mathcal{C}$ and a state $\omega$ -- a positive linear functional-- and produces a Hilbert space on which the algebra acts.  The main features are 
\begin{enumerate}
\item Vectors are algebra elements $\ket{a}, a \in \mathcal{C} $
\item The inner product is $\braket{a|b} = \omega(a^*b)$
\item The algebra acts by left multiplication $\hat{a}\ket{b} =\ket{ab}$
\item The algebra unit plays the role of the cyclic vector $\ket{1}$. Expectation values in this state produces the functional $\omega$: 
\begin{align}
    \braket{1|\hat{a}|1} = \omega(a)
\end{align}
\end{enumerate} 
In combinatorial quantization, we apply the GNS construction to Fun$_{q}(G)$ using the Haar state:
\begin{align}\label{h}
    h_q :  R_{ab } (U) \mapsto \delta_{R,0} ,
\end{align}
which is non-zero only on the trivial representation. It is evaluated on a general element by using the Clebsch--Gordan decomposition 
\begin{align}\label{clebsh}
    R_{ab}(U)\,R'_{cd}(U) = \sum_{R'', i,j} C^{R R'R''}_{aci}\, \overline{C_{bdj}^{RR'R''}}\; R''_{ij}(U),
\end{align}
where  $C^{R R'R''}_{aci}$ are the q-deformed Clebsch Gordon coefficients.  This reduces products of matrix elements to linear combinations of single ones. A general state in the GNS Hilbert space is given by
\begin{align}\label{wf}
    R_{a_{1}b_{1}} (U)\cdots  R_{a_{n}b_{n}} (U) \ket{1} ,
\end{align} 
and the inner product is
\begin{align} \label{hq}
 h_{q} \big(R^*_{ab}(U)\, R'_{cd}(U) \big)= \frac{ \delta_{RR'}\,\delta_{ac}\, \delta_{bd}\, D^{R}_{bb} }{\dim_{q}R } .
\end{align} This is the q-deformation of the Haar integral of matrix elements on a group $G$.
 Completion with respect to this inner product produces the GNS Hilbert space, which we  denote as $L^{2}_{q}(G)$. 
 
 \paragraph{Peter Weyl theorem and co-product}
 The q-deformed version of the Peter Weyl theorem states that 
\begin{align}
    L^{2}_{q}(G) \simeq \bigoplus_RV_R\otimes V_R^*,
\end{align}
where  $V_{R}$ and $V_{R}^*$ are irreps of $U_{q}(\mathfrak{g})$.  This is equivalent to saying that  matrix elements $R_{ab}(U)\ket{1}$ span the Hilbert space $L^{2}_{q}(G)$, on which $U_{q}(\mathfrak{g})$ acts by left and right multiplication (by ordinary matrices).  This is the regular representation of the quantum group. Applying $h_q$ to every link defines a state $\omega$ on $\mathcal{B}$.  To define factorization of this Hilbert space, we will use the co-product on the co-representation matrix elements:
\begin{align}
    \Delta : R_{ac}(U) \to  \sum_{b} R_{ab} (U) R_{bc} (U),\label{coprodR}
\end{align}
This is the analogue of the representation property
\begin{align}
    R(U V)_{ab} =  \sum_{c} R(U)_{ac}R(V)_{cb},
\end{align}
for ordinary matrices.
\subsubsection{The gauge-invariant algebra $\mathcal{A}$.}
The gauge-invariant observables are generated by quantum Wilson loops around closed loops $C$ of the fatgraph,
\begin{equation}
\widetilde{W}_R(C) = \widetilde{\mathrm{tr}}_R\big(U(C)\big),
\end{equation}
which are invariant under \eqref{qgauge} because the quantum trace is invariant under the adjoint action. Around a single plaquette they obey the Verlinde fusion algebra,
\begin{equation}
\widetilde{W}_{R_1}(C)\, \widetilde{W}_{R_2}(C) = \sum_{R_3} N^{R_3}_{R_1 R_2}\, \widetilde{W}_{R_3}(C).
\end{equation}
Flatness and puncture data are imposed by attaching to each plaquette a projector built from the modular $S$-matrix,
\begin{equation}\label{eq:projector}
P_R(U) = S_{00}\sum_{R'} (\dim_q R)\, S_{RR'}\, \widetilde{\mathrm{tr}}_{R'}(U), \qquad P_R^2 = P_R = P_R^{*} ,
\end{equation}
the quantum group analogue of a delta function on a conjugacy class. For a fatgraph $\Gamma$ on $\Sigma_{g,n}$, the observable algebra is
\begin{equation}
\mathcal{A}(\Sigma_{g,n},\Gamma) = \mathcal{A}_{\text{loops}}\, f_{\Gamma,\Sigma_{g,n}}, \qquad f_{\Gamma,\Sigma_{g,n}} = \prod_{p \in H_0} P_0(p) \prod_{\nu=1}^n P_{R_\nu}(p_\nu),
\end{equation}
where $p$ runs over plaquettes without punctures and $p_\nu$ over those surrounding a puncture of type $R_\nu$. The mapping class group of $\Sigma_{g,n}$ acts on $\mathcal{A}$ by automorphisms.

On $\mathcal{A}$, there is a re-normalized Haar state $\omega_{CS}$, which is positive, tracial, and independent of the fatgraph \footnote{  The Haar state $\omega_{q}$ is not independent of the fat graph, but this is fixed by a normalization factor that depends on the topology of the surface}\cite{Alekseev:1994pa,Alekseev:1994au}. The algebra $\mathcal{A}$ is the full matrix algebra on the space $\mathcal{H}_{\Sigma_{g,n}}$ of conformal blocks, and $\omega_{CS}$ is its trace:
\begin{align}\label{verlinde}
\omega_{CS} (f_{\Gamma,\Sigma_{g,n}}) = \mathrm{Tr}_{\mathcal{H}_{\Sigma_{g,n}}} (1) = \sum_{R} (S_{0R} )^{2-2g-n}\prod_{i=1}^{n} S_{RR_{i}} ,
\end{align}
This is the Verlinde formula giving the dimension for the space of conformal blocks,  obtained from the path integral on $\Sigma_{g,n}\times S^1$.

\paragraph{Equivalence with other quantizations.}
Combinatorial quantization is equivalent to other methods of quantization in the sense that it produces the same gauge invariant operator algebra and abstract Hilbert space carrying  the same projective representation of the mapping class group\cite{Alekseev:1994pa,Alekseev:1994au,Alekseev:1995rn,ReshetikhinTuraev1990}.  Moreover, this algebra and Hilbert space does not depend different choice of $\mathcal{R}$-matrices whose classical limit shares the same symmetric part  \cite{BelavinDrinfeld1982} . This is because these different $R$ matrices produce equivalent representation categories that determine the invariant algebra  \cite{Etingof:1999ess}.   Finally, the q-deformed entanglement entropies in this work are expressed in terms of gauge invariant operators, it is also independent of these $\mathcal{R}$ matrix ambiguities. 
\subsection{Hilbert space factorization and the $q$-tracial state} \label{sec:factorization}

In lattice gauge theory, the Hilbert space is factorized by lifting the Gauss law constraints that relate degrees of freedom across the entangling surface. This leads to an extended Hilbert space with  subregion edge modes.  In combinatorial quantization,  the extended Hilbert space is identified with the GNS Hilbert space for the lattice $\mathcal{B}$, with flatness and gauge constraints imposed only away from the entangling surface.   For this reason, the operator algebra formalism is fundamental.  We illustrate this in the simple example of the twice-punctured sphere considered in section 1, and explain how to obtain a gauge-invariant reduced density matrix whose entanglement entropy agrees with $\log \dim_{q}R$.

 Consider a figure-8 fat graph  consisting of two loops, each encircling one puncture, based at a single vertex where the Gauss law is imposed.  
 Lifting the Gauss law at that vertex embeds $\mathcal{H}_{S^2,R}$ into the extended state space of two punctured disks.
\begin{equation} i_{R}:
\vcenter{\hbox{
\begin{tikzpicture}[scale=0.8]
  \pic {sphere};
  \draw[ribbon] (0,-0.45) .. controls (-1.2,0.6) and (-0.7,1.8) .. (0,1.8)
                         .. controls (0.7,1.8) and (1.2,0.6) .. (0,-0.45);
  \draw[ribbon] (0,-0.45) .. controls (-0.8,-1.0) and (-0.6,-1.7) .. (0,-1.7)
                         .. controls (0.6,-1.7) and (0.8,-1.0) .. (0,-0.45);
  \node at (0,0.95) {$\times$};   \node[font=\scriptsize] at (0.35,0.75) {$\bar R$};
  \node at (0,-1.2) {$\times$};   \node[font=\scriptsize] at (0.35,-1.35) {$R$};
  \node[marked] at (0,-0.45) {};

  \draw[hookarrow] (2.6,0) -- (4.2,0);

  \begin{scope}[xshift=6.8cm]
    \pic {upper cap};
    \pic {lower cap};
    \draw[ribbon] (0,-0.15) .. controls (-1.0,0.7) and (-0.6,1.7) .. (0,1.7)
                           .. controls (0.6,1.7) and (1.0,0.7) .. (0,-0.15);
    \node at (0,0.95) {$\times$};  \node[font=\scriptsize] at (0.35,0.75) {$\bar R$};
    \node[marked] at (0,-0.15) {};
    \draw[ribbon] (0,-0.75) .. controls (-0.8,-1.2) and (-0.6,-2.0) .. (0,-2.0)
                           .. controls (0.6,-2.0) and (0.8,-1.2) .. (0,-0.75);
    \node at (0,-1.45) {$\times$}; \node[font=\scriptsize] at (0.35,-1.6) {$R$};
    \node[marked] at (0,-0.75) {};
  \end{scope}
\end{tikzpicture}}}\end{equation}

At the classical level  this reverses the gluing map from two punctured disks into a sphere discussed in \eqref{FRloop}.  Since the phase space on the sphere is trivial, it has a one dimensional Hilbert space.   However, as we saw previously, there is a non-trivial loop algebra on the punctured disk, leading to a non trivial  Hilbert space.
\paragraph{State space for the punctured disk}
This is determined by the anyon at the puncture
carrying the $U_q(\mathfrak{g})$ charge $R$.  The curvature constraint for the loop face containing the puncture  projects the loop algebra onto the representation $V_{R}$.
This has a finite basis
\begin{align}
\ket{R,i}, i =1, \cdots \, \dim R \end{align}
and transforms according to
\begin{align}
  \ket{R,i} \to \sum_{j} R_{ij}(O) \ket{R,j} ,\qquad O \in U_{q}(\mathfrak{g}) ,
\end{align} 
where $R_{ij}(\xi)$ are numbers. This contrasts with path integral quantization with a gapless boundary condition, where the punctured disk carries an infinite-dimensional Kac--Moody representation. It is precisely the reduction found by the naive quantization of section \ref{sec:shrinkable}, now with the correct q-deformed gauge symmetry.

Similarly, the $R^*$ punctured disk is a dual representation\footnote{We have abused notation  here by writing a covector in $V_{R}^*$ as a ket. } $V_{R}^*$ that transforms with the antipode:
\begin{align}
    \ket{R^*, i^*} \to R(S(O))_{ji}   \ket{R^*, j^*}  ,\qquad O \in U_{q}(G)
\end{align}

\paragraph{Extended Hilbert space}
Let us now write down the embedding $i_{R}$ explicitly.
Since the  Hilbert space of the sphere is trivial, it is a map
\begin{align}\label{singlet}
    i_{R}: \mathbb{C} \to V_{R} \otimes V_{R}^*
\end{align}
that picks out an element of  $ V_{R} \otimes V_{R}^*$.  Gauge invariance implies that up to normalization it must be
\begin{align}
    \ket{i_{R}} = \sum_{i=1}^{\dim R} \ket{R,i} \otimes \ket{R^*, i^*} .
\end{align}
As noted in \eqref{Uqcoproduct},  the transformation law for $V_{R} \otimes V_{R}^*$ involves the co-product. Thus, gauge invariance is the statement that 
\begin{align}
(R\otimes R^*)\big(\Delta(\xi)\big)\ket{i_R} = \varepsilon(\xi)\ket{i_R}, \qquad \xi\in U_q(\mathcal{G}),
\end{align}
where $\varepsilon$ is the counit, i.e. the trivial 1 dimensional representation.  This is satisfied for $\ket{i_{R}}$ due to an axiomatic property of the antipode $S$.    In particular, for  $U_{q}(sl(2))$, the co-unit is 1 on the group like element $K$, but 0 on $E$ and $F$.
Equation \eqref{singlet} is the precise sense in which entanglement  restores the quantum group gauge invariance along the entangling surface and glues the two subregion state spaces back together. 

\paragraph{Hilbert space description of entanglement }
How do we define entanglement entropy for  $\ket{i_{R}} $  ?  In the conventional Hilbert space approach, we first turn $\ket{i_{R}} $  into a genuine tensor product state by using a quantum group invariant epsilon symbol to map the $V_{R}^*$ factor to the conjugate vector space  $\bar{V}_{R}$.  This maps   $\ket{i_{R}}$ into an  entangled state in $V_{R}\otimes \bar{V}_{R}$.   For example, for the fundamental representation of $U_{q}(sl(2))$, the q deformed  epsilon symbol in the eigenbasis $\ket{\pm} $ of the generator $H$ (the Pauli Z operator)  is
\begin{align}
    \epsilon_{ij}(q) = q^{3/2- i} \epsilon_{ij}  ,\qquad \epsilon=\begin{pmatrix}
        0& q^{1/2} \\
        -q^{-1/2} &0
    \end{pmatrix}
\end{align}

It maps $\ket{i_{R}}$ into the quantum group singlet  
\begin{align}
    \ket{\text{singlet} }= \sum_{i,j} \epsilon_{ij}(q) \ket{i}\otimes \ket{j} = q^{-1/2}\ket{+-} -q^{1/2}\ket{-+}
\end{align}
This makes explicit that a state invariant under quantum group  must depend on $q$.
Next, we need a notion of Hermitian conjugation to define  the bra $\bra{\text{singlet}} $, which allows us to take expectation values.   For $q$ real, this is the standard Hermitian conjugation on $\mathbb{C}^2$, leading to 
\begin{align} \label{qbra}
\bra{\text{singlet}} = q^{-1/2}\bra{+-} -q^{1/2}\bra{-+}
\end{align}
However, note that the norm of this state is the quantum dimension
\begin{align}
\braket{\text{singlet}|O\otimes 1|\text{singlet}} &= q^{-1} + q^{1} \\
&= \widetilde{\tr}(1)
\end{align}
Here the trace is taken over a subregion of a single punctured disk. 
More generally, quantum group gauge invariance implies that  \begin{align}\label{qtracialR}
 \braket{i_{R}|O\otimes 1|i_{R}} = \tilde{\tr}_{V_{R}}(O), 
\end{align}
because the quantum trace is the unique adjoint invariant functional on $\mathrm{End}(V_{R})$ (up to normalization) 
This is in contrast to the ordinary singlet, which would lead to an ordinary trace.  We call a state satisfying \eqref{qtracialR} a \emph{q-tracial } state. 

In combinatorial quantization, all traces are promoted to quantum traces: this includes the partial tracing over $\bar{V}_{R}$ needed to produce the density matrix \cite{Couvreur_2017}.  The quantum trace normalized reduced density matrix on $V_{R}$ is   
\begin{align}
\rho_{R} = \frac{1}{\dim_{q}R}\, 1_{V_{R}}, \qquad \widetilde{\mathrm{tr}}_R(\rho_R)=1,
\end{align}
and the $q$-deformed entanglement entropy is
\begin{align}
S_{R}= - \widetilde{\mathrm{tr}}_R (\rho_R \log \rho_R) = \log \dim_{q}R .
\end{align}
 It coincides with the anyonic entanglement entropy of \cite{Bonderson:2017osr}, which has been used to characterize topological phases \cite{Bonderson:2017osr} and quantum group invariant spin chains \cite{Couvreur_2017, Quella:2020aa}.

Technically, the computation \cite{Couvreur_2017} of $\rho_{R}$ and $S_{R}$ involved a right and a left quantum partial trace, which inserts $D^{-1}$ and $D$ respectively.   This is a technical complication that  will become much more transparent in the diagrammatic approach discussed in the next section.
Note that with respect to the ordinary trace, the density matrix would have been $\frac{D}{\dim_{q}R}$. This is exactly the density matrix of the naive quantization in section \ref{sec:shrinkable}, where $D=e^{-Q(\mu)}$ came from the local boundary condition \eqref{local}.  

\paragraph{Operator algebra description of entanglement}  Since combinatorial quantization is formulated in terms of operator algebras, it is useful to understand the entangling of the two punctured disks from this point of view.  In this language, entanglement is a property of a state on a subregion
algebra: this is a subalgebra of $\mathcal{B}$, with gauge constraints lifted at the entangling  surface.   For the twice-punctured sphere, $\mathcal{B}$ is generated by the loop holonomies around each puncture, which gives 
\begin{align}
    \mathcal{B}=  \mathrm{End}(V_R) \otimes  \mathrm{End}(V^*_R).
\end{align}
The sub-algebra of the punctured disk containing $R$ is
\begin{align}
  \mathcal M_R = \mathrm{End}(V_R) .
\end{align}
There is a unique normalized, gauge invariant state $\omega_{q}$ acting as a linear functional on $\mathrm{End}(V_R)$: this is the normalized quantum trace
\begin{align}
    \omega_{q}(O) =  \frac{1}{\dim_{q}R} \widetilde{tr} O
\end{align}
This coincides with the normalized expectation value in $\ket{i_{R}}$ discussed previously, and it is the Haar state $\omega_{q}$ on  $\mathcal{B}$.   This becomes manifest if we first insert a link crossing the entangling surface and then factorize:  
\begin{equation}\label{linkcross}
\begin{tikzpicture}[scale=0.8]
  \pic {sphere};
  \draw[ribbon] (0,-0.45) .. controls (-1.2,0.6) and (-0.7,1.8) .. (0,1.8)
                         .. controls (0.7,1.8) and (1.2,0.6) .. (0,-0.45);
  \draw[ribbon] (0,-0.45) .. controls (-0.8,-1.0) and (-0.6,-1.7) .. (0,-1.7)
                         .. controls (0.6,-1.7) and (0.8,-1.0) .. (0,-0.45);
  \node at (0,0.95) {$\times$};   \node[font=\scriptsize] at (0.35,0.75) {$\bar R$};
  \node at (0,-1.2) {$\times$};   \node[font=\scriptsize] at (0.35,-1.35) {$R$};

  \node at (2.8,0) {$=$};

  \begin{scope}[xshift=5.6cm]
    \pic {sphere};
    \draw[ribbon] (0,0.55) -- (0,-0.75);
    \draw[ribbon] (0,1.0) circle (0.45);
    \draw[ribbon] (0,-1.2) circle (0.45);
    \node at (0,1.0) {$\times$};   \node[font=\scriptsize] at (0.22,0.85) {$\bar R$};
    \node at (0,-1.2) {$\times$};  \node[font=\scriptsize] at (0.22,-1.35) {$R$};
  \end{scope}

  \draw[hookarrow] (8.2,0) -- (9.8,0) node[midway,above] {$\iota$};

  \begin{scope}[xshift=12.4cm]
    \pic {upper cap};
    \pic {lower cap};
    \draw[ribbon] (0,0.55) -- (0,-0.15);
    \draw[ribbon] (0,1.0) circle (0.45);
    \node at (0,1.0) {$\times$};   \node[font=\scriptsize] at (0.22,0.85) {$\bar R$};
    \node[marked] at (0,-0.15) {};
    \draw[ribbon] (0,-0.75) -- (0,-1.1);
    \draw[ribbon] (0,-1.5) circle (0.4);
    \node at (0,-1.5) {$\times$};  \node[font=\scriptsize] at (0.2,-1.65) {$R$};
    \node[marked] at (0,-0.75) {};
  \end{scope}
\end{tikzpicture}
\end{equation}
With this splitting of the fat graph, the factorization map is the Fun$_{q}(G)$ co-product on a link  \eqref{coprodR},
\begin{align}\label{iota}
U_{ij} \;\mapsto\; \sum_{k} (U_A)_{ik}\otimes (U_{\bar A})_{kj},\qquad
R_{ij}(U) \;\mapsto\; \sum_{k} R_{ik}(U_A)\otimes R_{kj}(U_{\bar A}),
\end{align}
where $U_A$ and $U_{\bar A}$ are the holonomies along the two halves of the link, and the contracted index $k$ becomes the edge modes on the entangling surface. The Haar state on the crossing link is invariant under the coproduct,
\begin{align}\label{haarinv}
(h_q\otimes\mathrm{id})\,\Delta = h_q\,1 = (\mathrm{id}\otimes h_q)\,\Delta ,
\end{align}
so the Haar state on the dumbbell is the Haar state  on the two lollipops, which is what we referred to as $\omega_{q}$.  Restricted to the edge mode algebra $\mathcal{M}_R$, it is the state $\ket{i_R}$.  The operator algebra definition of entanglement entropy using $\widetilde{tr}$ then coincides with the Hilbert space definition.
\paragraph{Modular operator for the quantum group algebra}
Although $\omega_q$ is the unique gauge-invariant state on $\mathcal{M}_R$, it is not a trace because the insertion of $D$ into the ordinary trace spoils the cyclic property.  However, it satisfies a twisted cyclicity
\begin{align}\label{twistedcyclicity}
\omega_q(y\,x) = \omega_q\big(x\,D\,y\,D^{-1}\big), \qquad x,y\in\mathcal{M}_R .
\end{align}
As observed in \cite{kustermans2000locallycompactquantumgroups}, this is a KMS condition $\omega_q(y\,x)=\omega_q\big(x\,\sigma_{-i}(y)\big)$ for the modular flow
\begin{align}\label{modflow}
\sigma_t(x) = D^{it}\,x\,D^{-it},
\end{align}
which identifies $D$ as the modular operator for  the Haar states $\omega_{q}$.  
Note that in defining such a modular operator, no choice of trace was used.

\emph{Positivity.} Whether the $q$-tracial state \eqref{qtracialR} is positive depends on $q$. For real $q$, the balancing element $D$ is a positive matrix, so $\widetilde{\mathrm{tr}}_R(O)=\tr_R(DO)$ is a positive functional and \eqref{qtracialR} is an ordinary positive state.

At a root of unity, $D$ is instead a diagonal matrix of phases. The quantum trace is then not positive on general operators in $\mathrm{End}(V_R)$. It is, however, positive on gauge-invariant operators. By Schur's lemma, these are multiples of the identity on $V_R$, and $\widetilde{\mathrm{tr}}_R(1)=\dim_q R $ is positive for every integrable representation $R$.

Physically meaningful quantities should therefore be computed from gauge-invariant data.  These are captured by the morphisms of the representation category Rep$U_{q}(\mathfrak{g})$, which we will discuss in the next section. On these morphisms, a positive inner product is provided by the structure of a unitary modular tensor category \cite{kirillov1995innerproductmodulartensor}, defined at $q$ a root of unity.   This abstract description will allow us to avoid the continuation of  $q$ to real numbers in our discussion above. 

\section{Categorical traces and q-deformed entanglement entropy} \label{sec:categorical}

We saw in the previous section that combinatorial quantization
provides a rigorous framework for a $q$-deformed extended Hilbert space formalism, which introduces subregion states that transform under a quantum group symmetry. This is fundamentally an operator formalism, and even for the simple case of a twice punctured sphere it involves complicated quantum group operations like the coproduct and the antipode. However, most of these operations can be captured by anyon worldline diagrams, which are much easier to manipulate than quantum group algebras.  

In this section we replace the quantum group computations with anyon diagrammatics. We focus on the categorical trace, given by closing an anyon loop, which replaces the quantum trace of the operator language. We will use the diagrammatic language to relate the q deformed entanglement entropy defined by the quantum trace \cite{Bonderson:2017osr} to the asymptotic entanglement entropy of \cite{Kato:2013ava}, which has an operational interpretation.

We emphasize that the \emph{edge-mode} anyons describing subregion algebras in this
work are a priori distinct from \emph{Wilson-line} anyons. A Wilson line carries only a
representation label $R$, or equivalently a conjugacy class of the holonomy, and gauge
transformations at the puncture are redundancies. Edge-mode anyons live on a subregion
boundary, where gauge transformations become symmetries, and carry a representation
$V_R$ of $U_q(\mathcal{G})$. 

Nevertheless, the diagrammatic rules for the edge mode anyons are the same as those of  Rep($U_{q}(\mathcal{G})$) described in  \cite{ReshetikhinTuraev1990,Bonderson:2007ci}; we state only what we need here and collect the remaining details in appendix \ref{app:anyon-diagrammatics}. In the next section, we generalize these rules to more general topological subregions.

\subsection{The categorical quantum trace} \label{sec:qtrace}
In combinatorial quantization, the physical gauge invariant state space carries an ordinary inner product (defined by $\omega_{CS}$) and a representation of an ordinary matrix algebra of observables. On the other hand, the GNS Hilbert space of a subregion  carries  representations of the quantum group $U_{q}(\mathcal{G})$, which has a more subtle algebraic structure.  These are objects in a ribbon category, which we can identify with the category of anyon ribbon worldlines via a functorial map 
\begin{align}\label{functor}
    \mathcal{F}: \text{Anyon Worldlines} &\Longrightarrow \text{Rep} (U_{q} (\mathcal{G}))\nn
          R &\longrightarrow V_{R}.
\end{align}
 Under this map, worldline processes like fusion and braiding correspond to algebraic operations like the tensor product and the linear map given by the $\mathcal{R}$ matrix. Closing an anyon loop maps to a quantum trace on $V_{R}$. This \emph{categorical} trace is the algebraic mechanism that assigns quantum dimensions to loops, and it underlies the definition of anyonic entanglement entropy.

We now explain how the categorical trace is realized as a partial trace on the maximally entangled state $\ket{i_{R}} \in V_{R}\otimes V_{R}^*$ in the extended state space of the twice punctured sphere. The braket notation should be interpreted with care here: in general we only have a vector space carrying an action of a quantum group, and not necessarily an inner product that distinguishes kets and bras.\footnote{For modular tensor categories, a Hermitian conjugation can be defined \cite{kirillov1995innerproductmodulartensor}, but it is not guaranteed to produce a positive inner product.} To emphasize this, we denote the basis of $V_{R}$ by $e_{i}$, $i=1,\cdots, \dim V_{R}$, and the basis of $V_{R}^*$ by $e^{*j}$. We keep the braket notation for the $q$-deformed tracial state $\ket{i_{R}}$, since the global, gauge invariant Hilbert space does carry a standard Hermitian conjugation and sesquilinear inner product.

In the category language, the invariant state $\ket{i_{R}}$ is the ``coevaluation'' morphism\footnote{This is a basis independent map, whereas the usual Bell pair $\sum_{ij} \delta^{ij} e_{i}\otimes e_{j}$ is basis dependent.} describing the pair creation of anyons:
\begin{align}
\ket{i_{R}}=  \mathtikz{\pairSAnyons{0cm}{1cm} ; \draw  (-0.8cm, 1.2cm) node{\footnotesize $V_{R}$};
    \draw (0.8cm, 1.2cm) node{\footnotesize $V^{*}_{R}$};}
    =\sum_{i,j=1}^{\dim V_R} \delta^{i}_{j}\, e_{i}\otimes e^{*j} .
\end{align}
In Rep$(U_{q}(\mathcal{G}))$ this morphism is an intertwiner\footnote{An intertwiner is a map that commutes with the action of the quantum group.} from the trivial representation to $V_{R}\otimes V_{R}^*$. Its natural dual is the evaluation map 
$\bra{E_R}: V_R^*\otimes V_R\to\mathbb{C}$ 
\begin{align} \label{ER}
    \bra{E_{R}} =  \mathtikz{ \copairSAnyons{0cm}{-1cm};\draw  (-0.8cm, -1.2cm) node{\footnotesize $V^{*}_{R}$};
    \draw (0.8cm, -1.2cm) node{\footnotesize $V_{R}$} ;}: e^{*i}\otimes e_{j} \mapsto e^{*i}(e_{j} ) .
\end{align}
 Together the two satisfy the zigzag identity, which is the categorical description of teleportation \cite{Wong:2023bhs}.

To take expectation values in $\ket{i_{R}}$, however, we need a bra $\bra{i_{R}}$ with the opposite ordering, $V_{R}\otimes V_{R}^*\to\mathbb{C}$. A general monoidal category supplies no such map compatible with $\ket{i_{R}}$. A ribbon category does: thickening the lines into ribbons, we can braid the evaluation map and insert a   double twist of the ribbon  to define:
\begin{align}\label{ibra-main}
\bra{i_{R}}  \equiv   \vcenter{\hbox{\includegraphics[scale=.2]{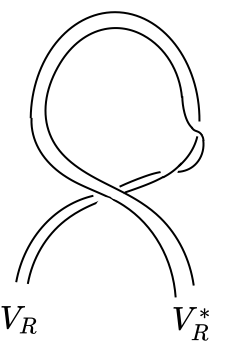}}}   =\vcenter{\hbox{\includegraphics[scale=.08]{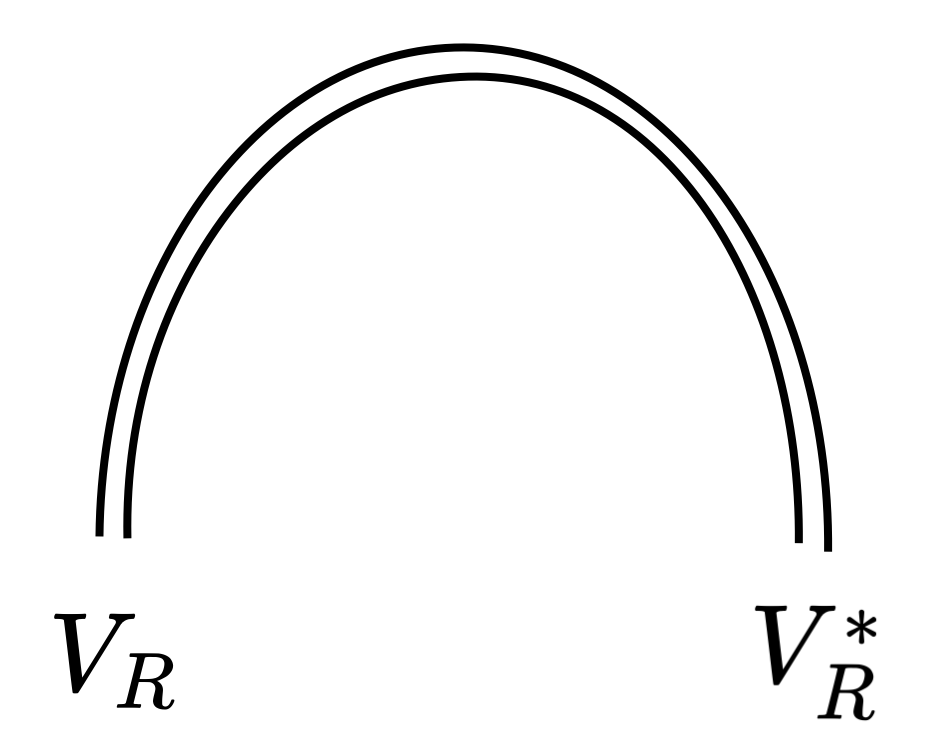}}} : V_{R}\otimes V_{R}^* &\to \mathbb{C},
\nn e_{i}\otimes e^{*j} &\mapsto e^{*j} (D e_{i}),
\end{align}
where $D=\theta\sum_i S(b_i)a_i$ is defined as the combination of ribbon element $\theta$ and the universal $\mathcal{R}$ matrix $\mathcal{R}=\sum_i a_i\otimes b_i$  that describes the double twist and braiding.  This provides a categorical definition of the balancing element.   Sandwiching an operator $O$ on $V_R$ between $\bra{i_R}$ and $\ket{i_R}$ closes the anyon loop and produces the quantum trace
\begin{align}
\braket{i_{R} |O| i_{R}} = \vcenter{\hbox{\includegraphics[scale=.2]{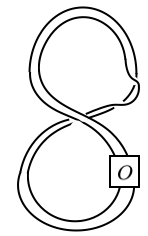}}} =\vcenter{\hbox{\includegraphics[scale=.2]{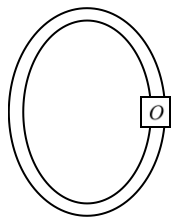}}} = \widetilde{\mathrm{tr}} _{R}\,O \equiv \tr_R(DO).
\end{align}
In particular, the norm of the state is the categorical dimension, which need not be an integer:
\begin{align}
    \braket{i_{R}|i_{R}} = \widetilde{\mathrm{tr}} _{R}1= \dim_{q}R .
\end{align}

\paragraph{Anyonic density matrix}
The categorical partial trace defines anyonic reduced density matrices and entanglement entropies. We use the four building blocks, where we draw each ribbon as a line and denote insertions of  $D$ explicitly by a dot: 
\begin{align}\label{morphisms}
    \ket{i_{R}} = \vcenter{\hbox{\includegraphics[scale=.35]{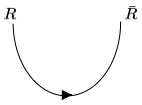}}},\quad   \bra{i_{R}} = \vcenter{\hbox{\includegraphics[scale=.35]{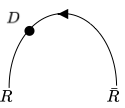}}},\qquad \ket{E_{R}} = \vcenter{\hbox{\includegraphics[scale=.35]{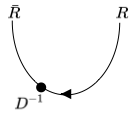}}} \qquad \bra{E_{R}} = \vcenter{\hbox{\includegraphics[scale=.35]{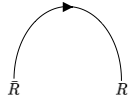}}}
\end{align}
where $\ket{E_R}$ is the coevaluation dressed with $D^{-1}$, chosen so that $\ket{E_R}$ and $\bra{i_R}$ obey the zigzag identity. The arrows on the worldline determine how the diagrams are glued.

The anyonic density matrix of the pair-created state is
\begin{align}\label{abpair}
  \tilde{\rho}= \frac{1}{\dim_{q}R}\ket{i_{R}} \bra{i_{R}} =   \frac{1}{\dim_{q}R}\vcenter{\hbox{\includegraphics[scale=.3]{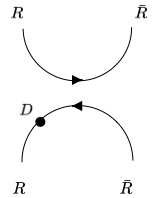}}} ,
\end{align}
where the tilde indicates normalization with respect to the quantum trace. We define the reduced density matrix $\tilde{\rho}_{L}$ on the left factor $V_{R}$ by a right partial trace over $V_{R}^*$ using $\bra{E_R}$ and $\ket{E_R}$:
\begin{align} \label{rho_L}
    \tilde{\rho}_{L}= \widetilde{\mathrm{tr}}_{R^*} \tilde{\rho} = \frac{1}{\dim_{q}R}\vcenter{\hbox{\includegraphics[scale=.3]{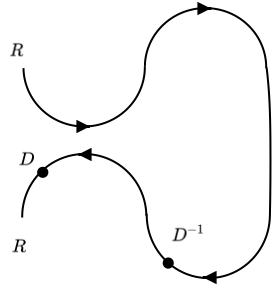}}}= \frac{1}{\dim_{q}R}\vcenter{\hbox{\includegraphics[scale=.3]{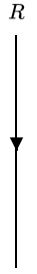}}} .
\end{align}
The balancing elements cancel to leave the identity operator, which is normalized with respect to the left quantum trace on $V_{R}$:
\begin{align}
\widetilde{\mathrm{tr}}_{R} \tilde{\rho}_{L}  =  \frac{1}{\dim_{q}R} \vcenter{\hbox{\includegraphics[scale=.2]{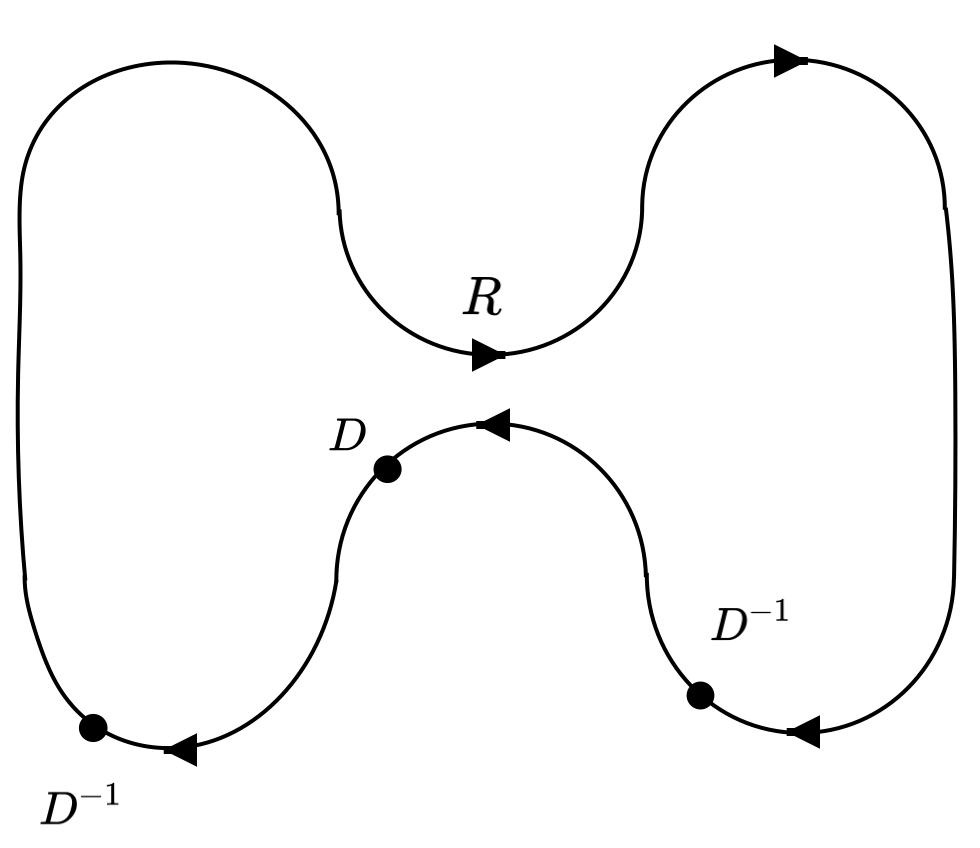}}} =1.
\end{align}
The anyonic entanglement entropy of $V_{R}$ \cite{Bonderson:2017osr}\footnote{The computation of anyonic EE directly from skein theory was done in \cite{Hikami_2008}.} is the von Neumann entropy of $\tilde\rho_{L}$ defined with the left quantum trace:
\begin{align}\label{Stilde-pair}
    \tilde{S}= -\widetilde{\mathrm{tr}}_{R}\, \tilde{\rho}_{L} \log \tilde{\rho}_{L}= \log \dim_{q} R .
\end{align}

\subsection{Fusion Hilbert spaces}
\label{section:operation}
To give the quantum dimension and the $q$-deformed entropy an operational meaning, we need a few facts about anyon fusion. The trivalent vertex
\begin{align}
\vcenter{\hbox{\includegraphics[scale=.2]{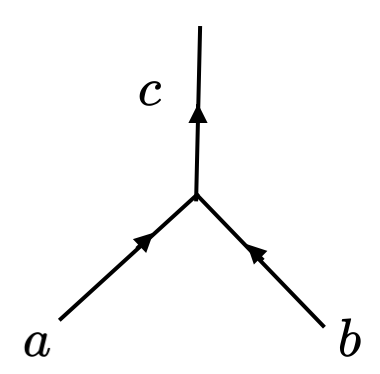}}}
    = V_{ab}^{c}
\end{align}
represents the fusion Hilbert space $V_{ab}^c$, which arises in combinatorial quantization as the Hilbert space of the thrice-punctured sphere. Its dimension is obtained by evaluating the identity operator on the Chern--Simons Haar state:
\begin{align}
\omega_{CS}(1_{V_{ab}^{c}}) = N_{ab}^{\,c},
\qquad
    R_a \times R_b = \sum_c N_{ab}^{\,c}\, R_c ,
\end{align}
where $N_{ab}^c$ is the multiplicity of $R_c$ in the Clebsch--Gordan decomposition of $U_q(\mathcal{G})$. Elements of $V_{ab}^c$ are intertwiners between $V_c$ and $V_a\otimes V_b$. In the isotopy-invariant normalization, a vertex is related to the orthonormal fusion basis by
\begin{align}\label{vertexnorm}
\vcenter{\hbox{\includegraphics[scale=.2]{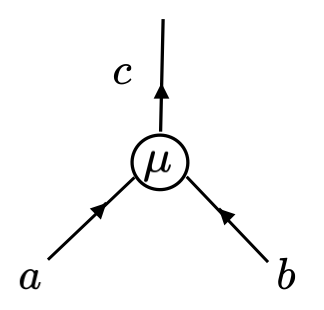}}}
    =
    \left(\frac{d_a d_b}{d_c}\right)^{1/4}
    \ket{a,b;c,\mu},
    \qquad \mu = 1,\dots, N_{ab}^{\,c},
\end{align}
where from here on we write $d_a\equiv\dim_q R_a$.
The associativity of fusion is governed by the \(F\)-matrix, which relates the Hilbert spaces corresponding to different fusion channels.  
It is the matrix implementing the change of basis 
\begin{align}
\includegraphics[scale=.3]{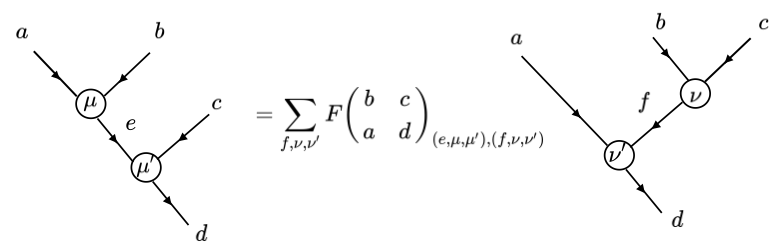}.
\end{align}

These \(F\)-symbols are precisely the quantum \(6j\)-symbols of the quantum group \(U_q(\mathcal G)\).  
They encode the associativity data of the modular tensor category \(\mathrm{Rep}(U_q(\mathcal G))\). We will make use of the special case:
\begin{align}\label{F1c}
F_{c1}\begin{pmatrix} b & a \\ \bar a & b \end{pmatrix} = \sqrt{\frac{d_c}{d_a d_b}} .
\end{align}

\subsection{Demystifying the quantum trace}
Heuristically, the quantum group $U_{q}(\mathcal{G})$ defines the internal symmetry of an anyon.    However, unlike an ordinary particle, an anyon never appears in isolation:  in the Hilbert space picture, this is because Gauss law on a closed surface requires zero total charge, which requires at least a pair of anyons punctures with opposite charge\footnote{In the operator formalism the non local feature of the representations  is manifest in the braiding property of the coproduct }. 

Thus, while representations $V_{a}$ of $U_{q}(\mathcal{G})$ are formally vector spaces with an integer dimension, this dimension has no relation to measurable quantities.   On the other hand, the quantum dimension $d_{a}$ does have an operational meaning that is related to the ordinary dimension on the fusion Hilbert space. 
Consider the fusion of an anyon $a$ with itself $n$ times. This produces a collective Hilbert space given by the fusion tree:
\begin{align}
    V_{aa\cdots a}^{c} = \bigoplus_{b_{i}} \vcenter{\hbox{\includegraphics[scale=.15]{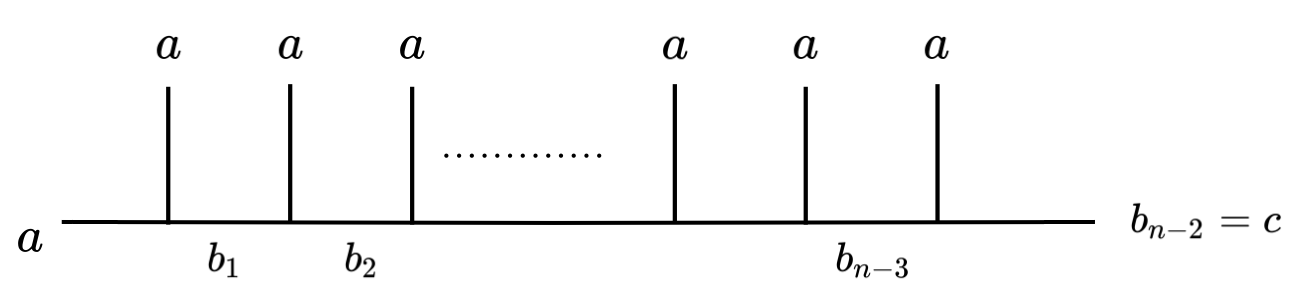}}}
\end{align}

The quantum dimension $\dim_{q}R_{a}$ of the anyon a can be related to the ordinary   dimension of this collective Hilbert space as $n \to \infty$ \footnote{the more precise version of this formula is \begin{align}
  \sum_{c_{i}} N_{aa}^{c_{1} }N_{c_{1} a}^{c_{2}} \cdots N_{c_{m-1}a}^{c_{n}}    \sim  \frac{d_{c}}{\mathcal{D}^2}d_{a}^{n}  \quad \text{as}\quad n\to \infty
\end{align} where $\mathcal{D} =\sum_{a} d_{a}^2$ is the total quantum dimension  }
\begin{align}\label{asympdim}
    \lim_{n\to\infty}\frac1n \log \dim \bigoplus_c V^c_{a^n} = \log d_a .
\end{align}
Similarly, there is a direct relationship between the categorical trace and the ordinary trace  on the physical Hilbert space, which is built out of the fusion spaces $V_{ab}^c$ \cite{Bonderson:2007ci}.  That is, consider the ordinary trace acting on a complete set of operators on  $\oplus_{c} V_{ab}^c$
\begin{align}
\tr \ket{a',b',c; \mu' } \bra{a,b,c; \mu} = \delta_{aa'} \delta_{bb'} \delta_{\mu\mu'}.
\end{align}
The quantum trace on $V_{ab}^c$ differs from the ordinary trace by a factor of  $d_{c}$: 
\begin{align}
    \tilde{\tr} \ket{a',b',c; \mu' } \bra{a,b,c; \mu} = d_{c} \delta_{aa'} \delta_{bb'} \delta_{\mu\mu'},
\end{align}
This can be taken as a definition of the quantum trace \cite{Bonderson:2007ci}. 
We can understand the consistency of this definition diagrammatically by considering the ordinary trace acting on to the projector $1_{ab} $ on $V_{ab}= \oplus_{c} V_{ab}^{c}$:

\begin{align}
     \tr (1_{ab})= \sum_{c, \mu }  \tr (\ket{a,b,c; \mu} \bra{a,b,c; \mu})  = \sum_{c} N_{ab}^c 
\end{align}
and comparing it to the diagrammatic quantum trace on the same operator, obtained by closing the $a,b$ anyon lines: 
\begin{align}
    \tilde{\tr} (1_{ab}) = \vcenter{\hbox{  \includegraphics[scale=.3]{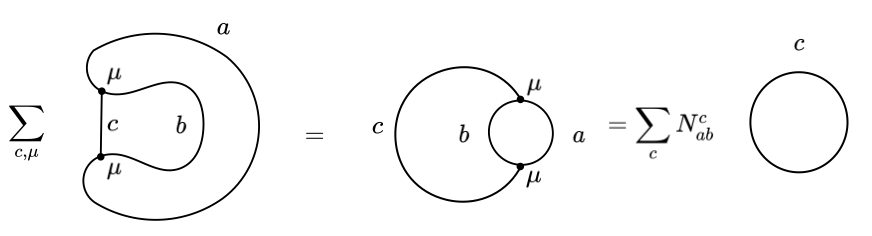}}} = \sum_{c} N_{ab}^c \dim_{q}c
\end{align}
In the third equality, we have used the diagrammatic orthogonality relation in  \ref{app:fusion}.
In the last equality, we see that the trace on each superselection sector labelled by $c$ is enhanced by a factor of $\dim_{q}c$\footnote{  
For completeness, we note that for a  general operator $O$ on the fusion Hilbert space , we can define a quantum partial trace by closing anyon lines on a region $B$ of the fusion graph for $O$.  This differs from the ordinary trace by a ratio of quantum dimensions \cite{Bonderson:2007ci}:
\begin{align}
\widetilde{\tr}_{B} X = \sum_{c,f} \frac{d_{c}}{d_f} [\widetilde{\tr}_{B}X_{c}]_{f}.
\end{align} Here $c$ labels the overall charges of the operator before the partial trace, and $f$ refers to the overal charges after the partial trace.} 
\subsection{Operational meaning of q deformed entanglement entropy }\label{sec:operational}
 \cite{Kato:2013ava} defined a notion of asymptotic entanglement entropy for anyonic systems, defined with ordinary traces, and provided an  operational meaning for it.  Here we adapt some of their ideas to give an operational interpretation to the q-deformed entanglement entropy for a subregion.  

 First let us recall the operational definition of ordinary entanglement entropy.
 Given an entangled state $\ket{\psi}$, its entanglement entropy $S(\psi)$ is the number of bell pairs that can be distilled from n copies  of $\ket{\psi}$, in the limit that $n \to \infty$:
  \begin{align}
    S(\psi)= \lim_{n\to \infty} \frac{\text{number of distillable bell pairs from  }  \ket{\psi}^{\otimes n} }{n}
 \end{align}

This interpretation is consistent with the additivity of entanglement entropy under
the ordinary tensor product: $n$ Bell pairs carry $n$ times the entanglement of one,
as any rate of distillation must. The $q$-deformed entropy has the same property with
respect to the tensor product of $U_q(\mathcal{G})$ representations, because the
quantum trace is multiplicative, $\widetilde{\mathrm{tr}}(X\otimes Y)=
\widetilde{\mathrm{tr}}X\,\widetilde{\mathrm{tr}}Y$.
Hence the $q$-deformed reduced density matrix for n anyon bell pairs is $\tilde\rho_L^{\otimes n}=d_a^{-n}\,1$, and 
\begin{align}\label{Stilde-additive}
\tilde S\big(\tilde\rho_L^{\otimes n}\big) = n\log d_a ,
\end{align}

However, the physical state space in Chern Simons theory does not factorize into a tensor product. Physical states live in the fusion space of Wilson-line anyons, and the fusion space of $n$ pairs is not the tensor product of the fusion spaces of single pairs. As a result, the ordinary entanglement entropy is not additive.  For a single pair, the Hilbert space is trivial.  For two pairs, a factorized  Hilbert space structure is obtained via the $F$-move 
\begin{align}\label{Fmove-2pairs}
\vcenter{\hbox{\begin{tikzpicture}[scale=0.9]
\draw[thick] (0,0) arc (180:360:1 and 0.9);
\draw[thick] (0.5,0) arc (180:360:0.5 and 0.45);
\foreach \x/\l in {0/a,0.5/a,1.5/\bar a,2/\bar a} \node[above] at (\x,0) {\footnotesize $\l$};
\end{tikzpicture}}}
\;=\;\sum_{b,\mu}\left(\frac{d_{b}}{d_a d_a}\right)^{1/2}
\vcenter{\hbox{\begin{tikzpicture}[scale=0.9]
\coordinate (L) at (0.25,-0.45); \coordinate (R) at (1.75,-0.45);
\draw[thick] (0,0) -- (L) -- (0.5,0);
\draw[thick] (2,0) -- (R) -- (1.5,0);
\draw[thick] (L) -- (0.25,-0.75) arc (180:360:0.75 and 0.45) -- (R);
\node[circle,draw,fill=white,inner sep=0.6pt] at (L) {\tiny $\mu$};
\node[circle,draw,fill=white,inner sep=0.6pt] at (R) {\tiny $\mu$};
\node[left] at (0.3,-0.95) {\footnotesize $b$};
\node[right] at (1.7,-0.95) {\footnotesize $\bar b$};
\foreach \x/\l in {0/a,0.5/a,1.5/\bar a,2/\bar a} \node[above] at (\x,0) {\footnotesize $\l$};
\end{tikzpicture}}}
\end{align}
So an entanglement cut that separates the $a$ and $\bar{a}$ charges goes through the $b$ line.  This cut defines a factorized state
\begin{align}\label{psi2}
  \ket{\psi_2} = \sum_{b} \frac{\sqrt{d_b}}{d_a}\sum_{\mu=1}^{N_{aa}^{b}} \ket{a,a;b,\mu}\ket{\bar a,\bar a;\bar b,\mu} \in V_{aa}^{b} \otimes V_{\bar{a}\bar{a}}^{\bar{b}},
\end{align}
and the reduced density matrix on $V_{aa}^{b}$ is
\begin{align}\label{rho2}
  \rho(\psi_2) = \bigoplus_b \frac{d_b}{d_a^2}\, 1_{V^b_{aa}} .
\end{align}
It is maximally mixed within each fusion channel $b$, but the channels are weighted by their quantum dimensions, with probabilities $p_b=N_{aa}^b d_b/d_a^2$.  The ordinary entanglement entropy is
\begin{align}\label{S2}
  S(\psi_2)&=  -\sum_{b} p_{b}\log p_{b} + \sum_{b} p_{b} \log N_{aa}^b\nn
  &= 2\log d_a - \sum_b p_b \log d_b .
\end{align}
The first expression is the familiar Shannon entropy for the probability $p_{b}$ over fusion channels, plus the entropy of the fusion space in each channel.  The second expression comes from plugging in the expression for $p_{b}$
Denoting the q-deformed entropy of a single bell pair by $\tilde{S}$, the total q-deformed entropy of both pairs is \emph{bigger} than the ordinary entanglement entropy.
\begin{align}\label{Stilde2}
  2\tilde S = S(\psi_2) + \sum_b p_b \log d_b .
\end{align}
The extra term $\sum_b p_b\log d_b$ is the conditional entropy of the internal states of the charge $b$ crossing the cut. The quantum trace counts these states with weight $d_b$, but they are invisible in the physical fusion space. 

For $n$ pairs, iterating the $F$-move fuses all the charges on each side into a single fusion tree, joined across the cut by the total charge $c$ (appendix \ref{app:ncopies}):
\begin{align}\label{psin}
\ket{\psi_n} \;\propto\; \vcenter{\hbox{\includegraphics[scale=.3]{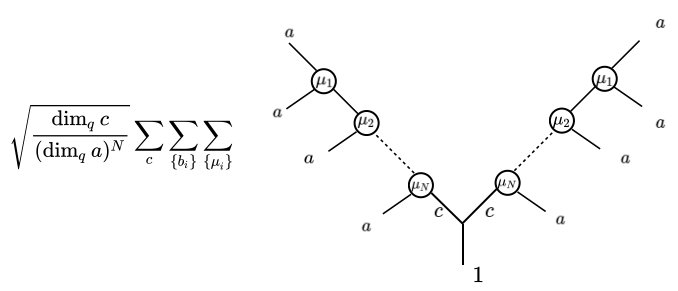}}}
\end{align}
The ordinary entropy is then
\begin{align}\label{Sn-main}
S(\psi_n) = n\log d_a - \sum_c q_c\log d_c ,\qquad q_c=\frac{d_c\dim V^c_{a^n}}{d_a^n},
\end{align}
which again differs from $n\tilde S$ by the conditional entropy of the charge crossing the cut.  

This term is bounded by $\log\max_c d_c$ independently of $n$, so additivity is restored per copy:
\begin{align}\label{AsympS}
  \lim_{n\to \infty } \frac{S(\psi_{n})}{n}=  \tilde{S}= \log d_a 
\end{align}
The $q$-deformed entropy is therefore the asymptotic growth rate of the ordinary entanglement entropy of physical anyon pairs, as well as of the fusion space itself \eqref{asympdim}.

\eqref{AsympS} is the asymptotic entanglement entropy
$E^\infty$  defined in \cite{Kato:2013ava}.  Defining
anyonic LOCC as local operations on each party's anyons that respect charge
superselection, together with classical communication, they showed that the rates of
entanglement distillation and dilution under anyonic LOCC both equal
$E^\infty$. Thus, under
these assumptions, the $q$-deformed entropy defined on a non-physical extended Hilbert space provides a measurement of the
entanglement that can  be distilled from physical anyon pairs.

\section{Topology from entanglement} \label{sec:gluing}

So far, we have applied our framework for factorization and $q$-deformed entropy to the twice-punctured sphere, with a punctured disk as a subregion. In this section, we explain the generalization to subregions of arbitrary topology. Each boundary circle of a general subregion supports edge-mode anyons transforming under $U_{q}(\mathfrak{g})$, and the subregion Hilbert space is built from the anyon state spaces $V_{R}$ of the punctured disk. The simplest example is the annulus, whose classical configuration space and phase space were described in \eqref{FRlink} and \eqref{FRbigon}. The single-link fat graph \eqref{FRlink} connecting the two boundaries forces the edge-mode anyons into dual representations, resulting in the annulus Hilbert space\footnote{The full subregion algebra is the quantum Heisenberg double, which includes both $\mathrm{Fun}_{q}(G)$  and the edge-mode symmetry $U_{q}(\mathfrak g)$ acting on the two boundary circles.}
\begin{align} \label{EREPR}
\mathtikz{\fatannulus{0}{0}} &= \bigoplus_{R}\; \mathtikz{\fatdisk{0}{0}{R}} \otimes \mathtikz{\fatdisk{0}{0}{R^{*}}} \nn
\mathcal{H}_{\rm annulus} &= \bigoplus_{R} V_{R}\otimes V_{R}^{*} .
\end{align}
This is a geometric realization of the $q$-deformed Peter--Weyl theorem. It is the quantum group counterpart of the annulus Hilbert space with gapless boundary conditions\cite{Elitzur:1989nr}, in which $V_{R}$ is replaced by a Kac--Moody module.

Similarly, the Hilbert space of a three-holed sphere is a direct sum over all ways of fusing $V_{R_{1}},V_{R_{2}},V_{R_{3}}^{*}$ into the vacuum, labelled by the fusion spaces of section \ref{sec:categorical}:
\begin{align}
\mathcal{H}_{\rm pants} = \bigoplus_{R_1,R_2,R_3} V_{R_{1}} \otimes V_{R_{2}} \otimes V_{R_{3}}^{*} \otimes V_{R_1R_2}^{R_3} , \qquad \dim V_{R_1R_2}^{R_3} = N_{R_1R_2}^{R_3} .
\end{align}
By entangling the Hilbert spaces for the annuli and three--holed spheres, one can assemble the Hilbert space for arbitrary two-dimensional spatial manifolds.   This implies that the Hilbert space on arbitrary spatial topologies emerges from the entanglement of edge-mode anyons associated to the punctured disk.  Finally, the full 3d topology can be obtained by quantum traces on  topological subregions \footnote{In contrast, the standard TQFT formalism computes amplitudes for a 3D topology from inner products of states on spatial slices, which corresponds to gluing handlebodies}.  It is in this sense that the quantum trace implements the shrinkable boundary condition discussed in \ref{sec:shrinkable}.  In this section, we will make explicit all of the algebraic structure of $L_{q}^2(G)$ needed to reconstruct spacetime topology in this manner, and apply them to the factorization of the torus Hilbert space.  

\subsection{Spacetime ribbons and the algebraic structure of $L^{2}_{q}(G)$} \label{sec:ribbons}
As in the case of the punctured disk, the quantum group operations on a general subregion are best described by anyon diagrammatics.  For the purposes of studying factorization and entanglement, it suffices to focus on the annulus, which  describes the single link of a fat graph crossing the entangling surface as in \eqref{linkcross}.   We represent the annulus by its reduction along $S^1$, which gives an oriented interval. Its time evolution is a spacetime ribbon representing a 3-manifold: 
\begin{align}\label{strip}
 \text{annulus} \times  \mathbb{R} =\mathtikz{\anyonstrip}
\end{align}
    
 We will show that operations on these 2D ribbons suffice to capture the algebraic structure of $L_q^{2}(G)$ needed to produce the full 3d spacetime by quantum traces.    Note that  since the Peter Weyl theorem tells us that $L_q^{2}(G)$ is an object in the representation category of $U_{q}(\mathcal{G})$, these spacetime ribbons are also elements of the ribbon category in the functorial equivalence \eqref{functor}, so they follow the same diagrammatical rules.  This becomes manifest when we decompose the ribbon into an entangled sum over anyon worldlines according to the Peter Weyl theorem:
 \begin{align}
     \mathtikz{\anyonstrip}= \bigoplus_R\mathtikz{\anyonpair{R}}
 \end{align}

\paragraph{Pointwise multiplication}
 $L_q^{2}(G)$ has a Hopf algebra structure defined by the pointwise multiplication of matrix elements $R_{ij}(U)$.  The 2D ribbon  diagram for this is \cite{donnellyhopf}
\begin{align}
    \vcenter{\hbox { \includegraphics[scale=.2]{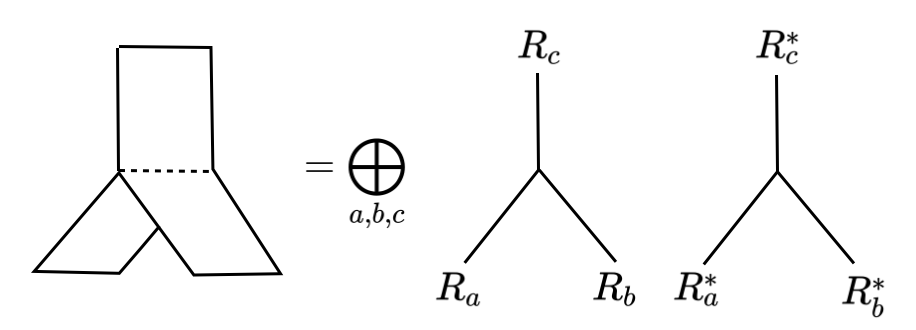}}}
\end{align}
In the representation basis, the multiplication rule is quite complicated, involving an entangled sum of Clebsch--Gordan coefficients in each allowed fusion channel.  A much simpler expression is obtained in the string basis, which is generated by pointwise products of matrix elements in the fundamental (co)representation  
\begin{align}\label{CP}
U_{IJ} = U_{i_{1}j_{1}} \cdots  U_{i_{n}j_{n}} 
\end{align} 
It is useful to treat each $U_{ij}$ as an open string\footnote{In the large N limit, this interpretation gives rise to the A model topological string \cite{Wong:2026ftn,Donnelly:2020teo, Jiang:2020cqo}. } with Chan Paton indices $i,j$, so the string basis consists of a stack of open strings, and the strip in \eqref{strip} is a stack of open string worldsheets.   The Hopf multiplication corresponds to the stacking of these worldsheets, and the unit element $1$ deletes a stack by annihilating the anyon-pairs 
\begin{align}
     \mathtikz{\nablaAflip{0}{0} }: U_{IJ}  \otimes U_{KL} \to  U_{IJ} U_{KL} ,\qquad \qquad \mathtikz{\etaAX{0}{0}}=1
\end{align}
The Hopf algebra structure of $L_q^{2}(G)$ also has a co-product  and an antipode. These satisfy algebraic axioms that capture the properties of $G$.   The ribbon diagrams introduced in this section make these axioms topologically manifest \cite{donnellyhopf}. 
\paragraph{Convolution product}
$L_q^{2}(G)$ also has a convolution product that fuses the string endpoints and dissolves the Chan Paton factors.  This product is part of a ``q-deformed'' Frobenius algebra,  and can be used to entangle Hilbert spaces in Chern Simons theory \cite{Donnelly:2020teo}.    For $q=1$, the Frobenius algebra product is given by the convolution integral:
\begin{align}
\mathtikz{ \deltaA{0cm}{0cm} } :&f_{1}\otimes f_{2} \to  f_{1}*f_{2}, \qquad f_{1}*f_{2}(U) :=\int dg \,f_{1}(g) f_{2}(g^{-1}U)\end{align}
In the representation basis elements, this is
\begin{align} \label{Rprod}
\mathtikz{ \deltaA{0cm}{0cm} } : R_{ij}(U_{V})\otimes R_{kl}(U_{\bar{V}} )\to \frac{ \delta_{jk}R_{il} (U)}{\dim R}.
\end{align}
In fact, the  convolution product just corresponds to group multiplication in the group basis $\ket{U}$.  This is the basis of delta functions on the group, which we can think of as a dual basis that evaluates a function $f$ at the point $U$, e.g.
\begin{align}
    \braket{U|f} =f(U)
\end{align}
In terms of the group basis, the convolution product is just group multiplication: 
\begin{align}
\mathtikz{ \deltaA{0cm}{0cm} } : \ket{U} \otimes  \ket{V } \to  \ket{UV}
\end{align}

For $q\neq 1$, the convolution product is defined by replacing the integral over $G$ with the Haar state $h_{q}$, the pull-back by group multiplication of $g$ with the coproduct \eqref{Rcoproduct}, and the inversion $g\to g^{-1}$ with the inverse antipode $S^{-1}$ \cite{VanDaele2006,Kahng2010}. On matrix elements this reads
\begin{align}\label{qconv}
R_{ij}*R'_{kl} = \sum_{m} h_{q}\big(S^{-1}(R'_{km})\,R_{ij}\big)\, R'_{ml} ,
\end{align}
and evaluating it with the orthogonality relations \eqref{hq} gives
\begin{align} \label{qRprod}
\vcenter{\hbox{\includegraphics[scale=.15]{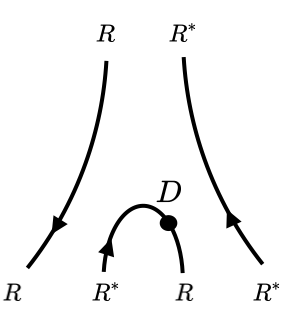}}}=
\mathtikz{ \deltaA{0cm}{0cm} } : R_{ij}(U_{V})\otimes R'_{kl}(U_{\bar{V}} )\to  \frac{\delta_{RR'}\,\delta_{jk}\, (D_{R})_{jj}}{\dim_{q}R}\,R_{il} (U).
\end{align}
In this formula,  we interpret $U_{V}=U\otimes 1$ and $U_{\bar{V}}= 1 \otimes U$.
At $q=1$, it reduces to \eqref{Rprod}; for $q\neq1$, we have to introduce an entry of the balancing element every time we fuse a pair of Chan--Paton indices.
\paragraph{Co-product}
$L_q^{2}(G)$ has a co-product which is shared by both the Hopf algebra and Frobenius algebra.   This defines the  factorization map that cuts the strings.  
 It is given by:
\begin{align} \label{Rcoproduct} \vcenter{\hbox{\includegraphics[scale=.15]{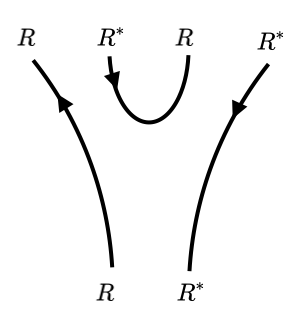}}}= \mathtikz{\muA{0}{0}}U_{ij} &\to  \sum_{k} (U_{V})_{ik} \otimes (U_{\bar{V}})_{kj}\nn
    R_{ij}(U) &\to R_{ij}(U_{V}U_{\bar{V}}) =\sum_{k} R_{ik}(U_{V}) \otimes R_{kj}(U_{\bar{V}}),
\end{align}
where $U=U_{V}U_{\bar{V}}$.
Iterating the coproduct cuts a wavefunction into many pieces, connected by contracted indices that transform under the quantum group. In section \ref{sec:tensor} we interpret this as a tensor network.

 Below, we will use the diagrammatic machinery developed in this subsection to compute q-deformed entanglement entropies of these wavefunctions.   
We will mainly make use of the Frobenius algebra structure associated to \eqref{Rprod} \eqref{Rcoproduct}, which includes a specification of the unit and co-unit: 

\begin{align} \label{units}
\vcenter{\hbox{\includegraphics[scale=.15]{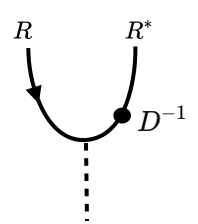}}}=
 \mathtikz{\epsilonA{0}{0}}= \sum_{R} \dim_{q}R\sum_{i} (D_{R}^{-1})_{ii}\, R_{i i}(U),\qquad \vcenter{\hbox{\includegraphics[scale=.15]{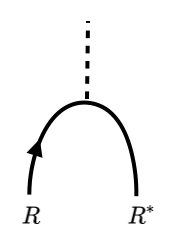}}}=\mathtikz{\etaA{0}{0}}  : R_{ij}(U) \to \delta_{ij}
\end{align}
The unit satisfies 
\begin{align}
\mathtikz{ \epsilonA{.5cm}{-1cm} \deltaA{0}{0}} = \mathtikz{\idA{0}{0}},
\end{align} 
because the $D^{-1} $ in the unit  cancels the $D$ in \eqref{qRprod}.

It is also consistent with the operation that closes up the interval into a loop, which produces quantum trace on the holonomy going around it.
\begin{align}
\mathtikz{\zipper{0}{0} \epsilonA{0}{-1cm}} = \mathtikz{\epsilonC{0}{0} }
\end{align}

Finally, we note that the combined structure of the Frobenius and Hopf algebras on $L_q^{2}(G)$ is a continuum and q-deformed generalization of ZX calculus\cite{coeckeInteractingQuantumObservables2008,coeckeInteractingQuantumObservables2011}, which is a complete categorical language for describing quantum information processing.   We will report on this in a separate publication \cite{wong2026ribbonzxcalculusgauge}.

\subsection{Haar measure and the factorization of the torus Hilbert space}
Consider the factorization of the torus Hilbert space obtained from combinatorial quantization. This is spanned by quantum characters, defined by taking the quantum trace of holonomies around the non contractible cycle of the torus. 
\begin{align}
    \mathcal{H}_{T^2}= \text{span} \{  \braket{U|R}= \widetilde{\tr}_{R} (U) \equiv \sum_{i} (D_{R})_{ii}^{-1} U_{ii} \} 
\end{align}
Note that we have defined the quantum trace with a $D^{-1}$ insertion to match our diagrammatic conventions.  While these quantum traces produce algebra elements, we can identify them with states via the GNS construction. 

\subsection{Factorization along the a-cycle}
Consider cutting the torus along two a-cycles  that link with the Wilson loop insertion. This produces two cylinders as the subregions on which there is  an extended state space $L_q^{2}(G)\times L_q^{2}(G)$. To obtain the corresponding factorization map
we first embed $\mathcal{H}_{T^2} $ inside $L_q^{2}(G)$.  The 2D diagram for this is a ``zipper''
\begin{align} 
\mathtikz{\cozipper{0cm}{0cm}: }:\tilde{\tr}_{R} (U)  \to  \sum_{i=1}^{\dim R} (D_{R})_{ii}^{-1}  R_{ii}(U)
\end{align} 
This diagram can be understood as follows.  The circle which makes up the initial slice of this surface is an anyon moving in a closed loop, which produces the quantum character  $\tilde{ \tr}_{R} (U)$.   The zipper represents a time evolution in which the circle is cut open into an interval, which produces a homomorphism from the algebra of quantum characters into the link algebra $L_q^{2}(G)$. In terms of formulas, this is a trivial unpacking of the quantum trace.  
We can then apply the co-product to factorize  this interval into two. 

This produces the desired factorization map, which is obtained by breaking the holonomy $U$ into two pieces by setting $ U = U_{1} U_{2} $:
 \begin{align} \label{Tfact}
       i=\mathtikz{\cozipper{0}{0} \muA{0}{1cm}}: \mathcal{H}_{T^2} &\to L_q^{2}(G) \otimes L_q^{2}(G)
    \nn
    \widetilde{\mathrm{tr}}_{R}(U)&\to  \sum_{i} (D_{R})^{-1}_{ii} R_{ij}(U_{1}) R_{ji}(U_{2})
    \end{align}

    If we define normalized\footnote{The subregion states $\ket{R ij} $ are normalized with respect to the Haar measure, but only up to a phase.} kets $\ket{R}$, $\ket{Rij}$ with wavefunctions $\braket{U|R}=\tilde{\tr}_{R}(U),$ and  $\braket{U|Rij} =\sqrt{\dim_{q}R} R_{ij}(U)$, then we can write the factorization map as\footnote{To make the diagrams match with 2D TQFT conventions, we have changed the placement of the balancing element $D$ in these diagrams relative to the conventions in \eqref{morphisms}.   }
    \begin{align} 
    i\ket{R} &=
    \frac{1}{\dim_{q}R}\sum_{i} (D_{R})^{-1}_{ii} \braket{U_{V} | R ij} \braket{U_{\bar{V}} R ji }     \nn
    &= \frac{1}{\dim_{q}R}\vcenter{\hbox{\includegraphics[scale=.2]{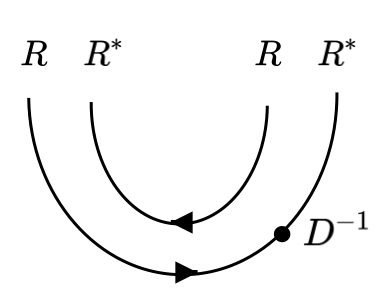}}
    }
    \end{align}
Let's check that this factorization map is an isometry  with respect to the Haar
functional defined by \eqref{h}. First we extend the Haar functional  to the tensor product  $L_q^{2}(G) \otimes L_q^{2}(G)$ componentwise:
\begin{align}
    h: L_q^{2}(G) \otimes L_q^{2}(G) &\to \mathbb{C} \nn
f\otimes g &\to h(f) h(g)
\end{align}If we denote the torus wavefunctions by $\ket{R}$, they satisfy the orthonormality condition
\begin{align}
\braket{R|R'} = \delta_{RR'}
\end{align}
We want to see that this orthonormality is preserved by the factorization.   To do so, we first generalize the definition of dual morphisms in \eqref{morphisms} to
\begin{align}
    \bra{R} i^{\dagger}&= \frac{1}{\dim_{q}R}  \vcenter{\hbox{\includegraphics[scale=.2]{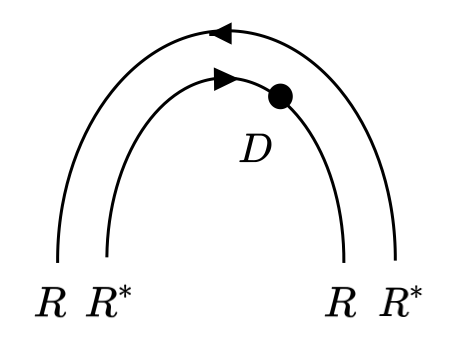}}}\nn
    &= \frac{1}{\dim_{q}R} \sum_{a,b} (D_{R})_{bb} \bra{R ab} \bra{Rba},
\end{align}
where the bra vectors are defined by the q deformed Haar measure \eqref{hq}
\begin{align}
    \braket{Rab|R ij}  = h(R_{ab}^* R_{ij})  = \frac{\delta_{ai}\delta_{bj} (D_{R})_{bb}}{\dim_{q}R}.
\end{align}
Then, the factorized wavefunctions satisfy 
\begin{align}
 \braket{ R| i^{\dagger}i|R'} &=  \sum_{a,b,i,j}  (D_{R})_{ii}^{-1} (D_{R'})_{bb}^{-1} h(R^{'*}_{ab} R_{ij}) h(R^{'*}_{ba}R_{ji}  )\nn 
 &=\frac{\delta_{RR'}}{\dim_{q}R \dim_{q}R'}\sum_{i,j,a,b }  (D_{R})_{ii}^{-1} (D_{R})_{aa}^{-1} \delta_{bj}\delta_{ai}  (D_{R})_{bb} (D_{R})_{ii} = \delta_{RR'}
\end{align}
Note that on a general state in $L_q^{2}(G)\otimes L_q^{2}(G) $, the Haar measure is not positive due to the phase introduced by the balancing element $D$, so it doesn't define a positive inner product.  However, the above calculation provides a  check that it is consistent with the positive inner product on the global, gauge invariant Hilbert space.

Given the state $i\ket{R}$ on the factorized Hilbert space, we can define the q deformed density matrix:

\begin{align}
\tilde{\rho}&= \frac{1}{(\dim_{q}R)^2 } \vcenter{\hbox{\includegraphics[scale=.25]{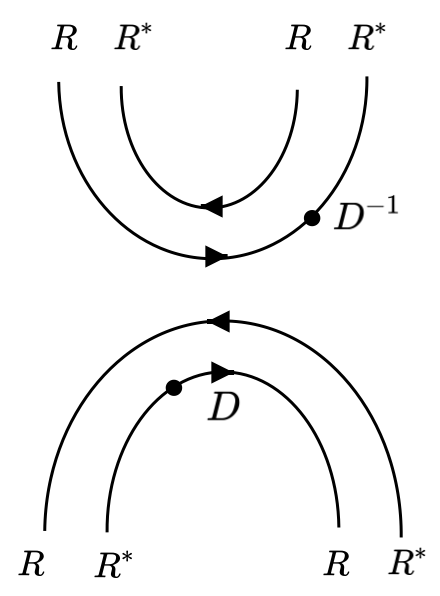}}}\nn
&=\frac{1}{(\dim_{q}R)^2} \sum_{i kab} D^{R}_{bb} (D^{R}_{ii})^{-1} \ket{R ik} \ket{Rki} \bra{Rab}\bra{Rba}
\end{align}
The diagram has the structure of two anyon bell pairs, one for each entanglement cut. 
Note the placement of the balancing elements for the co-evaluation, which has been chosen so that zigzag identity is satisfied by cancelling $D$ against $D^{-1}$. 
This implies that the reduced density matrix is proportional to the identity:
    \begin{align}
    \tilde{\rho}_{L} = \frac{1}{(\dim_{q}R)^2 } \vcenter{\hbox{\includegraphics[scale=.16]{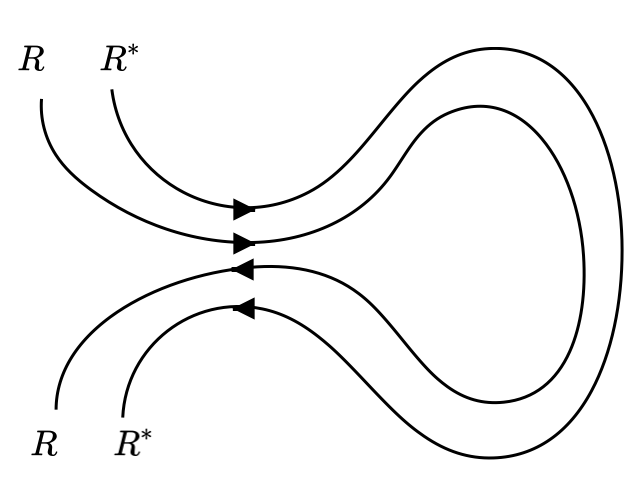}}}= \frac{1}{(\dim_{q}R)^2 } 1_{V_{R}\otimes V_{R}*}
\end{align}
which has a q-deformed entanglement entropy of 
\begin{align}
    \tilde{S} = - \tilde{\tr}_{L} \tilde{\rho_{L}} \log \tilde{\rho_{L}} =2 \log \dim_{q}R.
\end{align}
The factor of 2 corresponds to the 2 disconnected  components of the entangling surface.
\subsection{Factorization along the b-cycle   } \label{haarfact}
Here we consider factorization of the torus Hilbert space along the b-cycle, which illustrates the shrinkability of  our factorization  map as described in the introduction.
\begin{align}
  i: \vcenter{\hbox{
  \includegraphics[scale=.25]{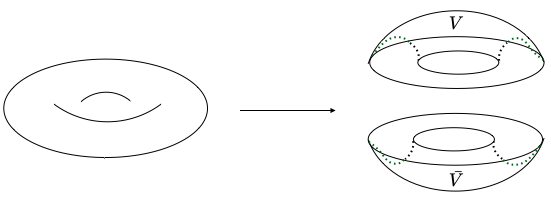}}}
\end{align}
The subregions $V$ and $\bar{V}$ are cylinders  with $L_q^{2}(G)$ state spaces as before.  Let us consider the vacuum state $\braket{U|0}  $, with no Wilson loop insertions.  We interpret this as the analogue of a Hartle Hawking vacuum, which is a maximally entangled state prepared by a Euclidean hemisphere.  In our example, this hemisphere is the one obtained by dimensional reduction along the b cycle, shown in the left diagram below.  To apply the same factorization map  we defined previously, we first go to the dual channel using the modular S matrix:
\begin{align}\label{q2dymS}
 \vcenter{\hbox{ \includegraphics[scale=.2]{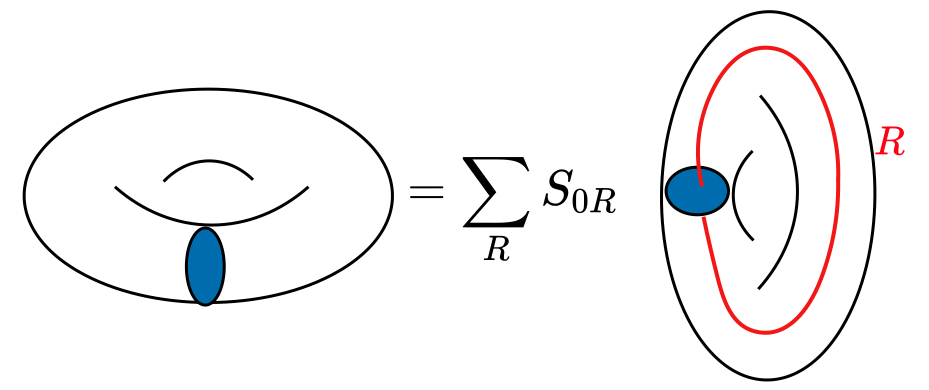}}} \nn \braket{U|0}  = \sum_{R} S_{0R}  \widetilde{\tr}_{R}(U)
 \end{align}
We can then apply the factorization map \eqref{Tfact} for the quantum characters in the dual channel, which produces the factorized state:
\begin{align} \label{BDfact}
(\bra{U_{1}} \otimes \bra{U_{2} }  )i\ket{0 } &=\sum_{R,i} S_{0R}  (D_{R})^{-1}_{jj} R_{ij}(U_{1}) R_{ji}(U_{2}) \nn
&= S_{00} \sum_{R,i,j}    (D_{R})^{-1}_{jj} \braket{U_{V} | R ij} \braket{U_{\bar{V}} R ji }     \nn
&=S_{00} \sum_{R}  \vcenter{\hbox{\includegraphics[scale=.25]{BunchDavies.png}}}
\end{align} 
In the second equality we used $S_{0R}=S_{00} \dim_{q}R$,  and cancelled this $\dim_{q}R$ factor with the  normalization of the kets.

In the conventional path integral picture described in the introduction, the factorization map is produced by taking the path integral $    Z(S^2\times S^1 )=\braket{0|0}$ and integrating out a neighborhood of the entangling surface:
\begin{align}
   \includegraphics[scale=.35]{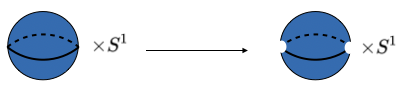}
\end{align}  
The resulting manifold on the right is $S^{2}\times S^1$ with two solid tori removed.   Cutting along the spatial surface then produces a path integral picture of the factorized state:
    \begin{align}\label{pathfact}
 \vcenter{\hbox{ \includegraphics[scale=.45]{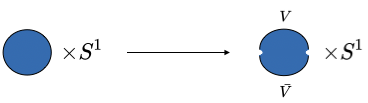}}} = \vcenter{\hbox{\includegraphics[scale=.1
  ]{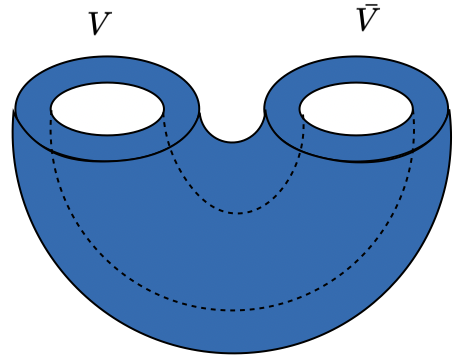}}}
\end{align}
Reducing along the $S^1$ shows a 2D description of the factorized Hartle Hawking state as a half-annulus.  In the quantum group language, this half-annulus is defined as the co-pairing of $L_q^{2}(G)$, produced by entangling anyons as in \eqref{BDfact}
\begin{align}
S_{00}\sum_{R}
\vcenter{\hbox{\includegraphics[scale=.25]{BunchDavies.png}}} = S_{00} \mathtikz{\pairA{0cm}{0cm}}
\end{align}

Since the original, unfactorized state  is given by $\ket{0} = S_{00} \mathtikz{\epsilonC{0}{0}}$, the ``path integral'' factorization of \eqref{pathfact}  corresponds to the following operations on ribbon diagrams:
\begin{align}
\mathtikz{\epsilonC{0}{0cm}} \to  \mathtikz{\epsilonC{0}{-1cm} \cozipper{0}{0} \muA{0}{1cm}} = \mathtikz{\pairA{0}{0}}
\end{align} 
which has purely algebraic interpretation in the quantum group algebra.

The computation of the reduced density matrix proceeds similarly as before, except that we have to sum over representations $R$, and include the appropriate normalization:
\begin{align}
    \tilde{\rho} =  S_{00}^2\sum_{R}  \sum_{i kab} D^{R}_{bb} (D^{R}_{ii})^{-1} \ket{R ik} \ket{Rki} \bra{Rab}\bra{Rba}  =  S_{00}^2 \mathtikz{\pairA{0cm}{.9cm} ;\copairA{0cm}{-.9cm} } 
\end{align}
Doing a quantum partial trace on the right gives us the anyonic maximally mixed density matrix on the left: 
\begin{align}
    \tilde{\rho}_{L} =\tilde{\tr}_{R} \tilde{\rho} = S_{00}^2 \mathtikz{\pairA{0cm}{1cm} ;\copairA{0cm}{-1cm};\pairA{1cm}{.-1cm} ;\copairA{1cm}{1cm}; \idA{1.5cm}{0cm} ; \idA{1.5cm}{1cm} } =S_{00}^2 \mathtikz{\idA{1.5cm}{0cm} ; \idA{1.5cm}{1cm} }=S_{00}^2 \bigoplus_{R} \left(1_{V_{R}\otimes V_{R}^*}\right)
\end{align}
In the last equality, we used a direct sum to emphasize that the reduced density matrix breaks into superselection sectors labelled by the representation of the quantum group.  As in lattice gauge theory, this is a direct consequence of the gauge invariance of the factorized states, which means that the quantum group edge mode symmetry acting at the entangling surface commutes with  $\tilde{\rho}_{L}$. 
For our chosen state and entanglement cut, shrinkability is the 2D diagrammatic relation 
\begin{align}
   Z(S^2\times S^1) =\tilde{\tr} _{L} \rho_{L}  \longleftrightarrow 
 \mathtikz{ \epsilonC{0cm}{0cm} \etaC{0cm}{0cm} }=
    \mathtikz{ \pairA{0cm}{0cm} \copairA{0cm}{0cm} } 
\end{align}

Since the normalization of the Chern Simons path integral gives  $ Z(S^2\times S^1)=1$,  there are factors of $S_{00}^2$ that we have cancelled out on both sides of the diagrammatic equation above.  
Explicitly, one finds that a match between 
\begin{align}
Z(S^2\times S^1) = S_{00}^{2} \mathtikz{\etaC{0}{0};\epsilonC{0}{0}} = 1,
\end{align}
and
\begin{align} 
\tilde{\tr} _{L} \rho_{L}=& S_{00}^2 \mathtikz{\pairA{0cm}{0cm} \copairA{0cm}{0cm} } \nn
    &= S_{00}^2\sum_{R}( \dim_{q}R)^2\nn
    &=1
\end{align}
In the last equality we used the fact that 
\begin{align}\label{S00}
S_{00}^{2}= \frac{1}{\sum_{R} (\dim_{q}R )^{2}}     
\end{align}
To compute the  q-deformed entanglement entropy, it is useful to write the reduced density matrix in terms of density matrices that are normalized with respect to the quantum trace in each fixed-$R$ sector: 
\begin{align}
 \tilde{\rho}_{L} &=   \bigoplus_{R} P(R) \left( \frac{1}{(\dim_{q}R)^2}  1_{V_{R}\otimes V_{R}^*} \right) \nn
 P(R)&= S_{00}^2 (\dim_{q}R)^{2} = (S_{0R})^2 
\end{align}
Here $P(R)$ is a probability distribution for being in the sector labelled by R. 
The q-deformed entropy then takes a standard form well known from ordinary gauge theory\footnote{This simplifies to $S= -\sum_{R} S_{0R}^2 \log S_{00}^2 = -\log S_{00}^2$} :
\begin{align} \label{entropy}
    S= -  \sum_{R} P(R) \log P(R) +   2\sum_{R}  P(R) \log \dim_{q} R 
\end{align}
The first term is a Shannon entropy for the probability distribution $P(R)$, and the second is the anyonic ``edge mode'' entanglement entropy. This comes from the anyon bell pairs that glue together the spacetime across the entangling surface.
\paragraph{Comment on the $S_{00}^2 $ normalization}

The quantity $\sum_{R} (\dim_{q} R)^2 $ appearing in the denominator of \eqref{S00} is the total quantum dimension of the quantum group $G$.   It plays the role of the volume of the quantum group, since for 
$q\to 1$ and a finite group, we have the well known group theory identity:
\begin{align}
   \sum_{R} (\dim R)^2 = \text{Number of elements in }  G\end{align}
This normalization factor is a feature of the interplay between three things:  the fact that $\braket{0|0} =1 $, that the modular S transform is unitary, and the shrinkability of the factorization map.  Finally, $S_{00}^2$ is exactly the  relative normalization factor in the Haar measure for Chern Simons as compared to q2DYM.  This is an instance of a general pattern, which allows us to interpret the 2D ribbon diagrammatics  in terms of q2DYM partition functions. 
\subsection{Disconnected subregions}
To give a further consistency check of the shrinkable factorization map, we consider the factorization of the torus along the b-cycle into disconnected subregions $V= A_{1}\cup A_{2}$ and $\bar{V} = B_{1}\cup B_{2} $ 
\begin{align}
    \includegraphics[scale=.3]{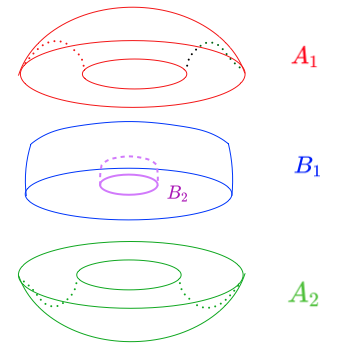}
\end{align}

The factorization map is defined by pulling back the quantum group wavefunction  $\tilde{\tr}_{R}(U)$ via  the multiplication map
\begin{align}
G \otimes G \otimes G \otimes G &\to G  \nn
    (U_{A_{1}} ,U_{A_{2}},U_{B_{1}},U_{B_{2}}) &\to U_{A_{1}} U_{A_{2}}U_{B_{1}}U_{B_{2}} 
\end{align}
This factorizes the quantum character
\begin{align}
    \tr_{R}(U) \to \sum_{i,j,k,l} (D_{R}^{-1})_{ii} R_{ij}(U_{A_{1}})R_{jk}(U_{A_{2}})R_{kl}(U_{B_{1}})R_{li}(U_{B_{2}})
\end{align}
Applying this factorization map to the state $\ket{0}$ gives:
\begin{align} \label{factket}
   i \ket{0}= S_{00}^2  \sum_{R} \frac{1}{(\dim_{q}R)}\sum_{i,j,k,l} (D_{R}^{-1})_{ii} \ket{R ij} \ket{Rjk} \ket{Rkl} \ket{Rli} 
\end{align}

In the path integral language, this is given by:
\begin{align}
    \vcenter{\hbox{
\includegraphics[scale=.5]{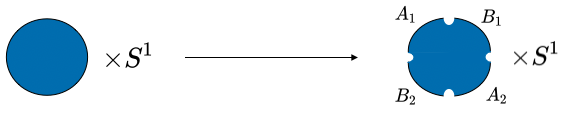}}}=\vcenter{\hbox {  \includegraphics[scale=.1]{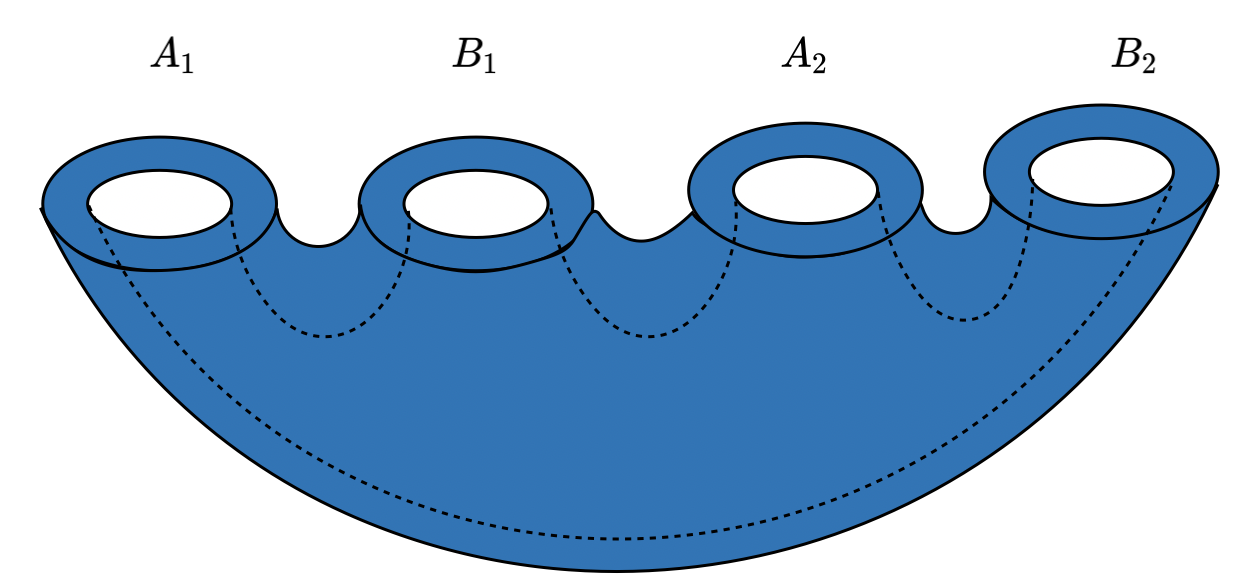}}}
\end{align}
To understand how to define the reduced density matrix, consider the path integral on the $S^{2}\times S^1$ geometry with 4 shrinkable toroidal holes introduced around the entangling surface
\begin{align}
  \vcenter{\hbox {  \includegraphics[scale=.4]{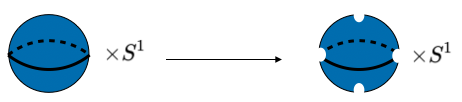}}}  
\end{align}
The reduced density matrix is the operator implementing the modular flow that sweeps out this geometry with the two annular regions $A_{1}\cup A_{2}$.    This flow can be understood in terms of a sequence of 2D diagrams describing the convolution product, co-product, and braiding.  We can visualize this by first considering half of the flow, which evolves the intervals $A_{1} \cup A_{2}$ to $B_{1}\cup B_{2}$.  This takes the form 
\begin{align}
\label{two} 
\vcenter{\hbox{
\includegraphics[scale=.30]{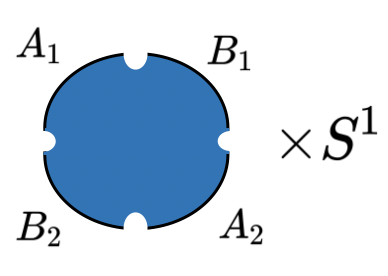}}}=S_{00}^2
\mathtikz{
\idA{0cm}{0cm} \muA{1.5cm}{0cm}
\idA{0cm}{1cm} \leftbraidA{1.5cm}{1cm}
\deltaA{0.5cm}{2cm} \idA{2cm}{2cm} 
} \times S^1
\end{align}
On the RHS, we represented the half modular flow in terms of subregion operations corresponding to the product, co-product, and braid.   In this operator representation of the factorized wavefunction, the 
 $\frac{1}{\dim_{q}R}$ factor  in \eqref{factket} comes from the normalization of the product and co product.  

In the full modular flow, we have to evolve $B_{1}\cup B_{2}$ back into $A_{1}\cup A_{2}$. This gives the normalized reduced density matrix: 
\begin{align}
\tilde{\rho}_{V}&=  \mathtikz{
\idA{1.5cm}{0cm} \muA{0cm}{0cm}
\idA{-.5cm}{1cm} \leftbraidA{1cm}{1cm}
\deltaA{0cm}{2cm} \idA{1.5cm}{2cm}
\deltaA{0cm}{-1cm} \idA{1.5cm}{-1cm}
\idA{-0.5cm}{-2cm} \rightbraidA{1cm}{-2cm}
 \muA{0cm}{-3cm} \idA{1.5cm}{-3cm}
} 
\nn
 &= S_{00}^2 \bigoplus_{R}  \frac{1}{(\dim_{q}R)^2} 
1_{V_{R}\otimes V_{R}^*\otimes V_{R}\otimes V_{R}^*} 
\end{align}
To see that the ribbon diagram reduces to the equality, one can trace out the worldline of the interval endpoints, and observe that all the braidings cancel out, leaving behind two anyon  pairs.  
The resulting anyonic entanglement entropy is
\begin{align}
     \label{entropydisc}
    S= -  \sum_{R} P(R) \log P(R) +   4\sum_{R}  P(R) \log \dim_{q} R 
\end{align}
where the probability factor is 
\begin{align}
 P(R)= S_{00}^2(\dim_{q} R)^2 
\end{align}
The factor of 4 on the edge mode term corresponds to the 4 disconnected components of the entangling surface.
\subsection{Iterated factorization and quantum group tensor networks}\label{sec:tensor}
In the introduction, we alluded to local edge modes in Chern Simons theory associated to a factorization map that produces an infinite entanglement entropy.  These edge modes transform under the representation of an ordinary symmetry described by a Kac--Moody algebra, which has an ordinary subregion trace.     In this work, we defined $q$-deformed factorization maps that produce a finite entropy, at the expense of introducing quantum group edge modes.  These edge modes seem a bit exotic, since they have nontrivial braiding properties described by the ribbon diagrams, and come with a categorical trace.   A useful insight into these edge modes can be gained by interpreting the quantum group factorization map from  the perspective of simulating a quantum state \footnote{We thank Will Donnelly for pointing out this analogy to us}.

Given a factorization map $i$ for a QFT we can iterate it to produce a simulation of the original quantum state on a lattice.   This can be given a graphical depiction by representing the quantum state as a matrix product state (MPS). The problem of efficient simulation of quantum states amounts to removing as much entanglement as possible between neighboring sites \cite{White1992}. In combinatorial quantization,  it is more natural to factorize operators: iterating the factorization map on a Wilson loop gives a matrix product operator (MPO), which produces an MPS when it acts on a product state.

The MPO description amounts to a tensor notation for quantum matrices.  We treat the non-commuting matrix elements $U_{ij}$ as  operators, represented by a  tensor with four legs:
\begin{align}\label{Utensor}
U_{ij} = \mathtikz{\tensorU{U}{i}{j}}
\end{align}
The horizontal legs carry the matrix (Chan--Paton) indices, and the vertical legs carry the operator index.  Pointwise products \eqref{CP} compose along the vertical leg,
\begin{align}\label{UUtensor}
U_{i_1j_1}U_{i_2j_2} = \mathtikz{\tensorUU{U}{i_1}{j_1}{i_2}{j_2}},
\end{align}
so vertical order is operator order.  Group multiplication, and hence the coproduct \eqref{Rcoproduct}, contracts horizontal legs:
\begin{align}
    U_{ij} = \mathtikz{\tensorU{U}{i}{j}}, \qquad  V_{ij} = \mathtikz{\tensorU{V}{i}{j}}, \qquad   (UV)_{ij} = \mathtikz{\tensorUV{U}{V}{i}{j}} .
\end{align}
Representation matrix elements $R_{ab}(U)$ will be drawn as boxes with the label $R$. This notation is a basis-dependent version of the spacetime ribbon diagrams, with the stacking operation represented by contracting vertical legs, and convolution product represented by contracting horizontal legs. 

The disconnected factorization \eqref{factket}  is an example of a process that produces a MPO representation of the quantum state.  Cutting the torus into four intervals splits the holonomy as $U=U_{A_1}U_{B_1}U_{A_2}U_{B_2}$, and in this notation the factorized quantum character is a ring of four tensors $R(U_{k})$, joined by contracted horizontal legs and closed by the balancing element.  For the Wilson loop operator $W_{R}\equiv\widetilde{\mathrm{tr}}_{R}(U)$, the unfactorized and factorized operators are
\begin{align}
W_{R}\;&= \mathtikz{\qMPStorus} = \mathtikz{\qMPSnet} ,\label{qMPSunfact}\\[0.5em]
i^{4}(W_{R})\;&= \mathtikz{\qMPStoruscut} = \mathtikz{\qMPSfactnet} .\label{qMPSfact}
\end{align}

   In the MPS language, reducing entanglement means reducing the bond dimension as far as possible, without violating the isometric constraint.  In contrast with the QFT factorization, which gives an infinite bond dimension, our factorization map provides a more efficient simulation by introducing a finite dimensional space of edge modes.  The price to pay is to allow these edge modes to transform under a quantum group.  But notice that the exotic degrees of freedom live only on the contracted, virtual edges of the MPO.  The same type of virtual quantum group edge modes have been applied to ordinary spin chains that transform under ordinary symmetries \footnote{One advantage of the MPS perspective is that the apparent non locality of the shrinkable boundary condition is moot: the state simulation only requires  the choice of a local tensor, and whether this lives on a Hilbert space obtained by quantizing a local boundary condition is irrelevant  } \cite{Couvreur:2022wyn}.
\subsection{Lens spaces }
A Lens space $L(p,q)$ is a 3-manifold that is a circle fibration over a two-sphere.  Here, $(p,q)$ are co-prime integers that specify the Dehn surgery which produces the manifold.  Consider the simplest case of  $L(p,1)$.  The surgery construction implies that the corresponding  Chern Simons partition function $Z(S^2,p)$ is given by an expectation value of modular $S$ and $T$ matrices:
\begin{align}\label{ZS3}
    Z(S^2,p) &= \braket{0| ST^{p} S|0}
\end{align}
The surgery operation $ST^{p}S $  glues the two solid tori $D^{2} \times S^1$ together along the boundary $T^{2}$, producing a manifold with an $S^2$ base and a $S^1$ fiber.  The T transform introduces a twist in the gluing: as we go around the equator of the $S^2$, the transition function that glues together the fiber $S^1$ has a nontrivial winding number $p$.   The $S$ transformation ensures that the nontrivial gluing is occurring on the fiber $S^1$ instead of along the equator of the  base $S^2$. 

To evaluate the matrix element in \eqref{ZS3}, we use the fact that in combinatorial quantization, $T$ acts as a phase $q^{C_{2}(R)} = e^{2 \pi i s_{R}}$ on the quantum characters \cite{Alekseev:1994au}, where $s_{R}$ is the topological spin . We then have
\begin{align}
    Z(S^2,p) &= \sum_{R} S_{0R}^2 q^{p C_{2}(R)}\nn
    &=S_{00}^2 \sum_{R} (\dim_{q}R)^2 q^{p C_{2}(R)}
\end{align}
Note that in path integral quantization, where the torus wavefunctions are Kac--Moody characters, the $T$ matrix element has an additional phase related to the chiral central charge.  As shown in the last equality, the matrix element in combinatorial quantization gives a match with the q2DYM partition function, up to the overall $S_{00}^2$ normalization. 

To see how we can obtain such manifolds via the quantum trace, 
consider the case of $p=1$
\begin{align}
    Z_{CS}(S^{2},1)= \sum_{R} S_{0R}^2 q^{C_{2}(R)}
\end{align}
We can describe the partition function  in term of a twisted ribbon graph:
\begin{align}
Z_{CS}(S^{2},p=1)= 
S_{00}^2\vcenter{\hbox{\includegraphics[scale=.15]{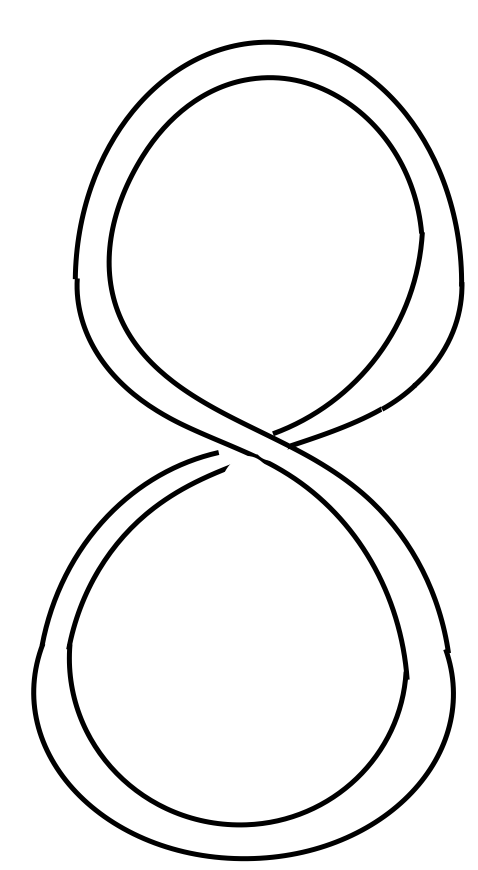}}} &=S_{00}^2 \,\,\vcenter{\hbox{\includegraphics[scale=.13]{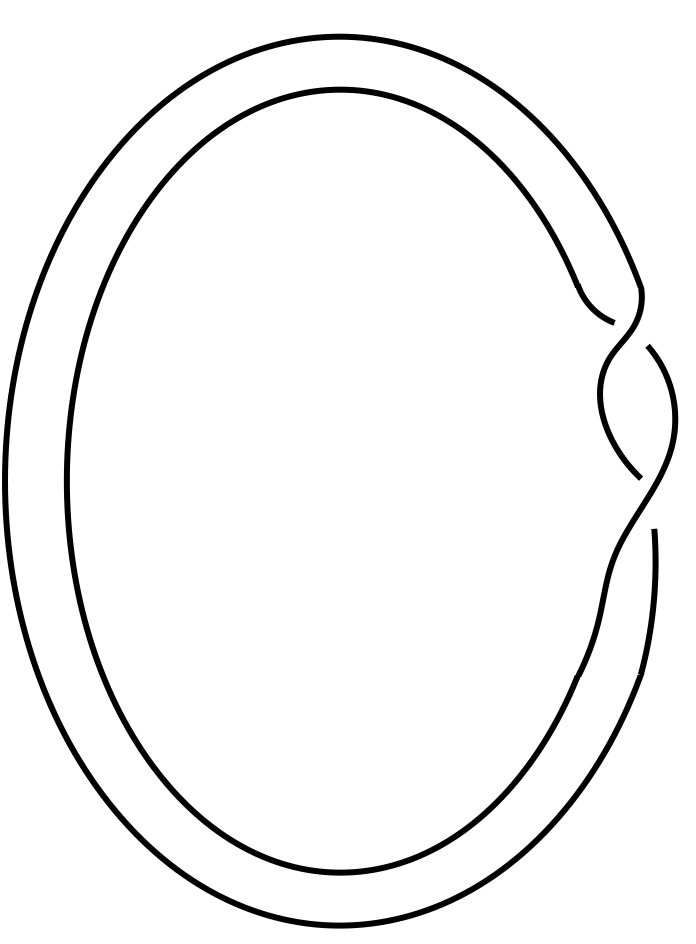}}}\nn
&=S_{00}^2 \widetilde{tr}_{\mathcal{H}_{annulus}} (T)
\end{align}
The ribbon graph describes our subregion perspective, in which the overlap  is interpreted as a quantum trace of the $T$ operator on the annulus. Note the Dehn twist only acts on one boundary of the annulus, reproducing the $q^{C_{2}(R)}$ factor.  Meanwhile, the $(\dim_{q}R)^2$ now comes from the sum over edge modes. 

We interpret $Z_{CS}(S^{2},p=1)$ as the norm of the following (unnormalized) state on the torus:
\begin{align}
   \ket{\psi_{p}}=  \sum_{R} S_{0R} q^{\frac{C_{2}(R)}{2}} \ket{R},
\end{align}
This is possible when q is a phase only if we define an adjoint operation that sends kets to bras, but doesn't conjugate $q$, as was done in \cite{Gabai:2024puk} .  Alternatively, we define the kets and bras with real $q$, then analytically continue the partition function.

Given this caveat, we can apply the same factorization map and quantum partial tracing as before, which results in a reduced density matrix given by the double twist.  
\begin{align}
 \tilde{\rho}_{L} &= S_{00}^2 \mathtikz{\twistA{0}{0} ;\twistA{0}{1cm}  } \nn
   &= \bigoplus_{R} P(R) \left( \frac{1}{(\dim_{q}R)^2}  \theta_{V_{R}\otimes V_{R}^*} \right) , 
\end{align}
where the probability factor is now 
\begin{align}
  P(R)=  \frac{S_{0R}^2q^{2 C_{2}(R)}}{\sum_{R}S_{0R}^2 q^{2C_{2}(R) }} 
\end{align}
The entanglement entropy takes the same generic  form as in \eqref{entropy}: only the  probability factor $P(R)$ is modified.

\section{Applications to topological string theory} \label{sec:topstring}
In this section, we would like to comment on the relevance of our quantum group factorization map in Chern Simons theory to the entanglement entropy computations in the A model topological string \cite{Donnelly:2020teo,Jiang:2020cqo}.   

In section \ref{sec:ribbons}, we alluded to a stringy interpretation of spacetime ribbon diagrams.  Each ribbon is a stack of open string worldsheets, describing the evolution of open strings with Chan Paton factors forming the representation space for $G$.   When $G=U(N)$ is an ordinary group, this produces the Gross--Taylor string  dual to ordinary 2DYM at large N, with a 2D target space equal to the base space $\Sigma_{g,n}$ of 2DYM.   When $G=U(N)_{q}$ with  $q= e^{-g_{s}}$ related to the string coupling $g_{s}$, the large N limit produces the A model topological string living on a Calabi Yau that is a fiber bundle over $\Sigma_{g,n}$ \cite{Aganagic:2004js}.  

In  \cite{Donnelly:2020teo,Jiang:2020cqo}, we leveraged this categorical formulation of topological strings to study the factorization problem in string theory. In particular, we defined a factorization of the A model closed string Hilbert space into open string Hilbert spaces defined by the large $N$ limit of $L_q^{2}(G)$. 
The entanglement cut is implemented by   ``entanglement branes'', which is a large N stack of brane-negative brane pair that cuts the closed strings into open strings.   
\begin{align}
    \includegraphics[scale=.2]{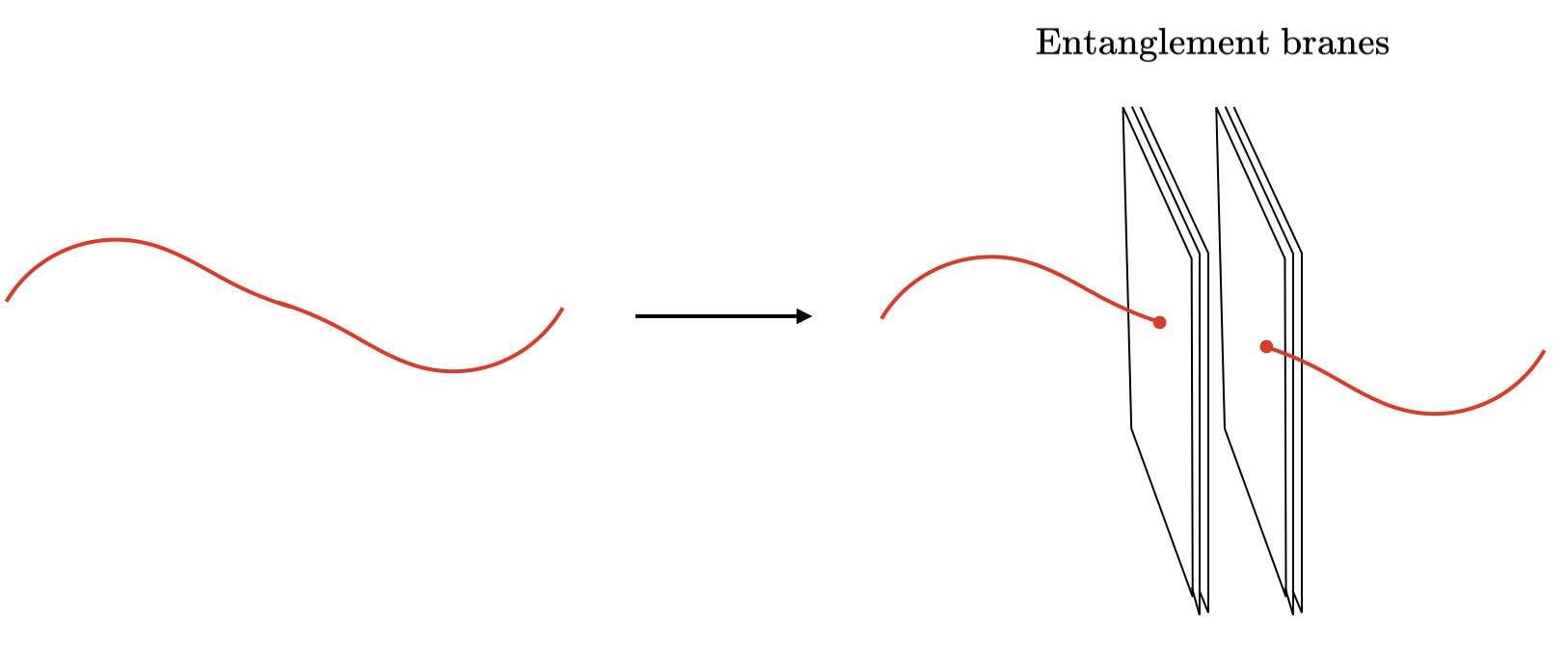}
\end{align}

The entanglement entropy computations in \cite{Donnelly:2020teo,Jiang:2020cqo} made heavy use of spacetime ribbon diagrams, interpreted as a type of $q$-deformed TQFT related to the large N limit of q2DYM.   One of the motivations for this work is to provide an alternative interpretation of this q-deformed TQFT in terms of standard anyon diagrams of a ribbon category.  From this point of view, the string itself is made up of entangled anyons in an appropriate large N limit. 

\begin{figure}[h]
    \centering
    \includegraphics[width=0.8\linewidth]{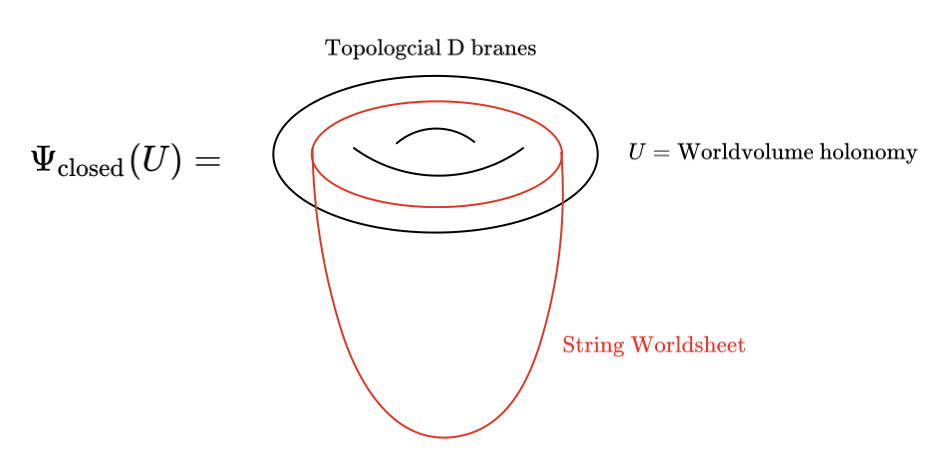}
    \caption{A closed string wavefunction is produced by the topological string amplitude for worldsheets that end on a stack of D branes with the topology of a solid torus.  This is a state in the large $N$ Hilbert space of the worldvolume Chern Simons theory. Combinatorial quantization of the worldvolume theory leads to quantum holonomies which couple to the boundary of the string worldsheet}
    \label{fig:closedwf}
\end{figure}

Furthermore, there is a direct relationship between the topological  string wavefunctions studied in \cite{Donnelly:2020teo,Jiang:2020cqo}, and the Chern Simons theory wavefunctions studied in this paper, which are both formulated in terms of quantum characters.   In \cite{Donnelly:2020teo,Jiang:2020cqo}, the  closed string wavefunctions are constructed via string amplitudes for  worldsheets that end on a stack of $N$
topological D branes, with a boundary that wraps a non trivial cycle on the branes (see figure \ref{fig:closedwf}). The worldvolume theory is $U(N)$ Chern Simons theory at large N, and it was proposed that the coupling of the worldvolume gauge field to the boundary of the worldsheet should produce quantum characters rather than the ordinary  characters. 

In this work, we explained how these quantum characters naturally arise in the combinatorial quantization of the worldvolume theory.   Similarly, the factorization of closed strings can be formulated as a factorization problem on the  worldvolume Chern Simons theory. 
\cite{Donnelly:2020teo,Jiang:2020cqo} showed that the conventional QFT factorization map on the branes does not give the appropriate edge modes consistent with the shrinkable boundary condition in topological string theory.  
Instead, they proposed a factorization map that produces quantum group edge modes transforming under $U(\infty)_{q}$, leading to a ``subregion'' open string Hilbert space   $L^{2}(U(\infty)_{q})$.   

Our work in this paper identified combinatorial quantization as a rigorous mathematical framework behind these stringy edge modes, at finite N.  Indeed, the finite-$N$ version of the closed-string Hartle Hawking state studied in  \cite{Donnelly:2020teo,Jiang:2020cqo}  has essentially the same entanglement structure as the Hartle Hawking state considered in section \ref{haarfact}.    The main difference between the computations in this paper and \cite{Donnelly:2020teo,Jiang:2020cqo}  is the large N limit.      In this work, we developed the operator algebraic perspective $L_q^{2}(G)$, which characterizes the Hartle Hawking state as a $q$-tracial state.  This perspective  will provide a useful starting point for understanding the large N limit taken in \cite{Donnelly:2020teo,Jiang:2020cqo} .
 In this limit, the  q-tracial property of the closed Hartle Hawking string  state  can be interpreted as a geometric transition in which the entanglement branes dissolve into closed string fluxes \cite{wonglocal,Wong:2026ftn}.   This provides the  operator algebraic interpretation of the closed string Gibbons--Hawking entropy computed in \cite{Donnelly:2020teo,Jiang:2020cqo}.

\section{Conclusion}
In this work, we explained how the entanglement structure of compact Chern Simons theory arises from the representation category data of the associated quantum group.   In  particular, we applied the framework of combinatorial quantization to define  a factorization map that leads to a finite entanglement entropy.   This quantization scheme produces a subregion operator algebra on a cylinder given by Fun$_q(G)$, the q deformation of the space of functions on a quantum group G.    We developed a spacetime ribbon calculus describing the algebraic structure of the associated GNS Hilbert space $L_q^{2}(G)$ and used this to describe the entangling of spatial subregions.   The resulting $q$-deformed entanglement entropy is defined in terms of the quantum trace, which is an operator realization of the categorical trace on the representation category Rep$U_{q}(\mathfrak{g})$.   We explained how this entropy gives an operationally meaningful measure of the entanglement between anyonic edge modes. 

The main motivation for this work is to prepare the background needed for generalizing this story to topological string theory and low dimensional gravity.   The former involves taking the large N limit of the representation category for $U(N)_{q}$, while the latter involves  non-compact quantum groups with certain positivity constraints.  We expect the quantum trace to play an important role in both cases.   

In topological string theory, the quantum group edge modes associated to  $U(\infty)_{q}$ are directly related to microscopic degrees of freedom corresponding to topological D branes at a particular value of the worldvolume holonomy \cite{Donnelly:2020teo,Jiang:2020cqo}.   In this story, a crucial role is played by a large N limit of the quantum trace, which implements a geometric transition of the entanglement branes.  This is a duality that realizes the shrinkable boundary condition in string theory.   Our work gives an operator-algebraic perspective on \cite{Donnelly:2020teo,Jiang:2020cqo},  which we will further develop in \cite{Wong:2026ftn}.

In 3d gravity, \cite{Mertens:2022ujr,Wong:2022eiu} defined a bulk factorization map and bulk entanglement entropy which agreed with the Bekenstein Hawking entropy and the holographic entanglement entropy formula for a single interval on the boundary.  We expect the machinery developed in this paper for compact quantum groups will elucidate the mathematical structure behind these computations.   In particular, it would be interesting to understand the role of the quantum trace in the gravitational context.

\appendix

\section*{Acknowledgements}

It is a pleasure to thank  William Donnelly, Yuhan Gai,  Shahn Majid, Thomas Mertens, Joan Simon, Jiaxin Qiao and Qifeng Wu for discussions on this work.    We also thank Thomas Mertens, Qifeng Wu, and Joan Simon for initial collaborations on this work, and William Donnelly for permission to use his tikz macros from previous collaborations.  Claude AI was used as a discussion partner, for the figures and to improve the presentation of the work.
GW is supported by STFC grant
ST/X000761/1, the Oxford Mathematical Institute, and Harvard CMSA.

\section{Balancing element from the Chern Simons path integral} \label{sec:CSholonomy}
In this section we explain how the balancing element of a quantum group arises when we integrate out an empty solid torus.   This arises from a quantum correction to the path integral measure, which leads to quantum shifted holonomies. We largely follow the original derivation in \cite{Elitzur:1989nr}.

In Chern Simons theory, the phase space on a torus consists of flat connections modulo gauge transformations.  The flat gauge field can be expressed as
 \begin{align}\label{AU}
A = -g^{-1} \tilde{d} g + g \theta (t) g^{-1},
 \end{align} 
 where $\tilde{d}$ is a spatial exterior derivative, and the \emph{zero mode} $\theta(t)$ is a spatially constant  Lie algebra valued 1-form
 \begin{align}
     \theta (t) =\theta^{i}_{a} H_{i} \,dx^{a} ,\qquad i=1,\cdots N,\, a=1,2
 \end{align}
Here we assumed that by a constant gauge transformation we  simultaneously diagonalized\footnote{ $\theta_{a}$,$\theta_{b}$ are  simultaneously diagonalizable because they commute as matrices, due to the abelian nature of the fundamental group of the torus. However, they do not Poisson commute. } $\theta_{a},\theta_{b}$,  so $H_{i}$ are generators of the Cartan subalgebra with dimension $N$. 

Since we are quotienting by gauge transformations, the reduced phase space just consists of the zero modes $\theta_{a},\theta_{b} $ which are canonically conjugate.  Naively quantizing these brackets on the reduced phase space gives
\begin{align}
    [\theta_{a}^{i},\theta_{b}^{j}] = \frac{2\pi i }{k}\delta_{ij}.
\end{align}
If this were the end of the story, then as shown in \cite{Elitzur:1989nr}, the insertion of a Wilson loop in the representation $R$ inside the $b$ cycle of the solid torus would induce a holonomy given by the highest weight vector $\mu_{R}$ of the representation:  
\begin{align}
   e^{ \frac{2 \pi i}{k} \mu_{R}} =\text{Diag} (\exp  \frac{2 \pi i}{k}  R_{1}, \exp  \frac{2 \pi i}{k} R_{2},\cdots \exp  \frac{2 \pi i}{k} R_{N})
\end{align}
where $R_{i}$ is the $i$ th column length  of a Young tableau labeling the representation $R$\footnote{ This formula is obtained by interpreting the Wilson line as a charged particle transforming in the representation $R$, and then solving the resulting Gauss law with the charge as a source.} .

However, this reduce-then-quantize method fails to give the correct quantum mechanical system.   This is  because the non-physical degrees of freedom parametrized by $g$ in\eqref{AU} have to be integrated out, taking care to include the appropriate Jacobian in the measure.  This is subtle: a naive computation of the solid torus path integral with Wilson loop insertion just gives infinity, because this is a trace over the infinite dimensional Hilbert space of a punctured disk, with zero Hamiltonian.   The vanishing of the Hamiltonian is a feature of real quantization, where we fix a spatial component $A_{a}$ of the gauge field.   However, a non-zero Hamiltonian can be obtained in holomorphic quantization, which introduces a complex coordinate $z=x_{a}+\tau x_{b}$ and fixes a holomorphic component $A_{z}= A_{a}+ \tau A_{b}$. Specifying $\tau$ introduces a length scale and breaks the topological invariance of Chern Simons theory.   
Integrating out non zero modes in holomorphic quantization then gives a finite Jacobian \footnote{This Jacobian can be computed from the inner product in holomorphic quantization } which has the effect of renormalizing $k \to k+N$.  The net result is equivalent to considering a reduced phase space where 
 \begin{align}
A=\theta_{a} \,dx_{a} +\theta_{b}  \,dx_{b} 
 \end{align} 
with quantum corrected commutators  
\begin{align}
    [\theta_{a}^{i},\theta_{b}^{j} ] = \frac{2 \pi i }{k+N} \delta_{ij}.
\end{align}

$\vec{\theta}_{a}$ and $\vec{\theta}_{b}$ satisfy standard commutators for  position and momenta except these are all periodic variables due to invariance under large gauge transformations, which shift $\vec{\theta}_{a,b}$ by an element of the root lattice $\Lambda_{\text{root}}$ .   Periodic position variable $\theta_{a}$ imposes a quantization on the momenta.  But now the momenta are also periodic, so the position is also quantized, which means the wavefunction has delta function support on a lattice.    

Explicitly, if $\ket{R}$ is the state with  Wilson loop  $R$ inserted, then the ``position space'' wavefunction is 

\begin{align}\label{Rwfn}
\braket{\theta_{a}| R} = \frac{1}{N!} \sum_{\sigma \in S_{N} } (-1)^{sgn(\sigma) } \delta^P \left(\theta_{a}+ \frac{2 \pi i}{k+N} \sigma(\alpha_{R})\right),
\end{align} where
$\alpha_{R}= \mu_{R} + \rho $ is the  highest weight vector shifted by the Weyl vector $\rho$ due to the quantum corrections. It has components 
\begin{align}
    (\alpha_{R})_{i} = R_{i} - i + N +1/2,
\end{align}
and  $\sigma(\alpha_{R})$ is this vector with the components permuted by $\sigma$.  These permutations are residual symmetries from gauge fixing the connection $A$ to belong to the Cartan.  The sum over $\sigma$ anti-symmetrizes the wavefunction under such gauge transformations\footnote{The antisymmetrization comes from the Jacobian factor that appears as an integration kernel in the inner product of the wavefunctionals $\Psi(A_{z})$ in holomorphic quantization. This factor is antisymmetric, and has been absorbed into the holomorphic wavefunction.}.   Finally, the periodic delta function $\delta^P$ is  a generalization of the Dirac delta comb for the root lattice of $U(N)$, which accounts for the periodicities  of $\theta^{a}_{i}$.  This is given by
\begin{align}
    \delta^{P}(\vec{\theta}_{a} - \vec{v})= \sum_{\vec{\alpha} \in \Lambda_{\text{root}}} \delta  (\vec{\theta}_{a} - \vec{v} + \vec{\alpha}))
\end{align}

A much simpler representation is obtained by switching to the holonomy variable $U_{a}= e^{i\theta_{a}}$, which are already invariant under shifting $\theta_{a}$ by root vectors. The wavefunction of $U_{a}$  can be conveniently written as a group theory  delta function: 
\begin{align}
    \braket{U_{a}|R} = S_{00} \delta ( U_{a}, U_{R})
\end{align}
where
\begin{align} \label{U_R}
   U_{R} = \text{Diag}  \{e^{-x_{1}}, e^{-x_{2}} \cdots  e^{ -x_{N}} \} ,\quad 
     x_{i}   = \frac{2\pi i}{k+N} (R_{i}- i + N +1/2)  
\end{align}

\begin{figure}
    \centering
\includegraphics[width=0.5\linewidth]{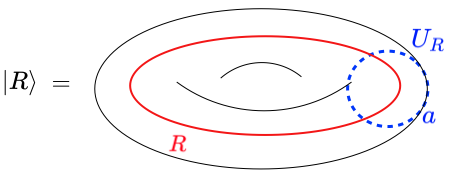}
    \caption{A state prepared with a Wilson loop insertion along the b cycle can also be specified by the holonomy along the dual a cycle }
    \label{fig:holonomy}
\end{figure}

The empty torus corresponds to $R_{i}=0$.  But notice that this doesn't give the identity holonomy.  Instead, it gives the balancing element for $U(N)$ : 
\begin{align}
   D_{ij} = q^{- i + N +1/2}
\end{align}

\section{The Fock--Rosly link bracket for $SL(2)$}
\label{app:FR_SL2}

We evaluate the bracket of a single link holonomy
\cite{Fock:1998nu,Alekseev:1994pa},
\begin{equation}\label{app:linkbracket}
  \{U_1,U_2\}=\kappa\,\big(r\,U_1U_2-U_1U_2\,r_{21}\big),
  \qquad \kappa\propto \frac{2\pi}{k},
\end{equation}
for $G=SL(2)$ in the fundamental representation. We use the basis
$(11,12,21,22)$ of $\mathbb C^2\otimes\mathbb C^2$, with the first label
referring to the first tensor factor, and write
\begin{equation}
  U=\begin{pmatrix}a&b\\c&d\end{pmatrix},\qquad ad-bc=1 .
\end{equation}
Then
\begin{equation}
  U_1=\begin{pmatrix}a&0&b&0\\0&a&0&b\\c&0&d&0\\0&c&0&d\end{pmatrix},
  \qquad
  U_2=\begin{pmatrix}a&b&0&0\\c&d&0&0\\0&0&a&b\\0&0&c&d\end{pmatrix},
  \qquad
  U_1U_2=\begin{pmatrix}a^2&ab&ab&b^2\\ac&ad&bc&bd\\ac&bc&ad&bd\\c^2&cd&cd&d^2\end{pmatrix}.
\end{equation}
The standard $r$-matrix, $r=\tfrac14 H\otimes H+E\otimes F$, and its
transpose are
\begin{equation}
  r=\begin{pmatrix}\tfrac14&0&0&0\\0&-\tfrac14&1&0\\0&0&-\tfrac14&0\\0&0&0&\tfrac14\end{pmatrix},
  \qquad
  r_{21}=\begin{pmatrix}\tfrac14&0&0&0\\0&-\tfrac14&0&0\\0&1&-\tfrac14&0\\0&0&0&\tfrac14\end{pmatrix}.
\end{equation}
Writing $r=t+r_a$ and $r_{21}=t-r_a$, the symmetric and antisymmetric
parts are
\begin{equation}
  t=\begin{pmatrix}\tfrac14&0&0&0\\0&-\tfrac14&\tfrac12&0\\0&\tfrac12&-\tfrac14&0\\0&0&0&\tfrac14\end{pmatrix},
  \qquad
  r_a=\begin{pmatrix}0&0&0&0\\0&0&\tfrac12&0\\0&-\tfrac12&0&0\\0&0&0&0\end{pmatrix}.
\end{equation}
Since $t$ is the Casimir element, $[t,\,U\otimes U]=0$, and
\eqref{app:linkbracket} reduces to
$\{U_1,U_2\}=\kappa\,(r_a\,U_1U_2+U_1U_2\,r_a)$. Explicitly,
\begin{equation}
  \{U_1,U_2\}=\kappa
  \begin{pmatrix}
    0&-\tfrac12 ab&\tfrac12 ab&0\\
    \tfrac12 ac&0&ad&\tfrac12 bd\\
    -\tfrac12 ac&-ad&0&-\tfrac12 bd\\
    0&-\tfrac12 cd&\tfrac12 cd&0
  \end{pmatrix},
\end{equation}
where the entry in row $(ik)$ and column $(jl)$ is $\{U^{ij},U^{kl}\}$.
That is,
\begin{equation}
  \{a,b\}=-\tfrac{\kappa}{2}\,ab,\quad
  \{a,c\}=\tfrac{\kappa}{2}\,ac,\quad
  \{b,d\}=\tfrac{\kappa}{2}\,bd,\quad
  \{c,d\}=-\tfrac{\kappa}{2}\,cd,\quad
  \{a,d\}=0,\quad
  \{b,c\}=\kappa\,ad .
\end{equation}
\section{Quantum groups and their co-representations}
It is useful to think of a quantum group $G$ as a set  of ``quantum matrices'' that represent a linear map
\begin{align} \label{cofund}
    f:V&\to G \otimes V \nn
     v_{i} &\rightarrow \sum_{j}U_{ji} \otimes v_{j}
\end{align}
Here, we have used the tensor product symbol to distinguish scalar multiplication on $V$ with the non commutative multiplication of matrix elements.  This is required because $v_{i}$ satisfy different multiplication rules than the matrix elements $U_{ij}$.   \eqref{cofund} is an example of a co-representation \cite{KlimykSchmudgen1997,Pressley}.

As an example, consider the quantum group $SL_{q}(2)$. Its  coordinate algebra is generated by the matrix elements $(a,b,c,d)$ of the quantum matrix 
\begin{align}
U= \begin{pmatrix} a & b\\ c & d\end{pmatrix}
\end{align} 
They satisfy the commutation relations
\begin{align} \label{sl2commutatorsapp}
ab&= q^{1/2} ba, \quad  ac= q^{1/2} ca,\quad  bd= q^{1/2} db,\quad cd= q^{1/2} dc\quad  \nn
bc&=cb,\quad ad-da = (q^{1/2}-q^{-1/2}) bc.
\end{align}
Additionally we impose the condition
\begin{equation} \label{sl2determinant}
ad-q^{1/2} bc = 1
\end{equation}
which is the $q$-deformed version of the condition $\det U = 1$.

The nontrivial commutators  imply that a,b,c,d should be viewed as operators, so they are matrices on a vector space $\mathcal{V}$ ( not the same as $V$ ! ).  Thus if we want to represent the quantum matrix $U$ as a tensor, it would have 4 legs instead of 2:
\begin{align}
    \includegraphics[scale=.4]{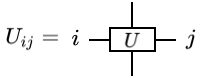}
\end{align}
Then pointwise multiplication looks like:
\begin{align}
     \includegraphics[scale=.4]{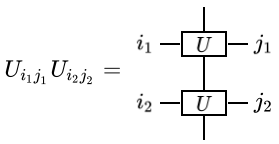}
\end{align}
The representation matrices obtained by anti-symmetrizing are:
\begin{align}
     \includegraphics[scale=.4]{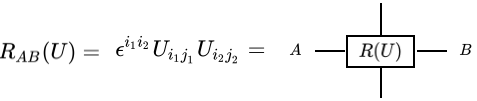},
\end{align}
where $A,B$ label the anti-symmetrized indices.
\paragraph{$\mathcal{R}$ matrix}
A very explicit way of defining $\mathcal{R}$ can be obtained from the tensor product of quantum matrices:
\begin{align}
    U_{1}= U \otimes I,\quad U_{2}= I \otimes U .
\end{align}
The commutation relations  of the  $U_{ij}$ are equivalent to

\begin{align}
\mathcal{R} U_{1} U_{2}  \mathcal{R}^{-1} = U_{2} U_{1}
\end{align}

For $SL_{q}(2)$, $U_{1}, U_{2}$ , and $\mathcal{R}$ are 4 by 4 matrices, with $\mathcal{R}$ given by

\begin{align} \label{Rmat}
\mathcal{R} = \begin{pmatrix} 
    q && 0       && 0       && 0 \\
    0 && q^{1/2} && 0       && 0 \\
    0 && q-1     && q^{1/2} && 0 \\
    0 && 0       && 0       && q
    \end{pmatrix}.
\end{align}
Finally, note that abstractly, we can define a universal R matrix as an element 
\begin{align}
   \mathcal{R} = \sum_{i} a_{i}\otimes b_{i} \in \mathcal{A}_{q}\otimes \mathcal{A}_{q}
\end{align}
which acts on all representations of the quantum group.  The matrix \eqref{Rmat} is just $\mathcal{R}$  acting in the fundamental representation, and can be decomposed as: 
\begin{align}
\tilde{R} &= q \amat \otimes \amat + \sqrt{q} \amat \otimes \emat \\
&\quad + (q - 1) \cmat \otimes \bmat + \sqrt{q} \emat \otimes \amat + q \emat \otimes \emat.
\end{align}

\section{Anyon diagrammatics for Rep$(U_q(\mathcal{G}))$} \label{app:anyon-diagrammatics}
This appendix collects the diagrammatic rules and derivations used in section \ref{sec:categorical}. Our conventions follow \cite{ReshetikhinTuraev1990,Bonderson:2007ci}, and we write $d_a\equiv\dim_q R_a$ throughout.

\subsection{Fusion rules} \label{app:fusion}
With the vertex normalization \eqref{vertexnorm}, the fusion basis is orthonormal and complete on $V_{ab}=\bigoplus_c V_{ab}^c$. Diagrammatically:
\begin{align}
\braket{a,b;c,\mu| a,b;c',\mu' } &= \delta_{\mu \mu' } \delta_{cc'},&
\vcenter{\hbox{\includegraphics[scale=.1]{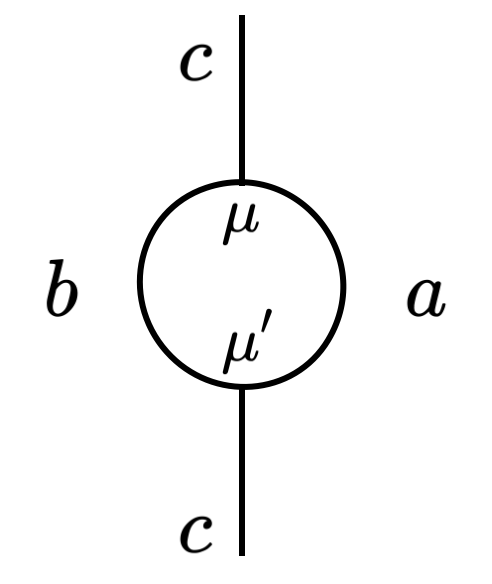}}} &= \delta_{\mu \mu' }   \delta_{cc'} \left( \frac{d_{a} d_{b}}{d_{c}} \right)^{1/2} \vcenter{\hbox{\includegraphics[scale=.1]{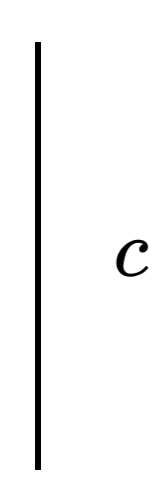}}} , \label{ortho}\\
\sum_{c,\mu} \ket{ a,b;c,\mu }\bra{a,b;c,\mu  } &= 1_{ab},&
\sum_{c,\mu}  \left( \frac{d_{c} }{d_{a} d_{b} } \right) ^{1/2}\vcenter{\hbox{\includegraphics[scale=.1]{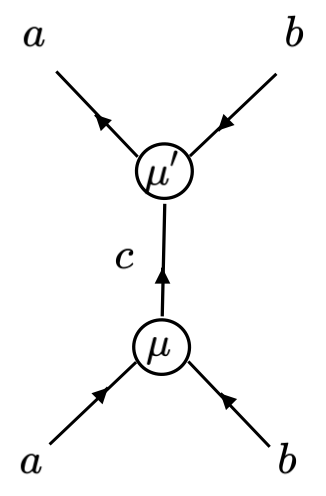}}} &=  \vcenter{\hbox{\includegraphics[scale=.1]{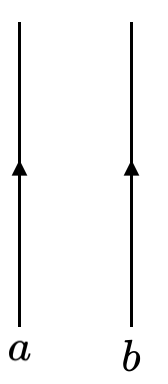}}} . \label{complete}
\end{align}

Associativity of fusion is governed by the $F$-matrix. For three anyons $a,b,c$ with total charge $d$, the fusion space decomposes in two equivalent ways:
\begin{align}
(V_a \otimes V_b) \otimes V_c
    &\supset \bigoplus_{e} V_{ab}^{e} \otimes V_{ec}^{d},\qquad
V_a \otimes (V_b \otimes V_c)
    \supset \bigoplus_{f} V_{bc}^{f} \otimes V_{af}^{d}.
\end{align}
The $F$-matrix implements the unitary change of basis between the two decompositions:
\begin{align}
\vcenter{\hbox{\includegraphics[scale=.3]{Fmove.png}}}
\end{align}
The $F$-symbols are the quantum $6j$-symbols of $U_q(\mathcal{G})$ and encode the associativity data of the modular tensor category Rep$(U_q(\mathcal{G}))$. The special case \eqref{F1c} used in the main text follows from \eqref{complete} applied to the two strands adjacent to a vacuum pair.

\subsection{Bi-partition of  $n$  anyon bell pairs } \label{app:ncopies}
Here we derive \eqref{psin}. By \eqref{vertexnorm} with $c=1$, the pair-creation diagram is $\sqrt{d_a}\ket{a,\bar a;1}$. A diagram of $n$ nested pair creations therefore equals $d_a^{n/2}\ket{\psi_n}$, with $\ket{\psi_n}$ normalized. Applying the $F$-move to the two innermost pairs gives
\begin{align}  \vcenter{\hbox{\includegraphics[scale=.4]{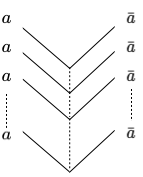}}} &=\sum_{b_{1}}F\begin{pmatrix} a&\bar{a} \\ a& \bar{a} \end{pmatrix}_{1,(b_{1},\mu, \nu)}\vcenter{\hbox{\includegraphics[scale=.4]{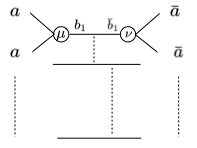}}}\nn
&=\sum_{b_{1},\mu_{1} } \left(\frac{d_{b_{1}}}{d_a d_a}\right)^{1/2} \vcenter{\hbox{\includegraphics[scale=.4]{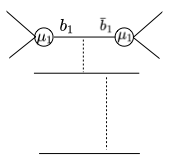}}}\nn
&=\sum_{b_{1},b_{2},\mu_{1},\mu_{2} } \left(\frac{d_{b_{1}}}{d_a d_a}\right)^{1/2}  \left(\frac{d_{b_{2}}}{d_{b_{1}} d_a}\right)^{1/2}\vcenter{\hbox{\includegraphics[scale=.4]{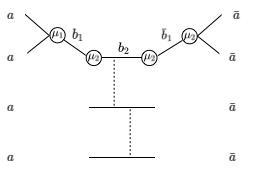}}}
\end{align}
Each further $F$-move fuses one more pair into the running total charge. After $n-1$ moves, the product of coefficients telescopes:
\begin{align}
\prod_{k=1}^{n-1}\left(\frac{d_{b_k}}{d_{b_{k-1}}d_a}\right)^{1/2} = \left(\frac{d_c}{d_a^{\,n}}\right)^{1/2},\qquad b_0\equiv a,\quad b_{n-1}\equiv c,
\end{align}
and we obtain a state in the factorized fusion Hilbert space,
\begin{align}
     \vcenter{\hbox{
      \includegraphics[scale=.3]{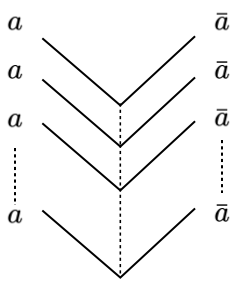}}}= \vcenter{\hbox{
      \includegraphics[scale=.3]{factorizedFH.png}}} \label{factorizedFH-eq}
\end{align}
with $N\equiv n$ copies. 
\bibliography{3Dgrav}
\bibliographystyle{JHEP}
\end{document}